\documentclass[preprint2]{aastex}  

\shorttitle{Dynamics and star formation in SB galaxies}
\shortauthors{Mart\'{\i}nez-Garc\'ia et al.}

\begin{document}

\title{THE RELATION BETWEEN DYNAMICS AND\\ STAR FORMATION IN BARRED GALAXIES}

\author{Eric E. Mart\'inez-Garc\'ia}
\affil{1 Instituto de Astronom\'ia, Universidad Nacional Aut\'onoma de M\'exico, AP 70-264, Distrito Federal 04510, M\'exico.\\
2 Centro de Investigaciones de Astronom\'ia, Apartado Postal 264, M\'erida 5101-A, Venezuela.}
\email{martinez@astroscu.unam.mx; emartinez@cida.ve}

\and

\author{Rosa A. Gonz\'alez-L\'opezlira\altaffilmark{3,4,5}}
\affil{Centro de Radioastronom\'ia y Astrof\'isica, UNAM, Campus Morelia,
     Michoac\'an, M\'exico, C.P. 58089}
\email{r.gonzalez@crya.unam.mx}

\altaffiltext{3}{Visiting astronomer at Kitt Peak National Observatory, National Optical
Astronomy Observatory, which is operated by the Association of Universities for
Research in Astronomy (AURA), under cooperative agreement with the National
Science Foundation.}
\altaffiltext{4}{Visiting astronomer at Cerro Tololo Inter-American
Observatory, National
Optical Astronomy Observatory, which is operated by the AURA, under contract
with the National Science Foundation.}
\altaffiltext{5}{Visiting astronomer at Lick Observatory, which is
operated by the
University of California.}

\begin{abstract}
We analyze optical and near-infrared data of a sample of 11 barred spiral galaxies, in order
to establish a connection between star formation and bar/spiral dynamics.
We find that 22 regions located in the bars, and
20 regions in the spiral arms beyond the end of the bar present 
azimuthal color/age gradients that may be attributed to star formation triggering.
Assuming a circular motion dynamic model,
we compare the observed age gradient candidates with stellar populations synthesis models.
A link can then be established with the disk dynamics that allows us 
to obtain parameters like the pattern
speed of the bar or spiral, as well as the positions of resonance radii. 
We subsequently compare the
derived pattern speeds with those expected from theoretical and observational
results in the literature (e.g., bars ending near corotation).
We find a tendency to overestimate bar pattern speeds derived from 
color gradients in the bar at small
radii, away from corotation; this trend can be attributed to 
non-circular motions of the young stars born in the bar region. 
In spiral regions, we find that $\sim$ 50\% of the 
color gradient candidates are ``inverse'', i.e., with 
the direction of stellar aging contrary to that of rotation. 
The other half of the gradients found in spiral arms have stellar
ages that increase in the same sense as rotation.
Of the 9 objects with gradients in both bars and spirals, 6 (67\%) 
appear to have a bar and a spiral with similar $\Omega_{p}$, while 3 
(33\%) do not.

\end{abstract}


\section{Introduction.}

Since the earliest {\it{N}}-body simulations~\citep[e.g.,][]{hoh71} established that bars
can form in spiral disks, much theoretical work has been undertaken in order to understand
their origin and evolution. Theoretical models show that bars can be formed 
as part of the natural development of the system~\citep{ath09}. On the classical view,
bars are mainly supported by elongated orbits (called {\it{x1}} orbits) along the
bar major axis~\citep[][and references therein]{sko02}.
Chaotic orbits, however, can also support bars in disk galaxies~\citep[see, e.g.,][]{vog06a,har11}.
Given that the stars that make up bars remain most
of the time within them, bars, unlike spiral arms, are not density waves
(Sparke \& Gallagher 2007; see also \S~\ref{spiral_origin}). 
Bars, however, can also be considered as long-lived waves and ``normal'' modes in the disk,
possibly driving spiral structure~\citep[][]{but96}.

It has been predicted that single large-scale (``fast'') bars commonly end inside corotation~\citep[CR,][]{con80a,con80b,sell81},
also called Lagrange radius~\citep[see, e.g.,][]{sell93}. This prediction has also been corroborated
by observations~\citep[e.g.,][]{merr95,ger99,deb02,ague03,fat09,gab09,cor10}.
From these studies, nonetheless, it can also be inferred that the bar's corotation radius\footnote{
\citet{cev07} show that corotation particles are actually located in a ``wide'' ring.}
lies between 1 and 1.4 times the bar's semimajor axis, and not exactly 
where the bar ends~\citep{ague03,but09,elm96}.
This may explain why some galaxies (e.g., NGC 1300, NGC 1365, NGC 5236) show dust lanes on the inside 
of the spiral arms and HII regions on the outside. {\it If bar and arms
have the same pattern speed,} this makes sense only while spirals
lie within corotation~\citep{rob79}. A crossing of the dust lanes from the inside to the outside 
of the arms, marking the corotation position, or an arm bifurcation\footnote{Bifurcations are not
a unique signature of CR, and can also be triggered by other resonances (e.g., 4:1).}
are also sometimes observed.

Another theory postulates that ``slow bars''\footnote{Those for which the CR radius
is larger than 1.4 times the bar's semimajor axis,~\citep[e.g.,][]{ague03}.}
actually end near the inner Lindblad resonance ~\citep[ILR;][]{pash94},
and that they are formed by the alignment of elongated and oscillating orbits along
the potential~\citep{lynd79}.
\citet{com93} on the other hand, based on the results of numerical models~\citep[see also][]{rau05},
propose that bars in early Hubble type galaxies (those with high bulge to disk mass ratio)
end near CR, whereas those in late type galaxies (with low bulge to disk mass ratio)
may end near the ILR, where the spiral arms begin~\citep{elm89}.
Nevertheless, ``slow'' rotating bars are not favored by the comparison of 
simulations and observations~\citep{sch84,sch85,wei01,dub09}, probably implying that
disks in barred galaxies are maximal~\citep{deb00}.
\citet{but09} apply the potential-density phase-shift method~\citep{zha07}
to a near-infrared (NIR) subsample of the Ohio State University Bright Galaxy Survey ~\citep[OSUBGS,][]{esk02}, and
also fail to find evidence of the existence of ``slow'' bars.

\citet{sell88} propose that multiple pattern speeds may be common~\citep[see also][]{sell93,pat09},
such that for some barred galaxies two corotation radii, one belonging to the bar
and the other to the spiral arms, may occur simultaneously in the disk.
This would imply that spiral arms are connected to bars only transitorily.
The possibility also exists that the bar and bulge mask the spiral arms at 
small radii; the arms would only start to be seen near the bar's end,
creating the impression of a physical connection where there is none.
Unsharp masking and/or Fourier-based methods 
can be used to search for spiral and bar perturbations in the  
inner parts of galaxies. However, ring-like features at the end of the bar
may interfere with the unambiguous identification of a physical connection 
between bar and arms in some objects.

\citet{rau99} perform simulations that confirm the scenario advanced by \citet{sell88}.
They also find cases, however, where the bar and spiral rotate with the
same pattern speed, as well as other cases where the bar
and spiral patterns have different speeds but coupled resonances~\citep{tag87,mass97}. 
In these cases, the bar's CR may overlap
with the spiral 4:1 resonance or ILR, and the bar
and spiral arms are indeed connected. Various simultaneous spiral modes are also possible.
According to~\citet{but09}, barred galaxies can have ``fast" bars 
(CR at 1 to 1.4 times the bar extent), and spirals rotating
at the same or with a different pattern speed.
They also consider ``super-fast'' bars,
for which the bar's CR is well inside the bar's end and spirals are decoupled,
although so far no theoretical models (e.g., n-body or response) have 
produced a consistent system with CR within the bar.

Observations also indicate that around 20\% of early spiral galaxies are
double-bar or nested bar systems~\citep{lai02,erw04,erw09}.
In these systems, an outer or primary bar harbors an inner or secondary bar in the nucleus.
The inner bar rotation is independent from that of the outer bar.
There have also been suggestions of triple-bar systems~\citep[e.g. NGC 2681,][]{erw99}.

\subsection{Star formation} \label{sec_SF}

It is commonly observed that giant HII region complexes occur at the end of
the bar, where the spiral arms {\it seem} to originate~\citep{rob79,ken91}.
However, not all barred spirals show this feature clearly, and they present
$H_\alpha$ emission in other regions as well, e.g., along the bar~\citep[see][]{gar96}.
\citet{phi93,phi96} proposes that barred galaxies can
be classified according to their star formation properties. Galaxies with Hubble type SBb 
and ``flat''\footnote{The light profile of flat bars decreases with radius
more slowly than the spiral arm profile. Conversely, the radial scale length 
of exponential bars is the same or shorter than that of the spiral arms~\citep{elm85}} bars
have scant star formation in the bar region
and a concentration of HII regions in 
inner rings.\footnote{Inner rings are commonly found at the end of the 
bar~\citep[e.g.,][]{atha09a}. Barred spirals may also show rings in the nucleus
and near the outer Lindblad resonance~\citep{sell93}.
Rings are commonly associated both with resonances~\citep{sch81,sch84,byr94},
and with active and recent star formation~\citep{but96}.}
Star formation activity can be very high in the circumnuclear region.
The second group corresponds to Hubble type SBc and ``exponential''
bar type. This group has luminous HII regions in the bars,
poorly defined rings, and less star formation in the circumnuclear region.
In both SBb and SBc galaxies, the star formation rates per unit area seem to be enhanced in the region where
the bar joins the spiral arms, more noticeably in the SBb group.
\citet{phi96} points out that SBa galaxies 
tend to show star formation in ring structures, but no star formation activity
in the bar and central regions~\citep[see also][]{gar96}.

According to \citet{ath92}, gas density enhancements perpendicular and at the
end of the bar are often seen in simulations. She also notices that, for bars with 
``straight'' dust lanes, star formation is difficult along the bar 
because of considerable shear (the situation may be different for curved lanes/shocks).
These ``straight'' dust lanes occur in strongly barred galaxies~\citep{com09}.
Regardless of this,~\citet{she00} argue that some stars may be formed between the bar's end and
the circumnuclear region. In their model, stars are born in ``dust spurs''
on the trailing side of the bar that feed the main dust lanes.
\citet{zur08}, based on $H_\alpha$, optical, and NIR observations, find evidence to support
this model in the barred spiral NGC 1530. They suggest that stars are formed in the dust spurs and
migrate across the bar to its leading side~\citep[see also][]{asi05,elm09}.

The existence of phase shifts between the $H_\alpha$ emission, the old stellar component of the bar, 
and the CO emission along it 
has been noticed by~\citet{mar97},~\citet{ver07}, and~\citet{she02}.
The $H_\alpha$ leads both the CO and old stars, 
although no systematic pattern with galactocentric radius is measured. These observations 
support the idea that star formation can be triggered in the bar region. 

If spiral arms in barred galaxies trigger star formation, then azimuthal color gradients
must be observed across them~\citep[e.g.][]{rob69,gon96,mar09a,mar09b}, due
to the aging of stars and their velocities relative to the spiral arms.
In the cases where spirals corotate with the bar,
inverse color gradients (i.e., the direction of stellar aging is contrary to the sense of rotation) are
expected.


Age gradients across bars have also been predicted by recent numerical
simulations~\citep{dob10}. Moreover,~\citet{maz08} have detected
azimuthal age gradients in $\sim$ half of a sample of 22 nuclear rings,
that they speculate may be related to enhanced star formation in the
contact points of bar and ring. In the bar region, the mechanisms
triggering star formation may be different from the one that sets off
star formation in spiral arms, due to the diverse dynamics and shock
conditions.

\begin{figure*}
\centering
\includegraphics[scale=0.80]{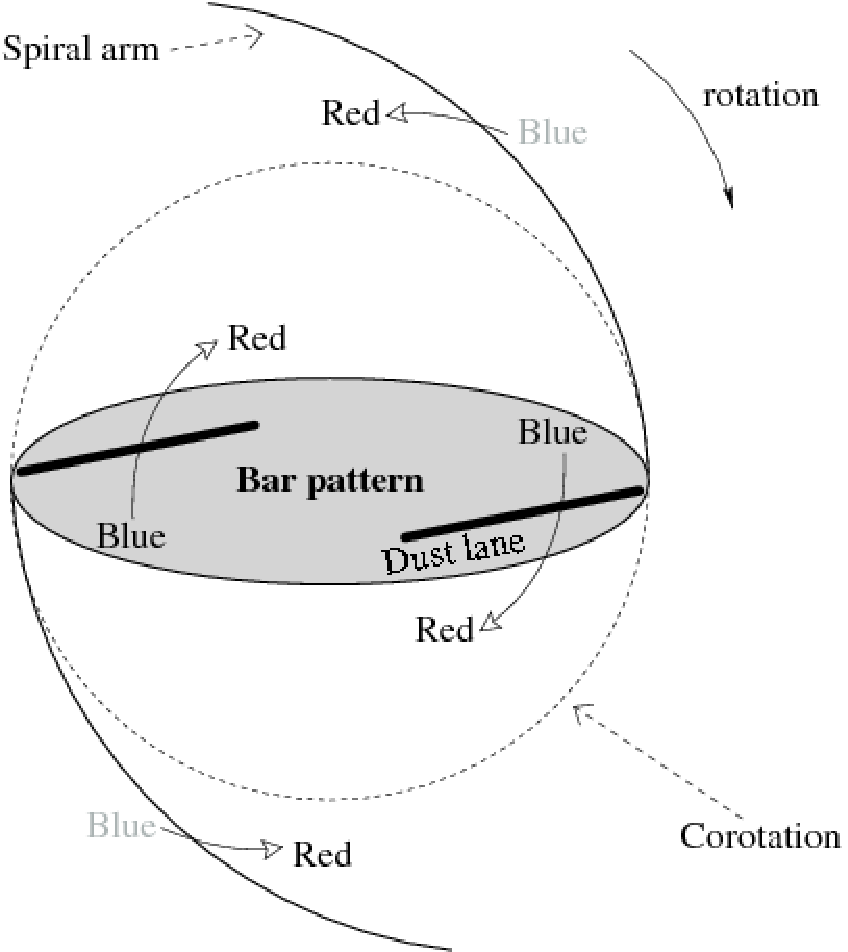}

\caption[f1.eps]{
Schematic of stellar age gradients in a barred galaxy. 
Gradients are shown by arrows that go from blue (young stars)
to red (old stars), across a strong bar with straight dust lanes, 
and through the spiral arms; they are produced by stars born in a shock that then drift away
as they age. In this example, the bar ends near CR and the arms corotate with the bar;
because the observed arms are completely beyond the CR radius, the spiral pattern
overtakes the gas in the disk and the gradients in the arms 
are ``inverse", that is, the stellar aging vector is opposite to the direction of
disk rotation.~\label{bar_grads}} 
\end{figure*}

Here, we aim to investigate azimuthal color (age) gradients across
bars and spirals in SB (or strong SAB) galaxies, and the connection
between these color gradients and bar/spiral dynamics. In
figure~\ref{bar_grads}, we show expected sites of color gradients; in
this example, the bar ends near corotation (CR) and spiral arms corotate
with the bar, which is not necessarily always the case.

\section{Observations}

Our sample consists of 11 barred spirals, 9 of them classified as SB, and two
as SAB in the Third Reference Catalogue of Bright Galaxies~\citep[RC3,][]{dva91}.
Statistics show that $\sim$30\% of spirals are strongly barred in the optical~\citep{deV63,sell93}.
In the NIR,~\citet{esk00} find that $\sim$56\% are strongly barred (SB) and $\sim$16\%, weakly barred (SAB).
Their study also shows that late type spirals (Sc-Sm) hold the same fraction of bars as early types (Sa-Sb).

Optical and NIR data were acquired during 1992-1995 
with four different telescopes in three observatories:
the Cerro Tololo
Interamerican Observatory (CTIO) 0.9-m and 1.5-m,
the Lick Observatory 1-m, and the Kitt Peak National Observatory (KPNO) 1.3-m
telescopes.
Deep photometric images were taken in the optical filters
$g$, $r$, and $i$, and in the NIR
$J$, $K_s$ or $K^\prime$.
Effective wavelengths and widths of all the filters are
listed in table~\ref{tbl-filters}; the observation log for the 11 galaxies
is shown in table~\ref{tbl-obslog}.

The CTIO 0.9-m optical telescope used a Tek $1024^{2}$
and a Tek $2048^{2}$ CCDs, both with a $0\farcs4$  pixel$^{-1}$
plate scale. The CTIO infrared observations were performed at the 1.5-m
telescope, with the Ohio State Infrared Imager/Spectrometer (OSIRIS) in February 1994, 
and with the CIRIM instrument in September 1994 and September 1995. 
Both CIRIM and OSIRIS used
$256^{2}$ NICMOS3 arrays;
the CIRIM focus was adjusted to give a $1\farcs16$ pixel$^{-1}$ plate scale,
whereas OSIRIS was used in the $f3$ image mode, that provided a plate scale of $1\farcs1$ pixel$^{-1}$.
The CCD at the Lick 1-m telescope was a Ford $2048^{2}$, with
a plate scale of $0\farcs185$ pixel$^{-1}$. For the infrared
observations at Lick, the same telescope was fitted with the LIRC-2 camera; it had
a $256^{2}$ NICMOS II detector, with a $1\farcs145$ pixel$^{-1}$ plate scale.
The KPNO infrared observations were made with the IRIM camera, that
employed a $256^{2}$ NICMOS3 array, with a $2\arcsec$ pixel$^{-1}$ plate scale.

\subsection{Data reduction}

The data were properly reduced with the image processing package IRAF\footnote{
IRAF is distributed by the National Optical Astronomy Observatories,
which are operated by the Association of Universities for Research
in Astronomy, Inc., under cooperative agreement with the National
Science Foundation.}~\citep{tody86,tody93} and fortran 77 routines.
In the optical we applied overscan and trimming corrections. We produced a combined
bias frame with the ``minmax'' rejection algorithm and subtracted it from the data.
Dark current frames were inspected, but their counts were found to be negligible.
Twilight flats were compared with dome flats, and the former were found to be better.
A ``master'' flat field image was obtained for each filter by scaling flat fields by their median,
averaging them with a sigma clipping algorithm, and dividing the result by its mean.
Objects were then divided by the appropriate ``master'' flat. 
Bad pixels masks were created with the assistance of both dome and twilight flat images.
Sky subtraction was achieved by masking bright objects and obtaining the median of the
remaining pixels. For some objects the sky emission was not uniform and we fitted a plane,
instead of a constant median value.
To produce mosaics, individual images were superpixelated by a factor of 2 in each
dimension. Each superpixelated frame
was inspected for artifacts (e.g., asteroid traces) and the respective pixels were masked.
The images were then registered to the nearest pixel (i.e., half an 
original pixel), and a median mosaic was obtained. Cosmic rays were located and
masked by comparing the median mosaic with each individual frame. A final mosaic was then obtained
by adding the superpixelated, registered, clean frames.

For the NIR data, we first corrected for the non-linearity of the detector. A polynomial
function was adjusted to each pixel of dome flats of increasing exposure times. 
Source variability and read-out delay time images
were used when available, or obtained via iterations. The flat-field correction
was done analogously to the optical case. 
Sky and object frames were taken in an alternating fashion, and with no more than
a few minutes difference; sky exposures were centered at different positions with
respect to the objects, and separated from them by $\sim$ 10$^\prime$, or about twice the 
linear size of the field-of-view. Bright objects in the sky frames were masked, and the four 
sky exposures closest in time to each object were
median scaled, averaged with a rejection of deviant
pixels, scaled to the mean sky value of the object, and subtracted from it; this process also takes care of dark
current removal. Flat field corrections were applied with master flats derived from dome flats.
Mosaic registering and cosmic ray masking were achieved with the same procedure
used for the optical data.

\subsection{Calibration}
The optical calibration was done in the Thuan-Gunn system \citep{thu76,wad79}.
The zero point of this photometric system is chosen such that
the standard star BD+17$\degr$4708 has $g=r=i=9.5$. Synthetic magnitudes
were obtained for other spectrophotometric standards\footnote{
Feige 15, 25, 34, 56, 92, 98; Kopff 27;
LTT 377, 7987, 9239; EG 21; BD+40$\degr$4032;
and Hiltner 600.}, using the spectral energy distributions in  
\citet{oke83}, \citet{sto77}, \citet{mas88},
\citet{mas90}, \citet{ham92}, and \citet{ham94}, 
and system response curves for each detector/filter/telescope/observatory
combination~\citep[for details, see][]{mar09a}.
The best photometric galaxy frame was selected for each object, and the 
final mosaic was scaled to it.
The NIR $J$, $H$ and $K_{s}$ data were calibrated with frames from the  
Two Micron All Sky Survey~\citep[2MASS,][]{skr97,skr06}.
The $K$ data were calibrated with the~\citet{car95} photometric standard stars, transformed to the CTIO
system~\citep{car90}. For the $K^\prime$ data, we adopt 
$K^\prime  = K + 0.2(H - K)$~\citep{wai92}, and use the
photometric standard stars of~\citet{haw01}.
The optical data were degraded to the resolution of the NIR data.

\begin{deluxetable}{cll}
\tabletypesize{\scriptsize}
\tablecaption{Filter characteristics \label{tbl-filters}}
\tablewidth{0pt}
\tablehead{
\colhead{Filter} & \colhead{$\lambda_{eff}$} & \colhead{FWHM}
}
\startdata
\emph{$g$}       & 5000\AA & ~830\AA \\
\emph{$r$}       & 6800\AA & 1330\AA \\
\emph{$i$}       & 7800\AA & 1420\AA \\
\emph{$J$}       & 1.25\micron & 0.29\micron \\
\emph{$H$}       & 1.65\micron & 0.29\micron \\
\emph{$K_s$} & 2.16\micron & 0.33\micron \\
\emph{$K^\prime$}     & 2.11\micron & 0.35\micron \\
\emph{$K$}     & 2.2\micron & 0.41\micron \\
\enddata
\end{deluxetable}

\begin{deluxetable}{llrccc}
\tabletypesize{\scriptsize}
\tablecaption{Observation Log\label{tbl-obslog}}
\tablewidth{0pt}
\tablehead{
\colhead{Object} & \colhead{Filter} & \colhead{Exposure(s)} & \colhead{Telescope}
& \colhead{Date (month/year)} & \colhead{Mean seeing}
}
\startdata


NGC 718  \dots\dots& $g$     & 8100. & Lick 1 m          & 10/92, 11/92, 9/93 & $2.1\arcsec$ \\
                   & $r$     & 7200. &    ''             & 10/92, 9/93, 11/93 & \\
                   & $i$     & 4200. &    ''             & 10/92, 9/93 & \\
                   & $J$     & 672.  & Kitt Peak 1.3 m   & 11/94 & \\
                   & $K_{s}$ & 336.  &    ''             &  '' & \\	

NGC 864  \dots\dots& $g$      & 3865. & Lick 1 m         & 9/93 & $1.7\arcsec$\\
                   & $r$      & 4695. &    ''            & 9/93, 11/93 & \\
                   & $i$      & 3600. &    ''            & 11/93 & \\
                   & $J$      & 1380. & Kitt Peak 1.3 m  & 9/93, 11/94 & \\
                   & $K_{s}$  &  360. &    ''            & 9/93 & \\

NGC 4314 \dots\dots& $g$     & 5400. & Lick 1 m          & 3/93, 4/93, 4/94 & $1.9\arcsec$\\
                   & $r$     & 7599. &    ''             & 4/93, 2/94, 4/94 & \\
                   & $i$     & 2100. &    ''             & 2/94, 4/94 & \\
                   & $J$     & 1252. & Lick 1 m          & 2/95 & \\
                   & $J$     &  840. & Kitt Peak 1.3 m   & 3/94 & \\
                   & $K_{s}$ &  540. &    ''             &  '' & \\  

NGC 266 \dots\dots & $g$     & 14646. & Lick 1 m         & 11/92, 9/93, 11/93, 10/94, 11/94 & $1.6\arcsec$\\
                   & $r$     & 10165. &    ''            & 11/92, 9/93, 10/93, 11/93, 10/94, 11/94 & \\
                   & $i$     &  6600. &    ''            & 11/92, 9/93, 11/93, 10/94, 11/94 & \\
                   & $J$     &  300.  & Kitt Peak 1.3 m  & 9/93 & \\
                   & $K'$    &  352.  & Lick 1 m         & 12/94 & \\  

NGC 986 \dots\dots & $g$     & 3600. & CTIO 0.9 m        & 9/94 & $1.5\arcsec$\\
                   & $r$     & 5100. &    ''             &   '' & \\
                   & $i$     & 3900. &    ''             &   '' & \\
                   & $J$     & 810.  & CTIO 1.5 m        & 9/94, 9/95 & \\
                   & $K_{s}$ & 420.  &    ''             &   '' & \\  

NGC 7496 \dots\dots& $g$     & 3600. & CTIO 0.9 m        & 9/94 & $1.6\arcsec$\\
                   & $r$     & 3600. &    ''             &   '' & \\
                   & $i$     & 3600. &    ''             &   '' & \\
                   & $J$     &  300. & CTIO 1.5 m        & 9/95 & \\
                   & $K_{s}$ &  330. &    ''             &   '' & \\  

NGC 5383 \dots\dots& $g$     & 2580. & Lick 1 m          & 4/94, 11/94 & $1.4\arcsec$\\
                   & $r$     & 2707. &    ''             &   ''  & \\
                   & $i$     &  780. &    ''             &   ''  & \\
                   & $J$     & 626.  & Lick 1 m          & 2/95 & \\
                   & $J$     & 840.  & Kitt Peak 1.3 m   & 3/94 & \\
                   & $K_{s}$ & 560.  &    ''             &   '' & \\  

NGC 4593  \dots\dots& $g$     & 3600. & CTIO 0.9 m       & 3/94, 3/95 & $1.5\arcsec$\\
                    & $r$     & 4500. &    ''            &  ''   & \\
                    & $i$     & 4200. &    ''            &  ''   & \\
                    & $J$     & 971. & CTIO 1.5 m       & 2/94  & \\
                    & $K$     & 868. &    ''            &  ''   & \\

NGC 3059 \dots\dots& $g$     & 4800. & CTIO 0.9 m        & 3/94, 3/95 & $1.6\arcsec$\\
                   & $r$     & 4200. &    ''             &   ''  & \\
                   & $i$     & 5400. &    ''             &   ''  & \\
                   & $J$     & 313.  & CTIO 1.5 m        & 2/94 & \\
                   & $H$     & 313.  &    ''             &   ''  & \\
                                                                 
NGC 7479 \dots\dots& $g$     &  10620.& Lick 1 m         & 8/92, 9/93, 10/94 & $1.8\arcsec$\\
                   & $r$     &  6960. &    ''            &   ''  & \\
                   & $i$     &  8649. &    ''            &   ''  & \\
                   & $J$     &  1080. & Kitt Peak 1.3 m  & 9/93, 11/94 & \\
                   & $K_{s}$ &   396. &    ''            &   '' & \\

NGC 3513  \dots\dots& $g$     & 4200. & CTIO 0.9 m       & 3/94, 3/95 & $1.5\arcsec$\\
                    & $r$     & 4500. &    ''            &  ''   & \\
                    & $i$     & 4200. &    ''            &  ''   & \\
                    & $J$     & 2719. & CTIO 1.5 m       & 2/94  & \\
                    & $H$     & 2718. &    ''            &  ''   & \\
                    & $K$     & 1042. &    ''            &  ''   & \\

\enddata

\end{deluxetable}

\section{Data analysis} \label{sec_analysis}

The objects were deprojected, by first rotating the frames 
to align the galaxy's major axis with the ``y" direction,
and then stretching the rotated image in the ``x" direction 
by a factor $1/\cos{\alpha}$, where $\alpha$ is the inclination angle. The 
position and inclination angles were taken from the literature, mostly from 
Hyperleda~\citep{pat03b} and the RC3 (see table~\ref{tbl-param}). 
The bar and spiral arm perturbations were traced in the NIR bands
(mainly $K_s$, $K^\prime$, and $K$, but other when specified), 
assuming that young stars do not contribute to the observed radiation~\citep[e.g.,][]{rix93}.
However, young stars and clusters may contribute locally up to $20\%-30\%$ to the observed
radiation~\citep{rix93,gon96,rho98,pat01,gros06,gros08}. 

\begin{deluxetable}{lllllcl}
\tabletypesize{\scriptsize}
\tablecaption{Galaxy parameters\label{tbl-param}}
\tablewidth{0pt}
\tablehead{
\colhead{Name} & \colhead{Type} & \colhead{PA\ (deg)}
& \colhead{$\alpha$ (deg)} & \colhead {$v_{\mathrm{max}}$ (km s$^{-1}$) }
& \colhead{Radial Velocity (km s$^{-1}$) }
& \colhead{Distance (Mpc) }
}
\startdata
NGC ~718	 &	SAB(s)a 	& ~45	                & $	29.4	\pm	10.4	$ & $	~58.5	\pm	14.0	$ & $	1733	\pm	10	$ & $	24.3	\pm	2.1	$ \\
NGC ~864	 &	SAB(rs)c	& ~20             	& $	40.7	\pm	~3.1	$ & $	~97.9	\pm	~4.1	$ & $	1560	\pm	~4	$ & $	22.0	\pm	1.9	$ \\
NGC 4314 &	SB(rs)a	        & 140\tablenotemark{a}	        & $	27.0	\pm	~5.3	$ & $	~70.6	\pm	~5.0	$ & $	~963	\pm	26	$ & $	17.7	\pm	2.1	$ \\
NGC ~266	 &	SB(rs)ab	& ~95\tablenotemark{b}	& $	12.2	\pm	16.5	$ & $	217.8	\pm	~7.5	$ & $	4661	\pm	~6	$ & $	64.8	\pm	5.5	$ \\
NGC ~986	 &	SB(rs)ab	& 150	                & $	40.7	\pm	~4.7	$ & $	~43.3	\pm	~8.4	$ & $	2005	\pm	~7	$ & $	28.0	\pm	2.4	$ \\
NGC 7496 &	SB(s)b	        & 169.7\tablenotemark{c}        & $	24.2	\pm	~9.2	$ & $	~65.0	\pm	~6.0	$ & $	1649	\pm	~6	$ & $	23.6	\pm	2.0	$ \\
NGC 5383 &	SB(rs)b    	& ~85	                        & $	31.7	\pm	~6.7	$ & $	142.6	\pm	13.1	$ & $	2250	\pm	~4	$ & $	39.2	\pm	3.4	$ \\
NGC 4593 &	RSB(rs)b	& ~56\tablenotemark{d}          & $	42.2	\pm	~4.4	$ & $	161.4	\pm	10.0	$ & $	2492	\pm	~6	$ & $	38.2	\pm	3.3	$ \\
NGC 3059 &	SB(rs)bc	& ~70.9\tablenotemark{c}	& $	27.0	\pm	~8.4	$ & $	~55.6	\pm	~5.5	$ & $	1260	\pm	~6	$ & $	15.0	\pm	1.3	$ \\
NGC 7479 &	SB(s)c	        & ~25	                        & $	40.7	\pm	~3.1	$ & $	162.2	\pm	~7.1	$ & $	2378	\pm	~3	$ & $	34.4	\pm	2.9	$ \\
NGC 3513 &	SB(rs)c	        & ~75	                        & $	37.4	\pm	~3.5	$ & $	~38.6	\pm	~4.7	$ & $	1194	\pm	~7	$ & $	16.0	\pm	1.4	$ \\

\enddata

\tablecomments{
Col.\ (2). Hubble types from RC3.
Col.\ (3). Position angles, mainly from RC3 and Hyperleda~\citep{pat03b}.
Col.\ (4). Inclination angle, $\alpha = cos^{-1}{(b/a)}$;
$a/b$ is the isophotal diameter ratio derived from the $R_{25}$ parameter in RC3.
Col.\ (5). Maximum rotation velocity obtained from the HI data of \citet{pat03}, uncorrected for inclination.
Col.\ (6). Heliocentric radial velocity from RC3.
Col.\ (7). Hubble distance obtained from the heliocentric radial velocity and the infall model of~\citet{mou00},
H$_0=71\pm6$ km s$^{-1}$ Mpc$^{-1}$.
}

\tablenotetext{a}{~\citet{ben02}.}
\tablenotetext{b}{~\citet{pat03b}.}
\tablenotetext{c}{~\citet{pat03b}; Lauberts A. \& Valentijn E.A. (1989).}
\tablenotetext{d}{~\citet{vau99}.}

\end{deluxetable}

\subsection{Azimuthal color gradient analysis with the GG96 method}~\label{GG96_method}

The search and analysis of azimuthal color gradients
are based on the four band, supergiant sensitive and reddening-free\footnote{For a foreground screen,
and for a mixture of dust and stars as long as $\tau_V < 2$
\citep{gon96,mar09a}. See also \S~\ref{Qdust}, and figure~\ref{cfall_model}.} photometric index

\begin{equation}
  Q(rJgi) = (r-J) - \frac{E(r-J)}{E(g-i)}(g-i),
\end{equation}

\noindent
where $\frac{E(r-J)}{E(g-i)}$ is the color excess term for a
foreground screen. $Q$ is a very good diagnostic of star formation, since its
value increases with the presence of blue and red supergiants.
Details of the method can be found in~\citet[][GG96 hereafter]{gon96},
and~\citet{mar09a,mar09b}. 
Briefly, the procedure involves ``unwrapping'' the spiral
arms by plotting them in a $\theta$ vs.\ ln$R$ map~\citep[e.g.,][]{iye82,elm92}.
Under this geometric transformation,
logarithmic spirals appear as straight lines with slope = cot(-$i$),
where $i$ is the arm pitch angle.
The search for color gradient candidates in the bar and arms can be done in the $Q(rJgi)$ ``unwrapped'' frame,
with the aid of a dust lane tracer (e.g., the $g-J$ color).
The arms are then ``straightened", by adding a different phase shift
at each radius, until 
the arms appear ``horizontal''.
In the straightened arms, selected regions can be collapsed
in ln $R$ to yield 1-D plots of intensity
vs.\ distance that can be compared with 
stellar population synthesis (SPS) models.

Stellar models give $Q(rJgi)$ as a function of stellar age ($t_{\rm{age}}$),
while observations provide $Q(rJgi)$ as a function of distance. 
Distance $d=0$ is fixed 
at the location with the highest extinction (i.e., the highest $g-J$ value)
in the dust lane.
The pattern speeds of the bar, $\Omega_{p}^{\rm {bar}}$, 
and of the spiral arms, 
$\Omega_{p}^{\rm {spiral}}$, are derived from candidates in, respectively, 
the bar and spiral arm regions, through equation~\ref{eqOMEGA}, by 
``stretching" the models to fit the observations \citep[GG96,][]{mar09a}:

\begin{equation}~\label{eqOMEGA}
  \Omega_{p} \cong \frac{1}{R_{\rm{mean}}} \left(v_{\rm rot} 
  - \frac{d}{t_{\rm age}}
  - \frac{v_{\rm rad}}{\tan{i}} \right);
\end{equation}

\noindent
$R_{\rm{mean}}$ is the mean radius of the studied region, $v_{\rm{rot}}$ is the
galactic rotation velocity\footnote{
For this research, we assume that $v_{\rm{rot}} \sim$ constant 
in the regions of interest. The mean value of $v_{\rm{rot}}$ 
for our sample (after inclination correction, and
excluding NGC 266 due to its highly uncertain inclination angle) is
$160 \pm 30$ km s$^{-1}$. The error in the velocity is comparable to expected 
deviations from a flat rotation curve~\citep[e.g., see the model rotation curves in][]{rom07}.
We also try to avoid the inner parts of the disk, where a flat rotation curve
should no longer be valid (the minimum value of $R_{\rm{mean}} / R_{\rm {bar}}$ in 
figure~\ref{SB_grads} is $\sim 0.4$).}
from the literature (corrected for inclination; see table~\ref{tbl-param}),
$d$ is the azimuthal distance from the shock (i.e., measured from $d=0$), 
$t_{\rm age}$ is the stellar model age at distance $d$,\footnote{For practical purposes, the $d$ and $t_{\rm age}$ quantities introduced in equation~\ref{eqOMEGA} 
are really $\delta d$ and $\delta t_{\rm age}$
within the same azimuthal region at a given radius (see also~\S\ref{chara_grads}).}
$v_{\rm rad}$ is the radial velocity,
and $i$ is the spiral shock pitch angle, assuming ``steady state''~\citep{rob69}.\footnote{
The term $v_{\rm rad} /\tan{i}$, and equation~\ref{del_theta} are only meaningful for spiral regions.}
This equation can be easily derived from the one presented in~\citet{mar09a},

\begin{eqnarray}
  \label{eqOMEGA_II}
  \Omega_{p} = \frac{1}{t} \left(\int_{0}^{t} \frac{\vec v(t') \cdot \hat{\varphi}(t')}{R(t')} \mathrm{d}t'
             - \left(\theta_{\rm shock} + \Delta\theta \right)\right), \\
 \label{del_theta}
  \Delta\theta = \cot(i) \ln \left(\frac{R(0)}{R(t)}\right).
\end{eqnarray}

Equation~\ref{eqOMEGA} assumes nearly circular motion for the involved stellar regions. 
For the pattern speed determinations we take $v_{\rm rad} \sim 0$,
although this term becomes important for tightly wound spirals~\citep[see also][]{gros09}.
Important deviations from circular orbits and velocity gradients are expected in the gas 
in the azimuthal direction, perpendicular to the bar~\citep{duv83}.
\citet{mar09b} have shown that, this notwithstanding, 
azimuthal color gradients across spiral arms can 
be detected, although assuming a circular motion dynamic 
model will result in a systematic trend to overestimate spiral
pattern speeds at small radii, away from CR, in non-barred or
weakly-barred galaxies.\footnote{In these galaxies, gas flows to some extent along the
arms after passing through a steady rotating spiral shock. The age
gradient is narrower than in the absence of these non-circular motions,
and hence an observer will infer a smaller difference between
the orbital velocity of the stars and the pattern speed. Inside corotation,
this will lead to an overestimate of the pattern speed that increases
as the radius decreases, and converges to zero at corotation,
where the shock strength and the non-circular motions are minimal.}
Pixel averaging due to image processing compensates 
in part for this effect~\citep{mar09b}, such
that the measured pattern speed will be correct within
the errors, but the systematic trend, whereby the 
difference between measured and real pattern speeds 
$\propto R^{-1}$, will still be observed.

\subsection{Bar extent} \label{sec_Bar_end}

According to~\citet{woz95}, the bar end is located after the radius of maximum ellipticity, 
and is marked by a change in the position angle (PA) of the isophotes. Isophotal PA, on the other hand, 
must remain roughly constant along the bar region, although spiral structure,
rings surrounding the bar, and stellar bar ansae~\citep[mainly in early-type galaxies,][]{marv07} may 
disturb the elliptical profiles.

\begin{figure*}
\centering
\includegraphics[scale=0.80]{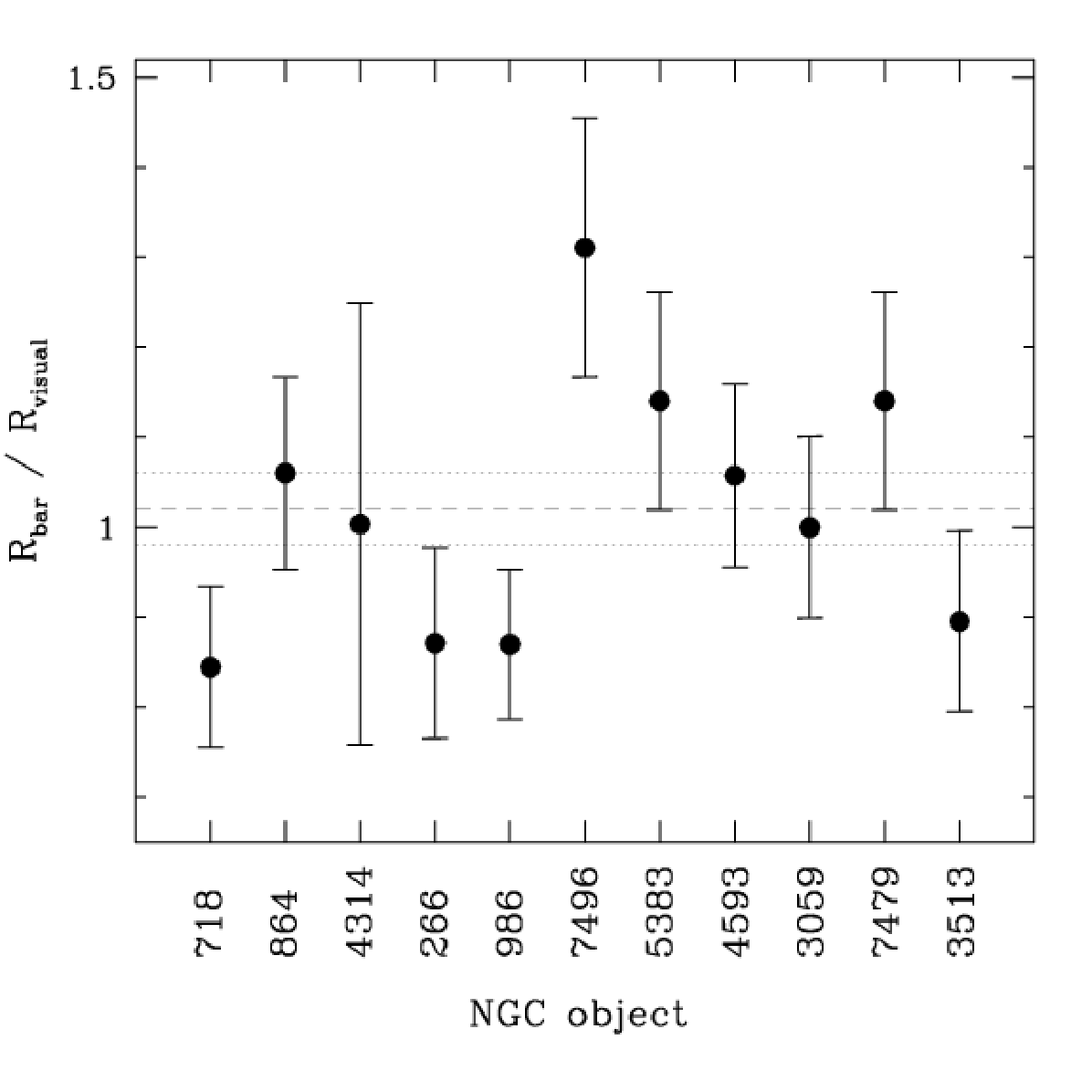}

\caption[f2.eps]{Comparison between the bar extent obtained with the bar's isophotes ($R_{\rm bar}$, see text),
and the estimate by sight ($R_{\rm visual}$). The dashed and dotted lines indicate, respectively, the mean
and the error of the ratio of both measurements, 1.02$\pm$0.04.
The object numbers in the ``New General Catalogue"~\citep[NGC,][]{dre888},
ordered by Hubble type, are shown
in the horizontal axis. 
\label{BARend_graph}}
\end{figure*}

Although the bar extent may be underestimated with the maximum ellipticity criterion~\citep{mic06},
this method provides a good, homogeneous, standard for our purposes. We determined most  
bar lengths using the maximum ellipticity criterion (see table~\ref{tbl-omegas}).
To this end, we masked bright stars and nearby objects, and
fitted ellipses to the bar's isophotes (in the NIR, mostly in the 2 $\mu$m, deprojected frames)
with the IRAF task ELLIPSE~\citep{jed87,bus96}. We then generated plots of 
isophotal ellipticity\footnote{
$\mathrm{Ellipticity} = \left( 1 - \frac{\mathrm{minor~axis}}{\mathrm{major~axis}} \right)$}
and PA versus radius~\citep{woz95,mul97,gad06}.
The brightest pixel near the nuclear region was given as an initial 
guess for the isophote center, while re-centering was allowed for 
outer isophotes.
A second ``mean isophote center'' was computed with these re-centered isophotes;
this second center was then used,
without allowing for further re-centering.
We also performed a visual estimate of the bar's extension. In figure~\ref{BARend_graph}, we compare 
the results of both measurements.

Finally, we compare the bar length with the bar CR radius; the latter is 
calculated based on the pattern speeds derived from
the color gradients in bar regions. 

\section{Results}

\begin{figure*}
\centering
\includegraphics[scale=0.80]{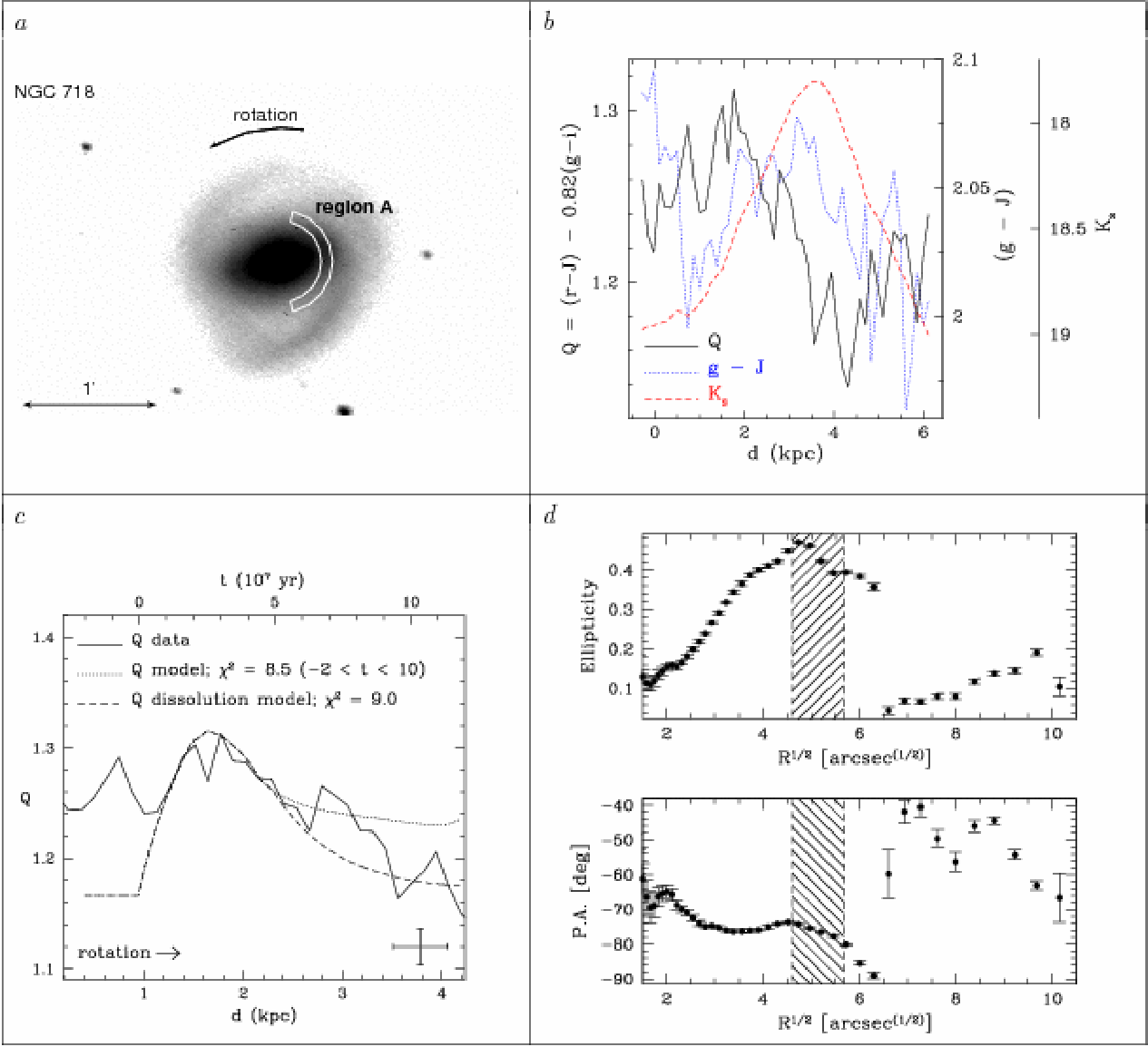}

\caption[f3.eps]{NGC~718, region A.
({\it a.}) $g$-band deprojected mosaic of the galaxy,  
in logarithmic scale. 
({\it b.}) {\it Solid black line and left y-axis:} observed $Q(rJgi)$ profile
vs.\ azimuthal distance, $d$, in
kpc; {\it dotted blue line and second-from-right y-axis:} observed ($g-J$) color vs.\ $d$;
{\it dashed red line and rightmost y-axis:} 
observed $K_s$ surface brightness (mag arcsec$^{-2}$) vs.\ $d$.
({\it c.}) {\it Solid line:} zoomed-in version of $Q(rJgi)$ vs.\ $d$ profile.  
{\it Dotted line:} stellar population model, 
``stretched'' in $t_{\rm age}$ to fit the data; IMF $M_{\rm upper} = 10 M_{\sun}$.
Model stellar age, in units of $10^7$ yr, is shown in the upper $x$-axis. 
{\it Dashed line:} model including ``dissolution of stellar groups'',
see~\S\ref{down_falls}. Reduced $\chi^{2}$ values cover the same time
interval for both models.
({\it d.}) {\it Upper plot:} isophote ellipticity vs.\ $R^{1/2}$;
{\it lower plot:} isophotal PA vs.\ $R^{1/2}$. 
{\it Hatched areas:} bar corotation region (see text). 
\label{REG_718_A}}
\end{figure*}

\begin{figure*}
\centering
\includegraphics[scale=0.80]{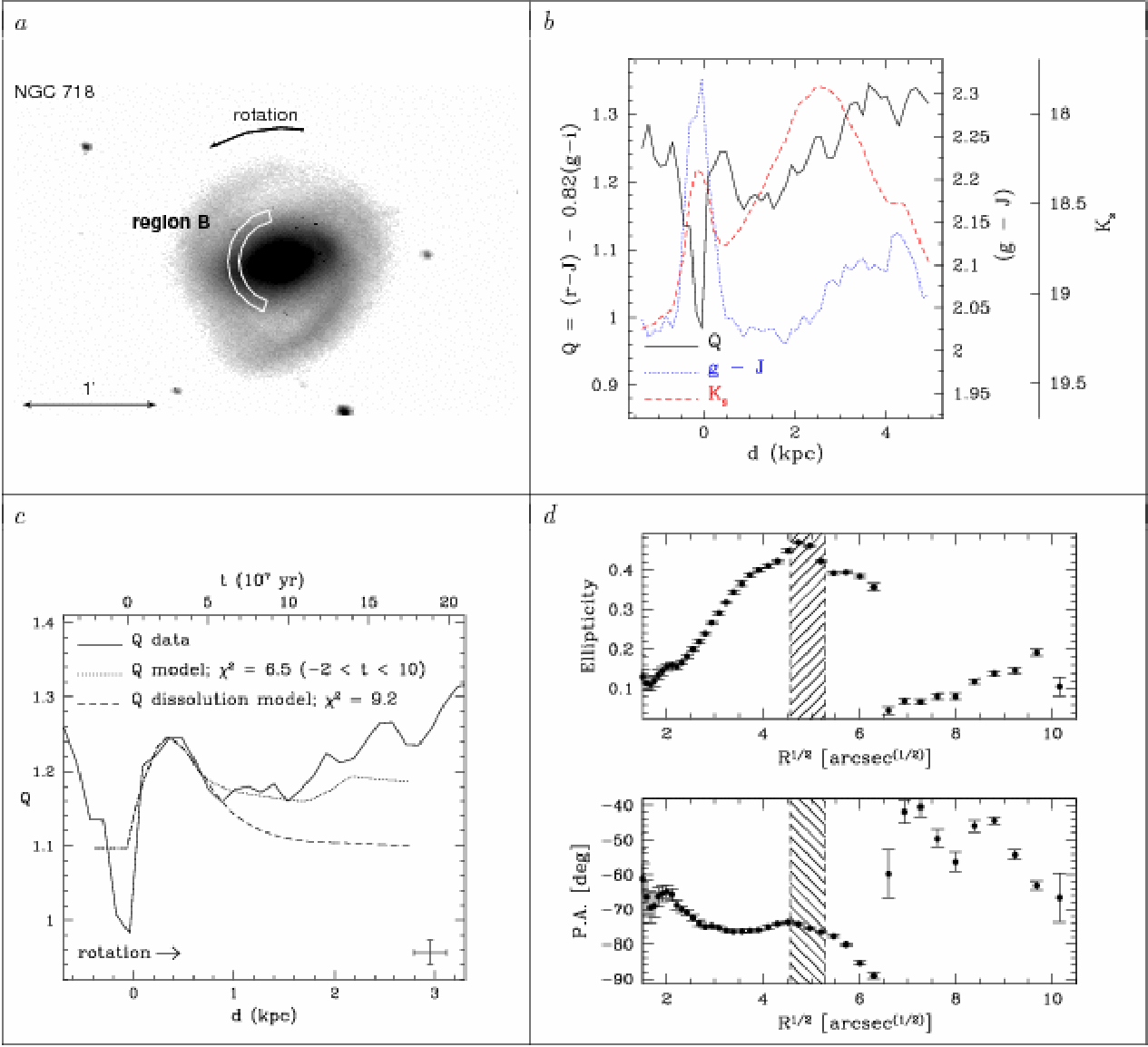}

\caption[f4.eps]{NGC~718, region B. 
({\it c.}) {\it Dotted and dashed lines:} stellar population models, IMF $M_{\rm upper} = 10 M_{\sun}$.
\label{REG_718_B}}
\end{figure*}

\begin{figure*}
\centering
\includegraphics[scale=0.80]{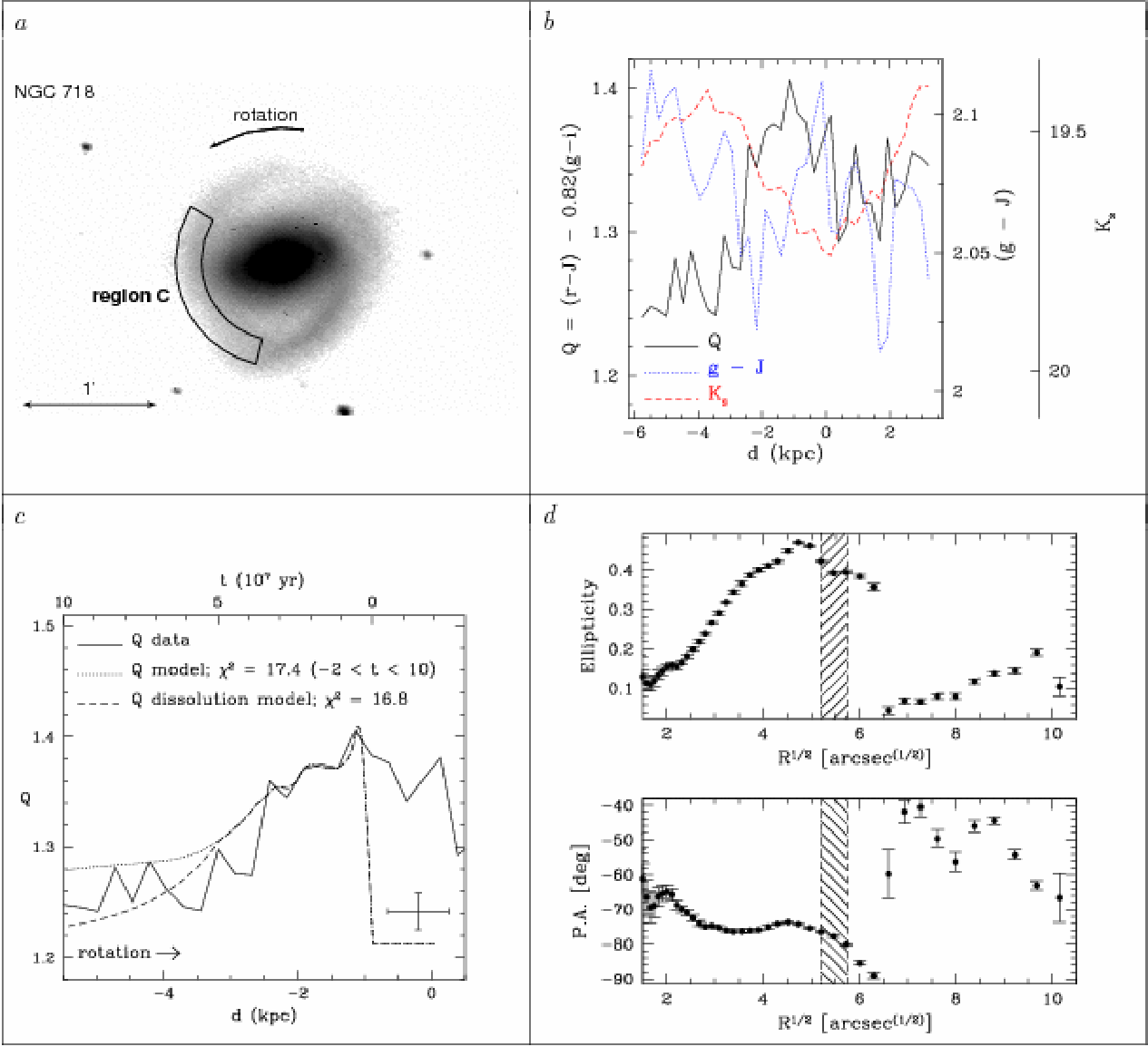}

\caption[f5.eps]{NGC~718, region C. 
({\it c.}) {\it Dotted and dashed lines:} stellar population models, IMF $M_{\rm upper} = 100 M_{\sun}$.
\label{REG_718_C}}
\end{figure*}

\begin{figure*}
\centering
\includegraphics[scale=0.80]{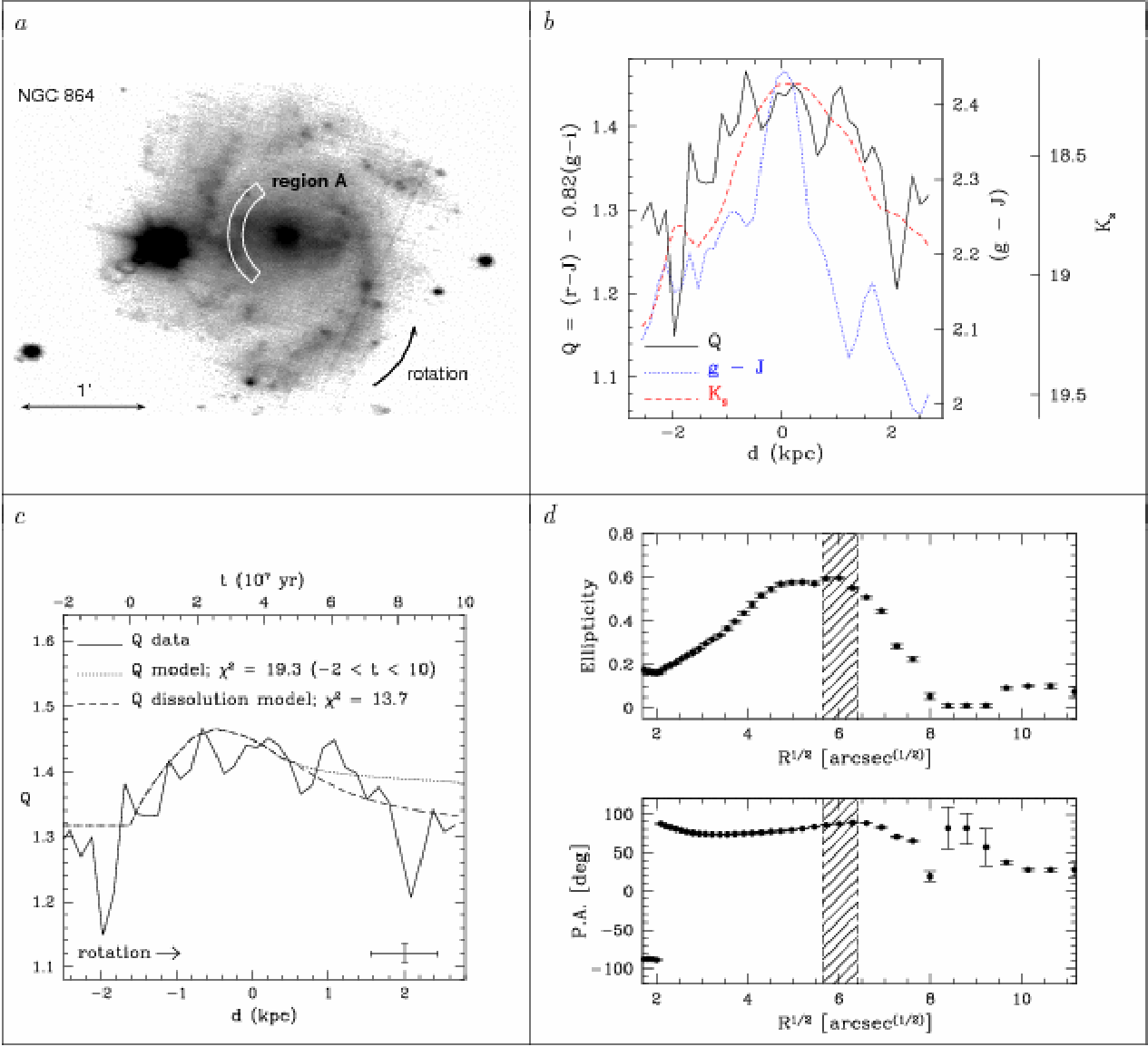}

\caption[f6.eps]{NGC~864, region A.
({\it c.}) {\it Dotted and dashed lines:} stellar population models, IMF $M_{\rm upper} = 10 M_{\sun}$.
\label{REG_864_A}}
\end{figure*}

\begin{figure*}
\centering
\includegraphics[scale=0.80]{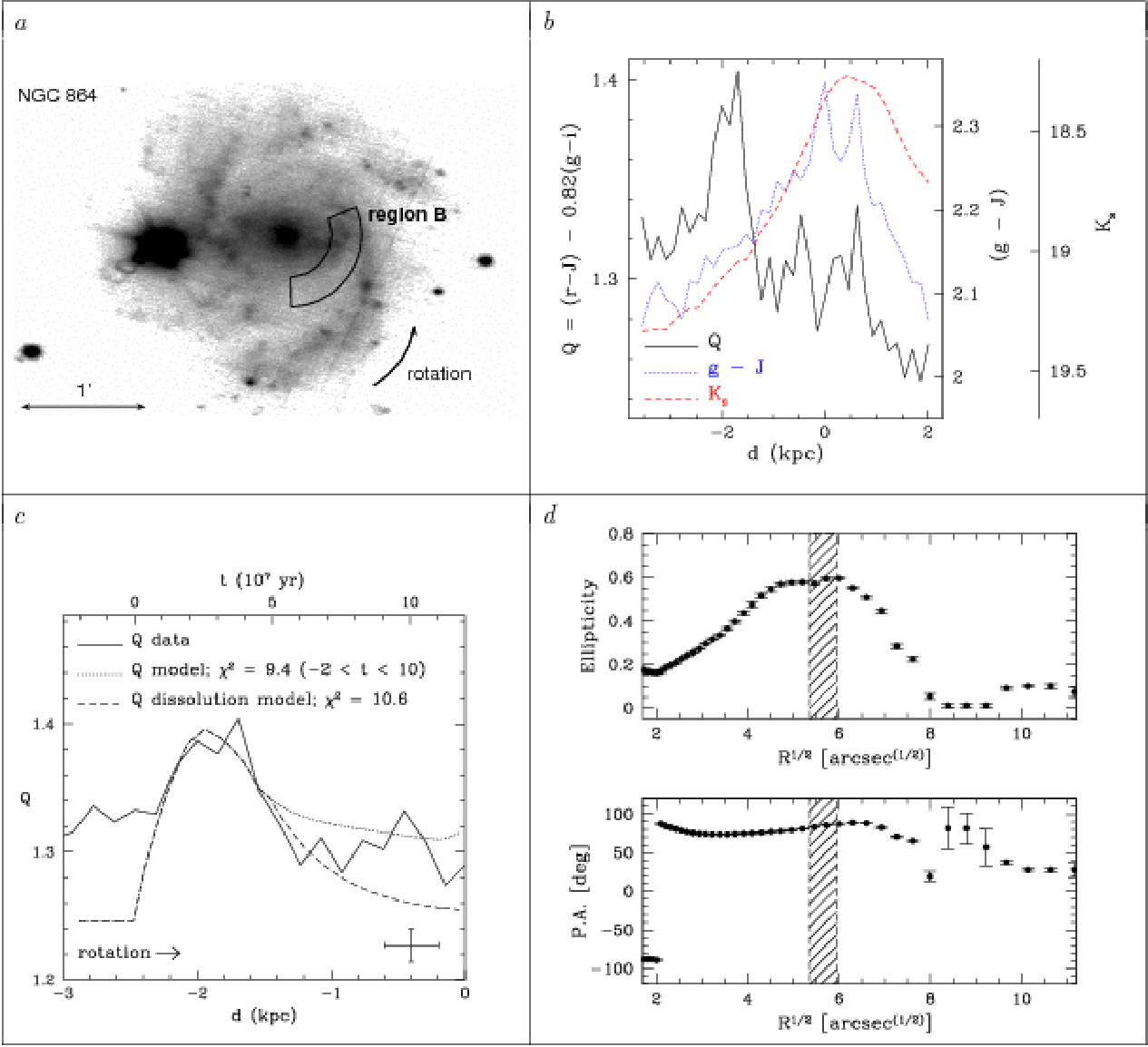}

\caption[f7.eps]{NGC~864, region B.
({\it c.}) {\it Dotted and dashed lines:} stellar population models, IMF $M_{\rm upper} = 10 M_{\sun}$.
\label{REG_864_B}}
\end{figure*}

\begin{figure*}
\centering
\includegraphics[scale=0.80]{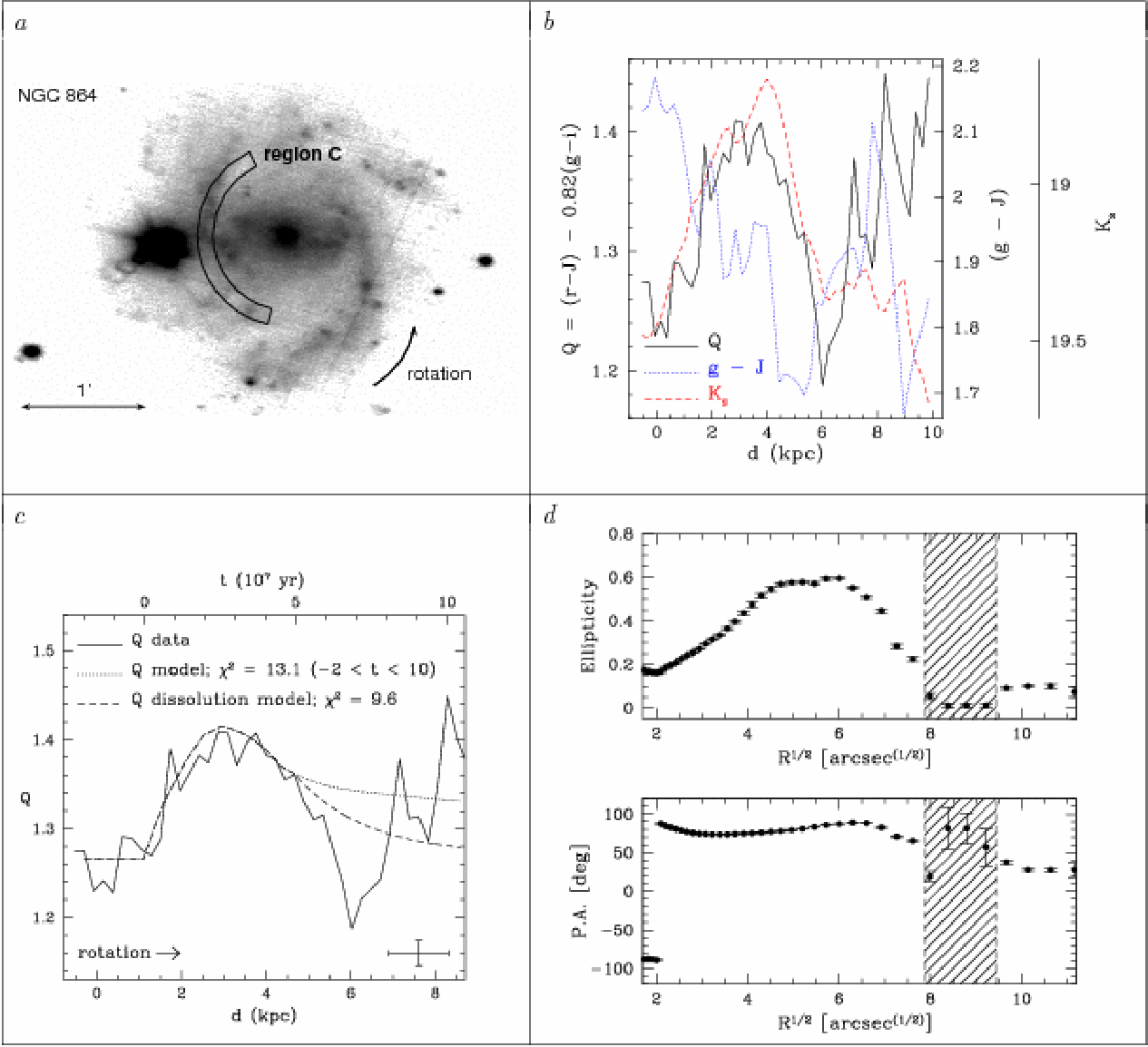}

\caption[f8.eps]{NGC~864, region C.
({\it c.}) {\it Dotted and dashed lines:} stellar population models, IMF $M_{\rm upper} = 10 M_{\sun}$.
\label{REG_864_C}}
\end{figure*}

\begin{figure*}
\centering
\includegraphics[scale=0.80]{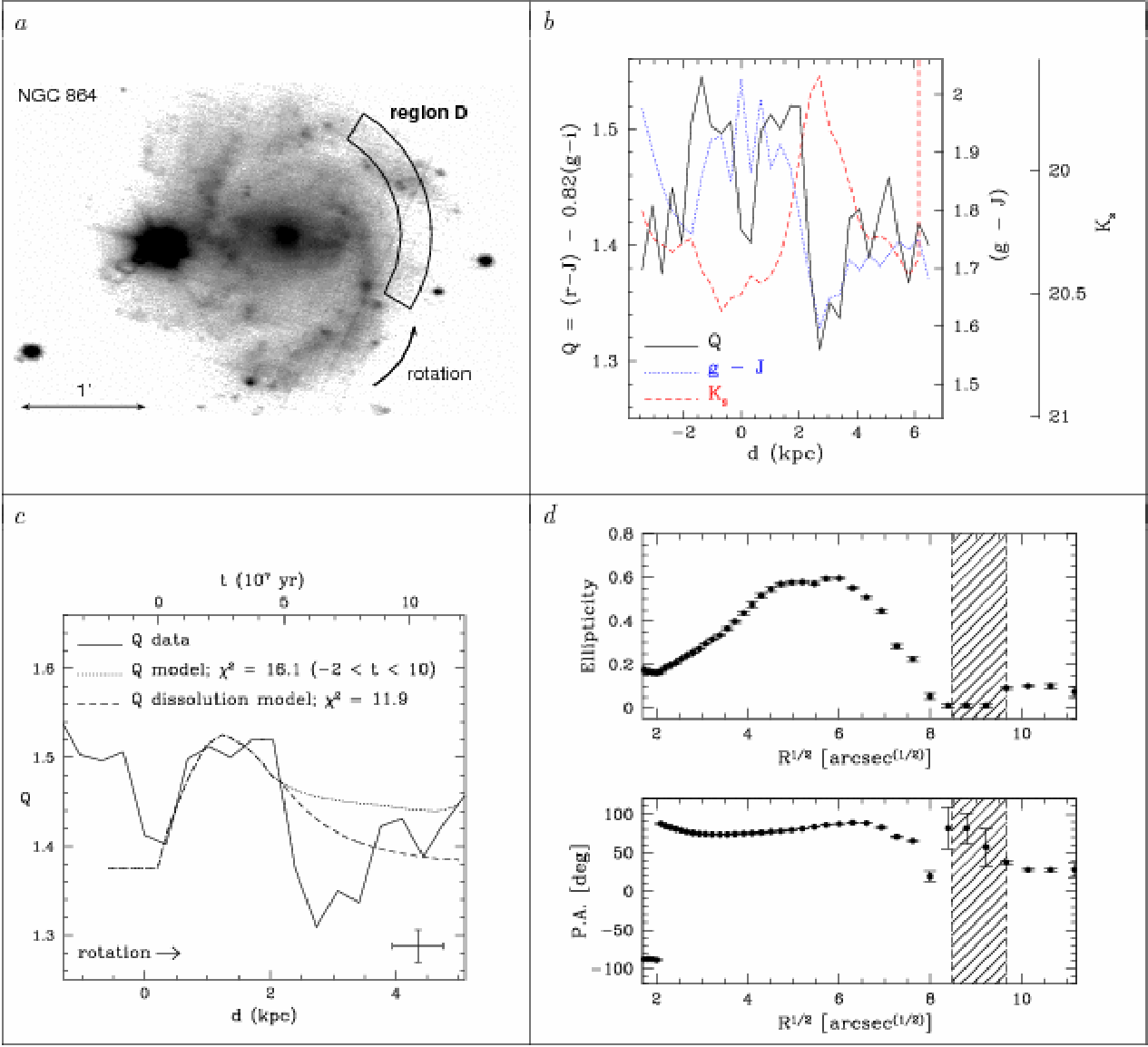}

\caption[f9.eps]{NGC~864, region D.
({\it c.}) {\it Dotted and dashed lines:} stellar population models, IMF $M_{\rm upper} = 10 M_{\sun}$.
\label{REG_864_D}}
\end{figure*}

\begin{figure*}
\centering
\includegraphics[scale=0.80]{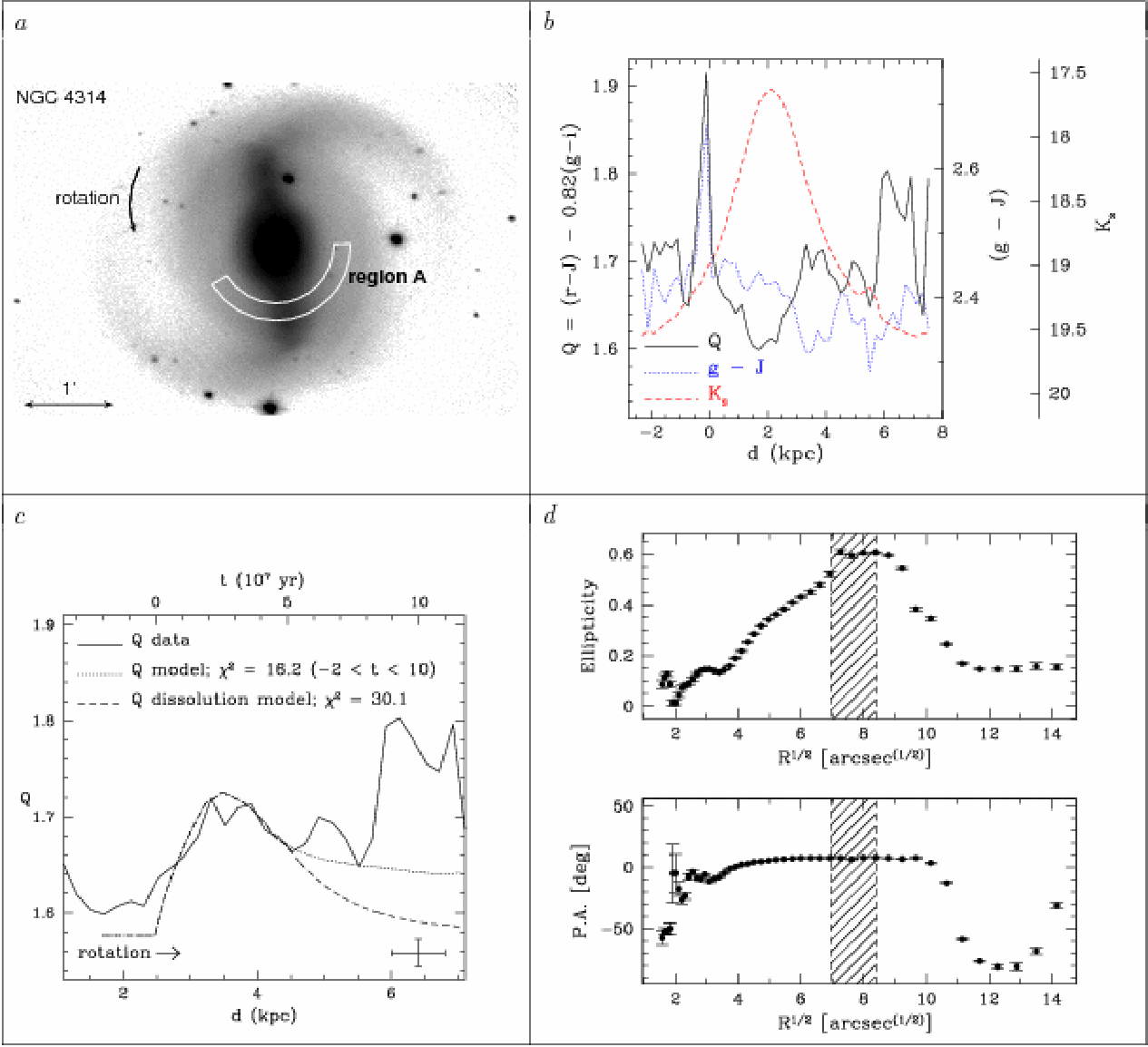}

\caption[f10.eps]{NGC~4314, region A. ({\it a}): optical $r$ mosaic.
({\it c.}) {\it Dotted and dashed lines:} stellar population models, IMF $M_{\rm upper} = 10 M_{\sun}$.
\label{REG_4314_A}}
\end{figure*}

\begin{figure*}
\centering
\includegraphics[scale=0.80]{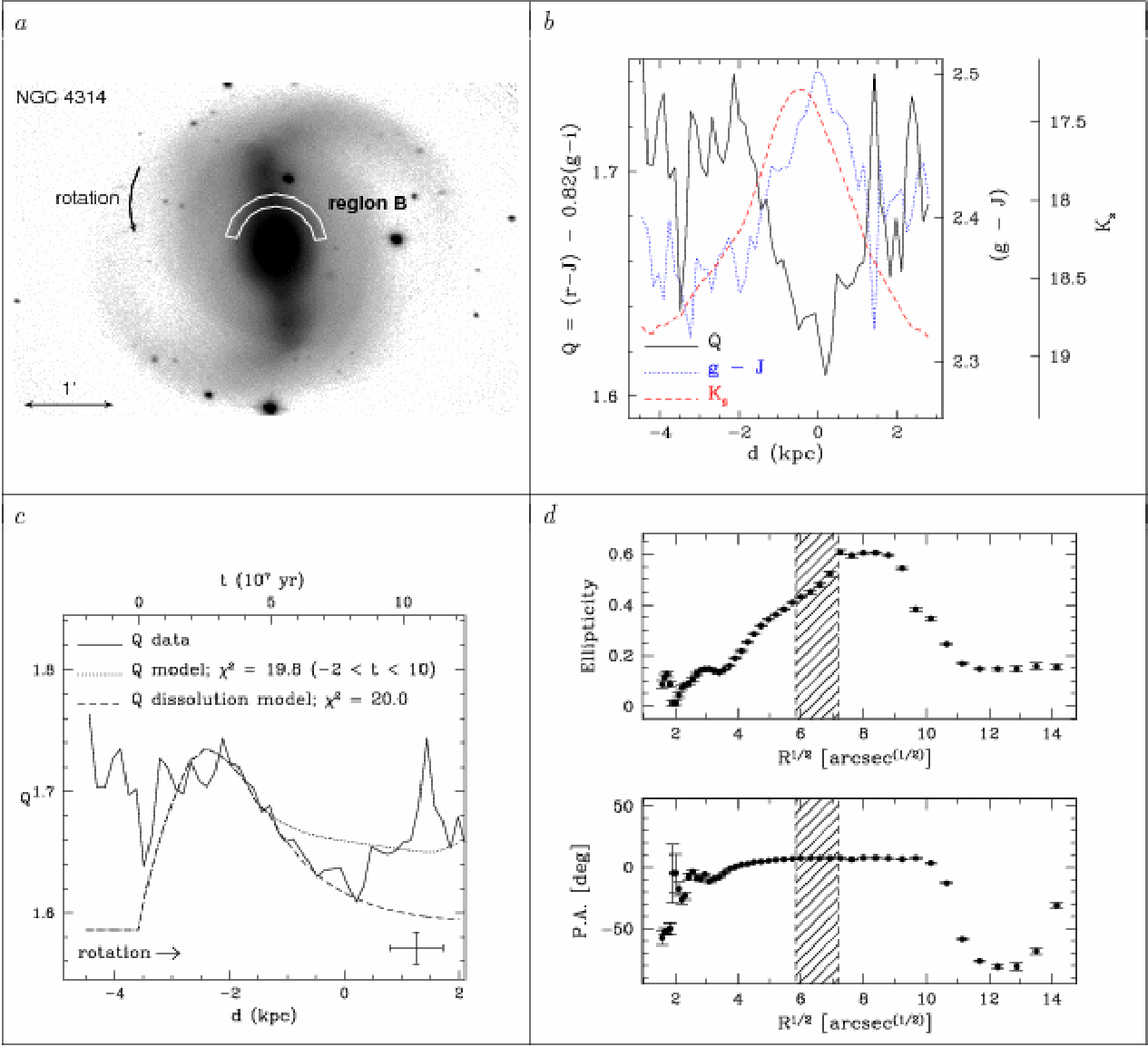}

\caption[f11.eps]{NGC~4314, region B. ({\it a}): optical $r$ mosaic.
({\it c.}) {\it Dotted and dashed lines:} stellar population models, IMF $M_{\rm upper} = 10 M_{\sun}$.
\label{REG_4314_B}}
\end{figure*}

\begin{figure*}
\centering
\includegraphics[scale=0.80]{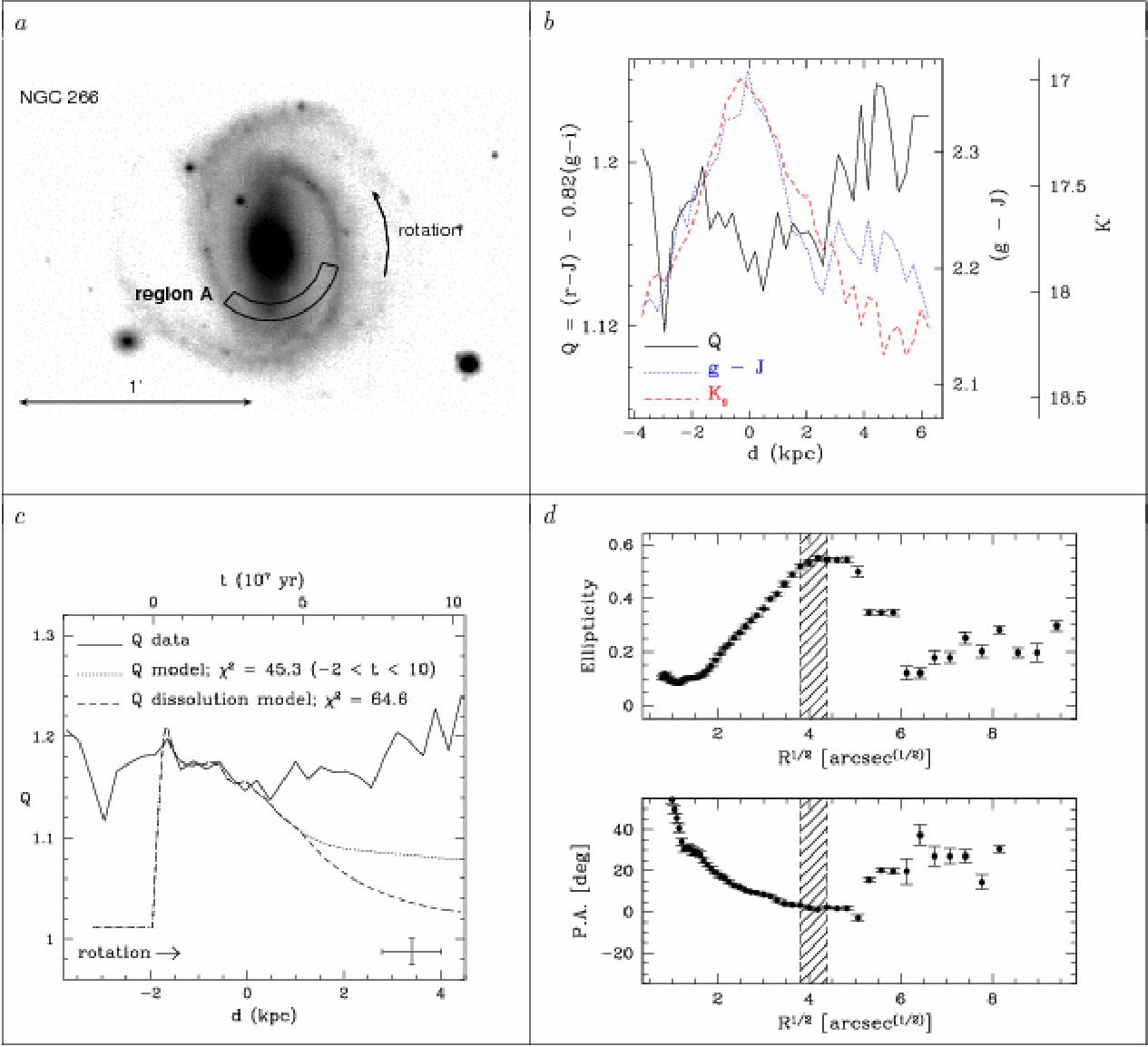}

\caption[f12.eps]{NGC~266, region A.
({\it c.}) {\it Dotted and dashed lines:} stellar population models, IMF $M_{\rm upper} = 100 M_{\sun}$.
\label{REG_266_A}}
\end{figure*}

\begin{figure*}
\centering
\includegraphics[scale=0.80]{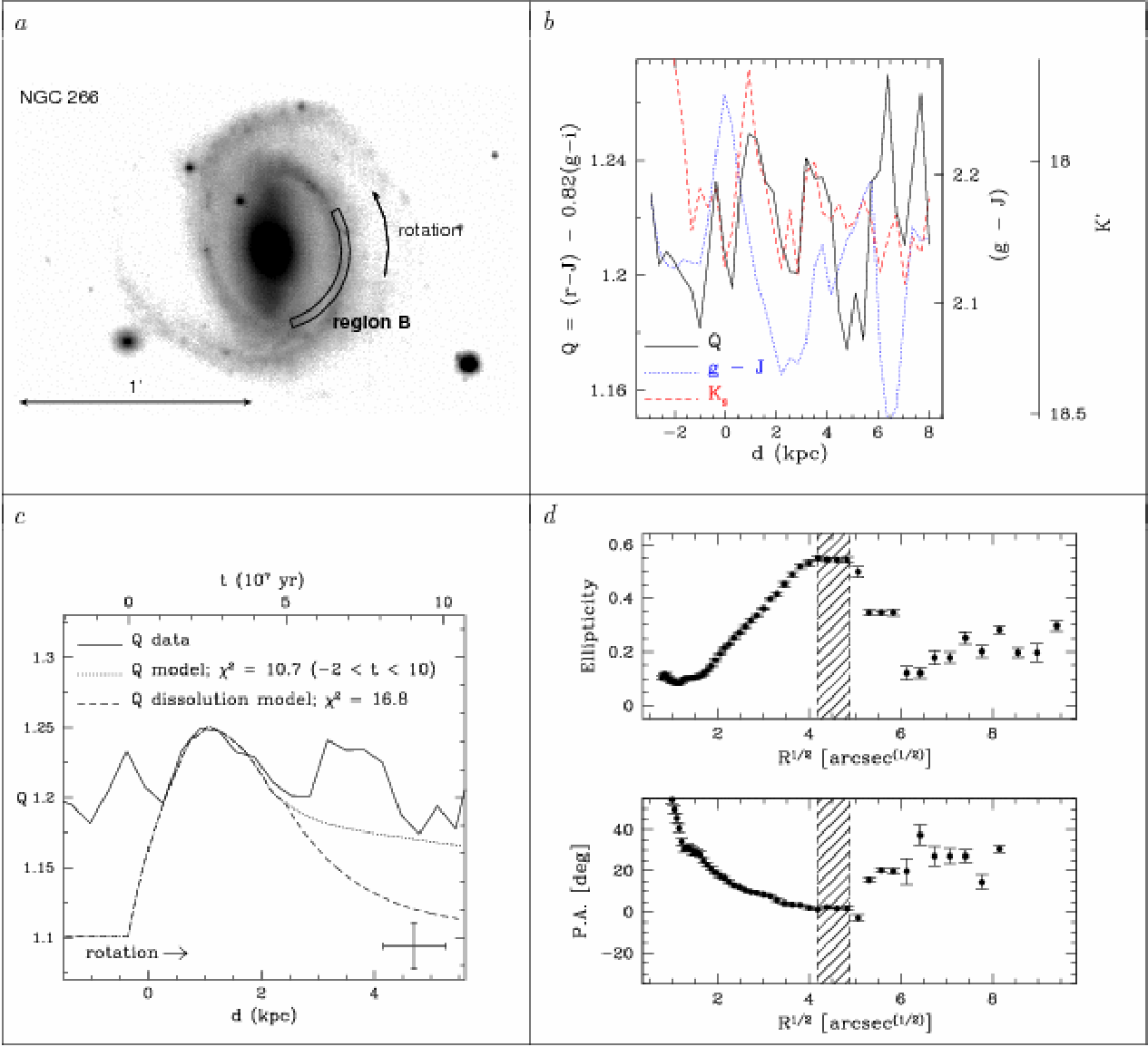}

\caption[f13.eps]{NGC~266, region B.
({\it c.}) {\it Dotted and dashed lines:} stellar population models, IMF $M_{\rm upper} = 10 M_{\sun}$.
\label{REG_266_B}}
\end{figure*}

\begin{figure*}
\centering
\includegraphics[scale=0.80]{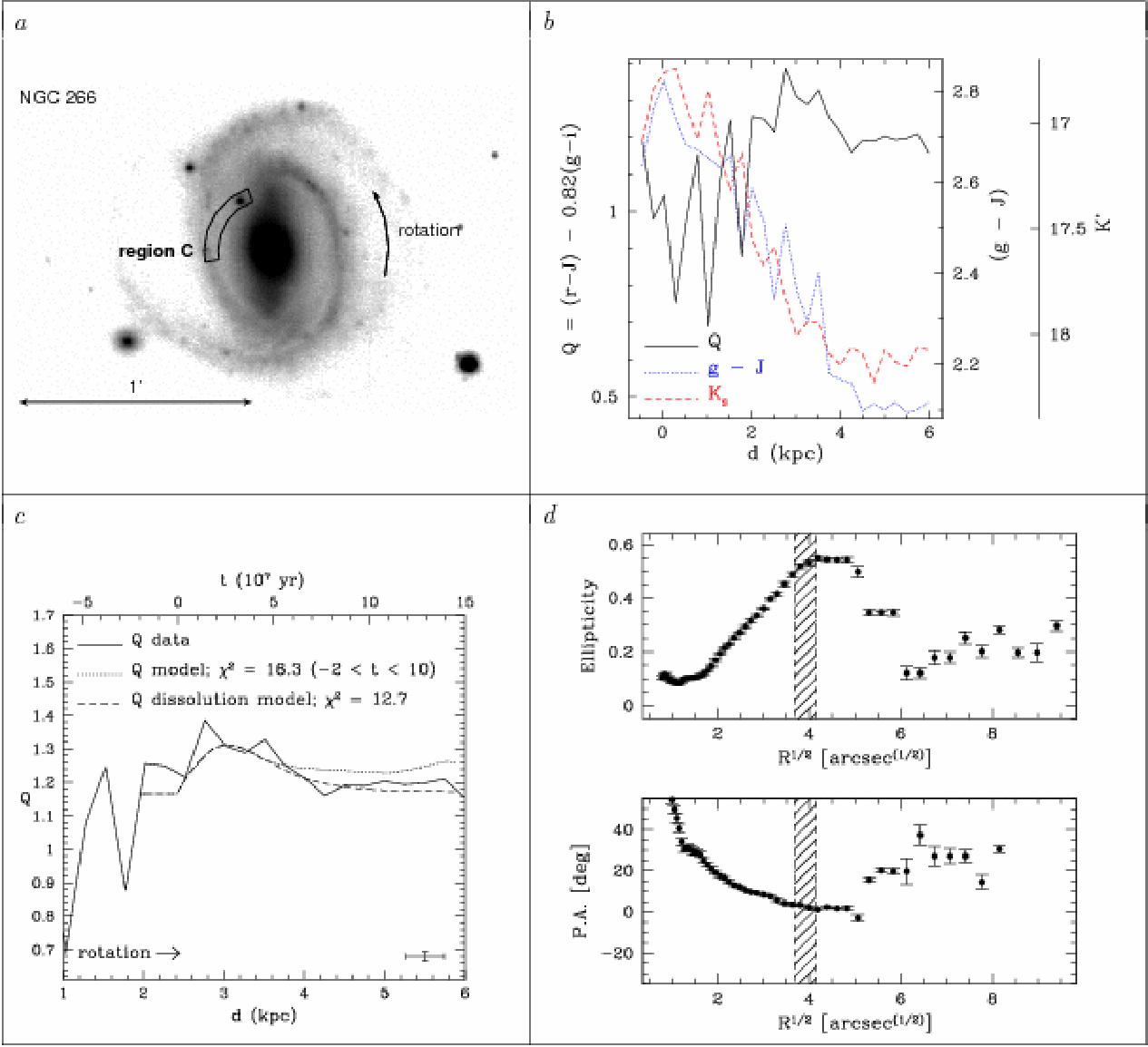}

\caption[f14.eps]{NGC~266, region C.
({\it c.}) {\it Dotted and dashed lines:} stellar population models, IMF $M_{\rm upper} = 10 M_{\sun}$.
\label{REG_266_C}}
\end{figure*}

\begin{figure*}
\centering
\includegraphics[scale=0.80]{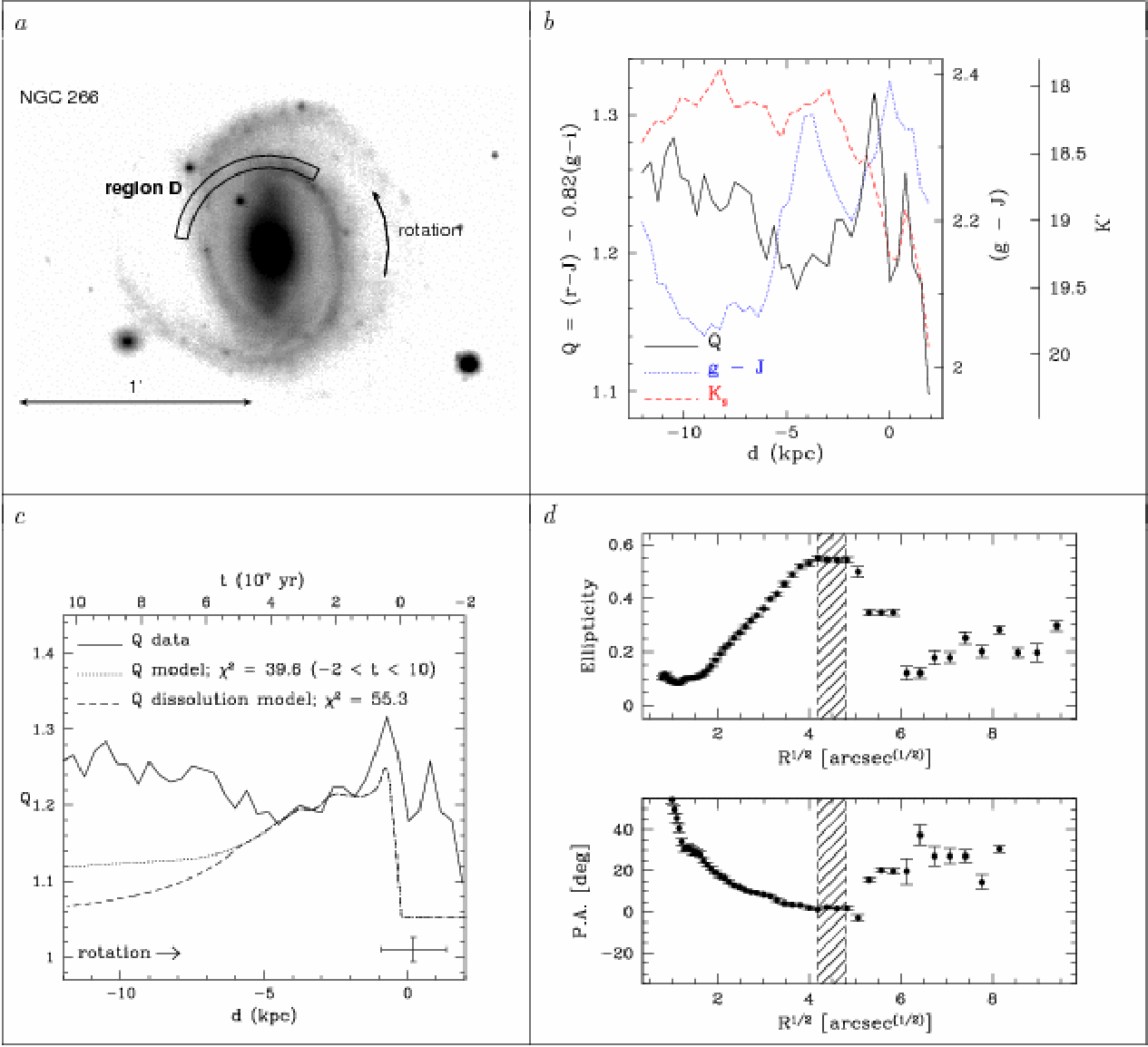}

\caption[f15.eps]{NGC~266, region D.
({\it c.}) {\it Dotted and dashed lines:} stellar population models, IMF $M_{\rm upper} = 100 M_{\sun}$.
\label{REG_266_D}}
\end{figure*}

\begin{figure*}
\centering
\includegraphics[scale=0.80]{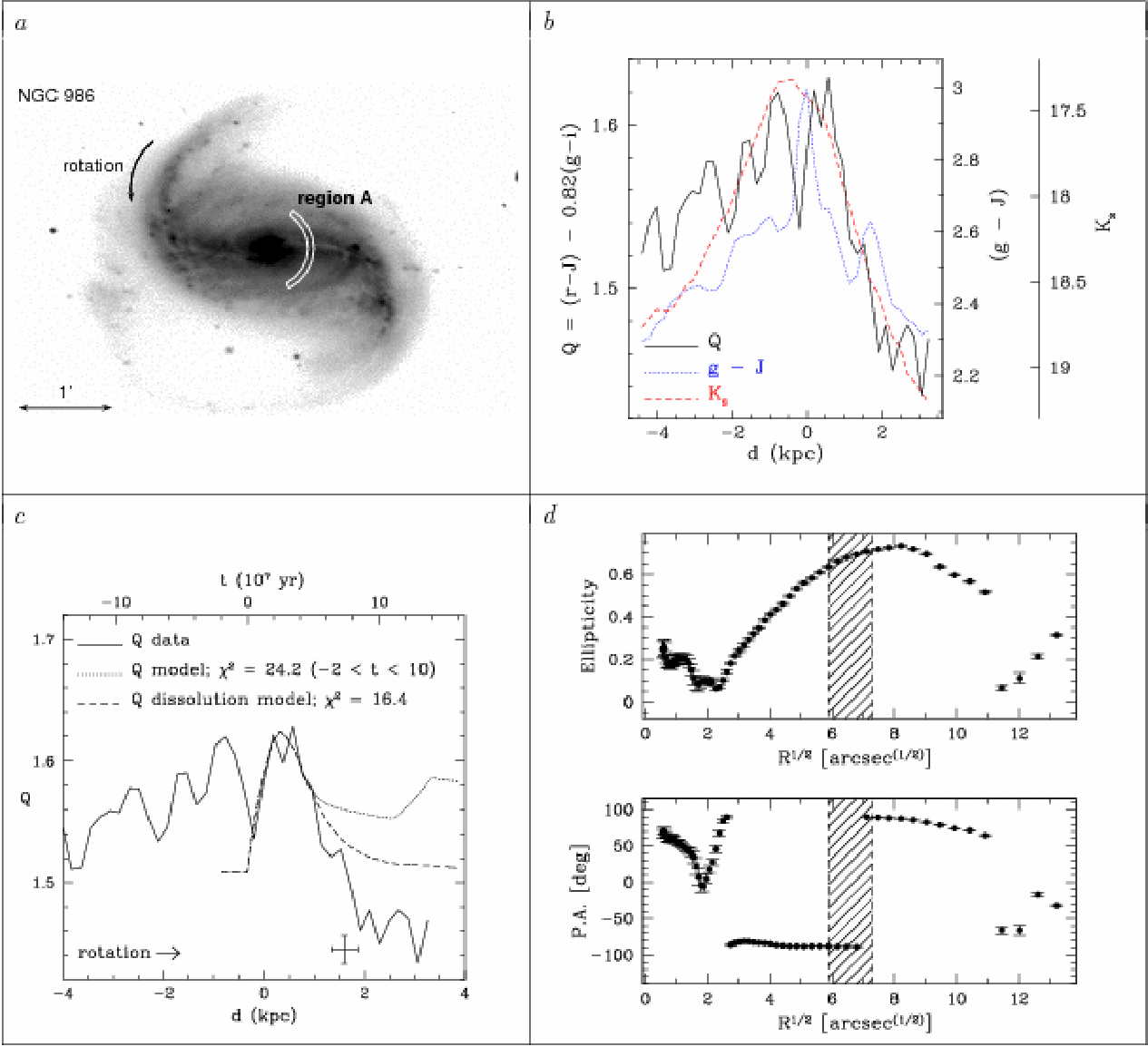}

\caption[f16.eps]{NGC~986, region A.
({\it c.}) {\it Dotted and dashed lines:} stellar population models, IMF $M_{\rm upper} = 10 M_{\sun}$.
\label{REG_986_A}}
\end{figure*}

\begin{figure*}
\centering
\includegraphics[scale=0.80]{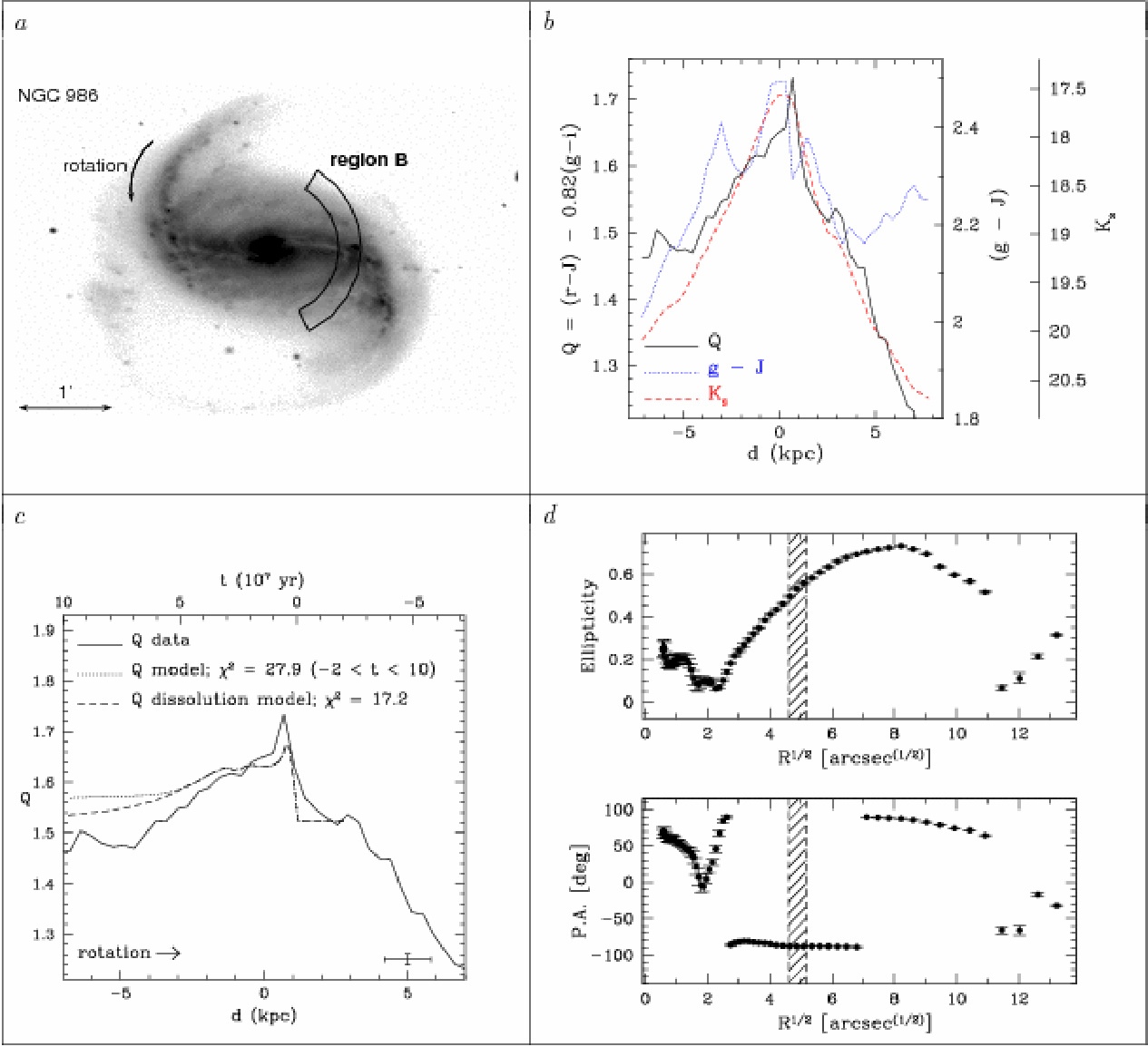}

\caption[f17.eps]{NGC~986, region B.
({\it c.}) {\it Dotted and dashed lines:} stellar population models, IMF $M_{\rm upper} = 100 M_{\sun}$.
\label{REG_986_C}}
\end{figure*}

\begin{figure*}
\centering
\includegraphics[scale=0.80]{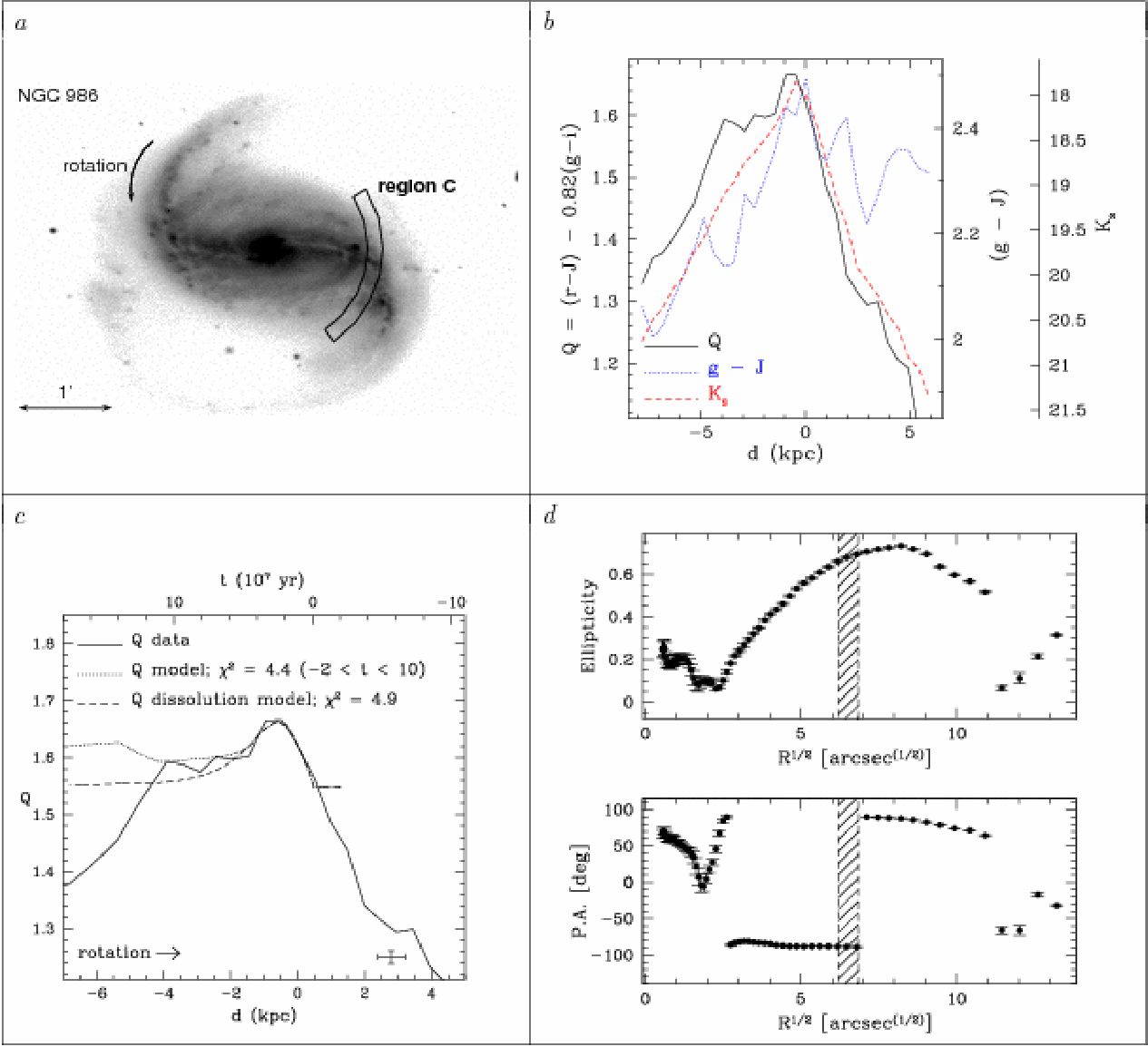}

\caption[f18.eps]{NGC~986, region C.
({\it c.}) {\it Dotted and dashed lines:} stellar population models, IMF $M_{\rm upper} = 10 M_{\sun}$.
\label{REG_986_D}}
\end{figure*}

\begin{figure*}
\centering
\includegraphics[scale=0.80]{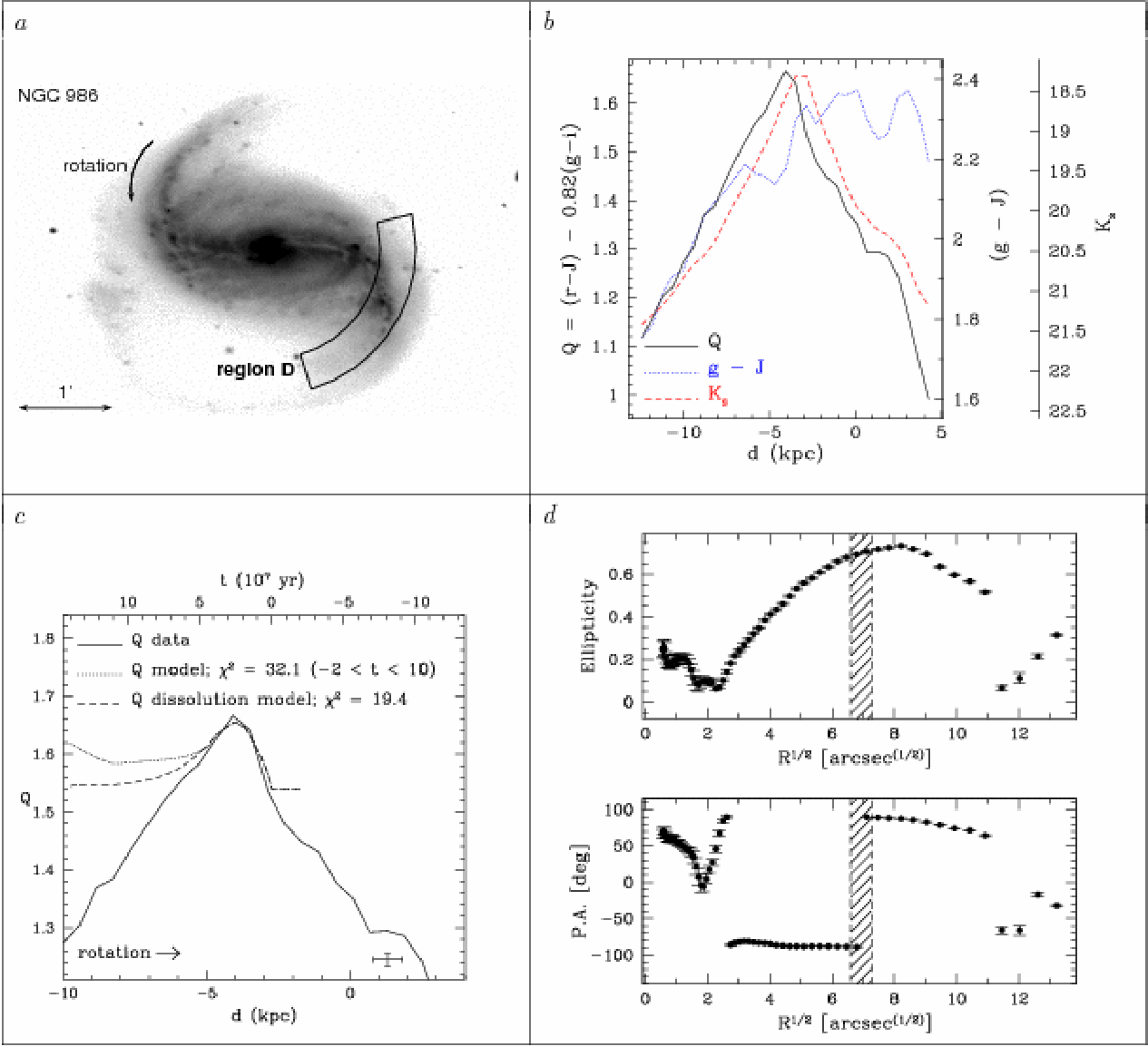}

\caption[f19.eps]{NGC~986, region D.
({\it c.}) {\it Dotted and dashed lines:} stellar population models, IMF $M_{\rm upper} = 10 M_{\sun}$.
\label{REG_986_E}}
\end{figure*}

\begin{figure*}
\centering
\includegraphics[scale=0.80]{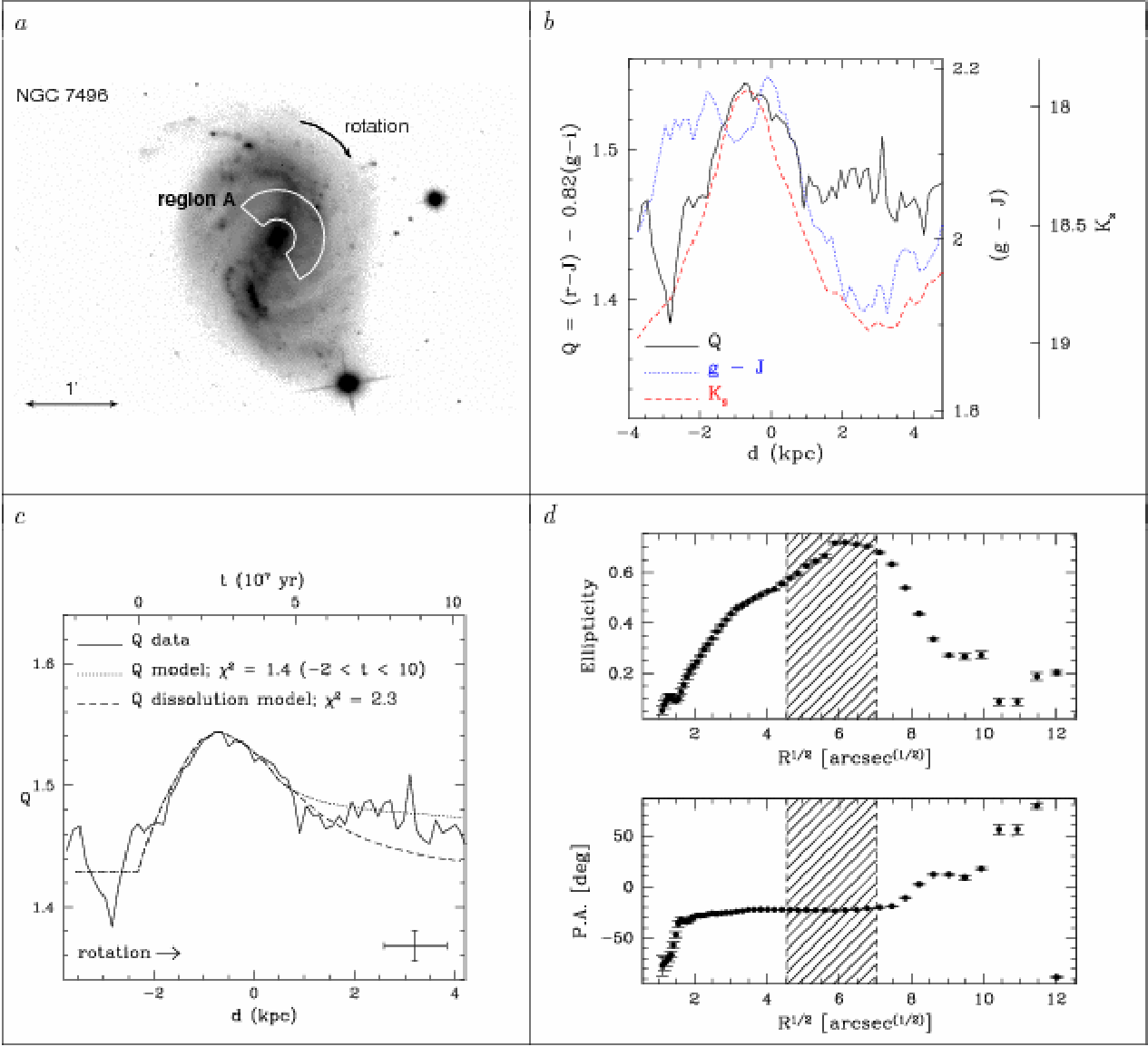}

\caption[f20.eps]{NGC~7496, region A.
({\it c.}) {\it Dotted and dashed lines:} stellar population models, IMF $M_{\rm upper} = 10 M_{\sun}$.
\label{REG_7496_A}}
\end{figure*}

\begin{figure*}
\centering
\includegraphics[scale=0.80]{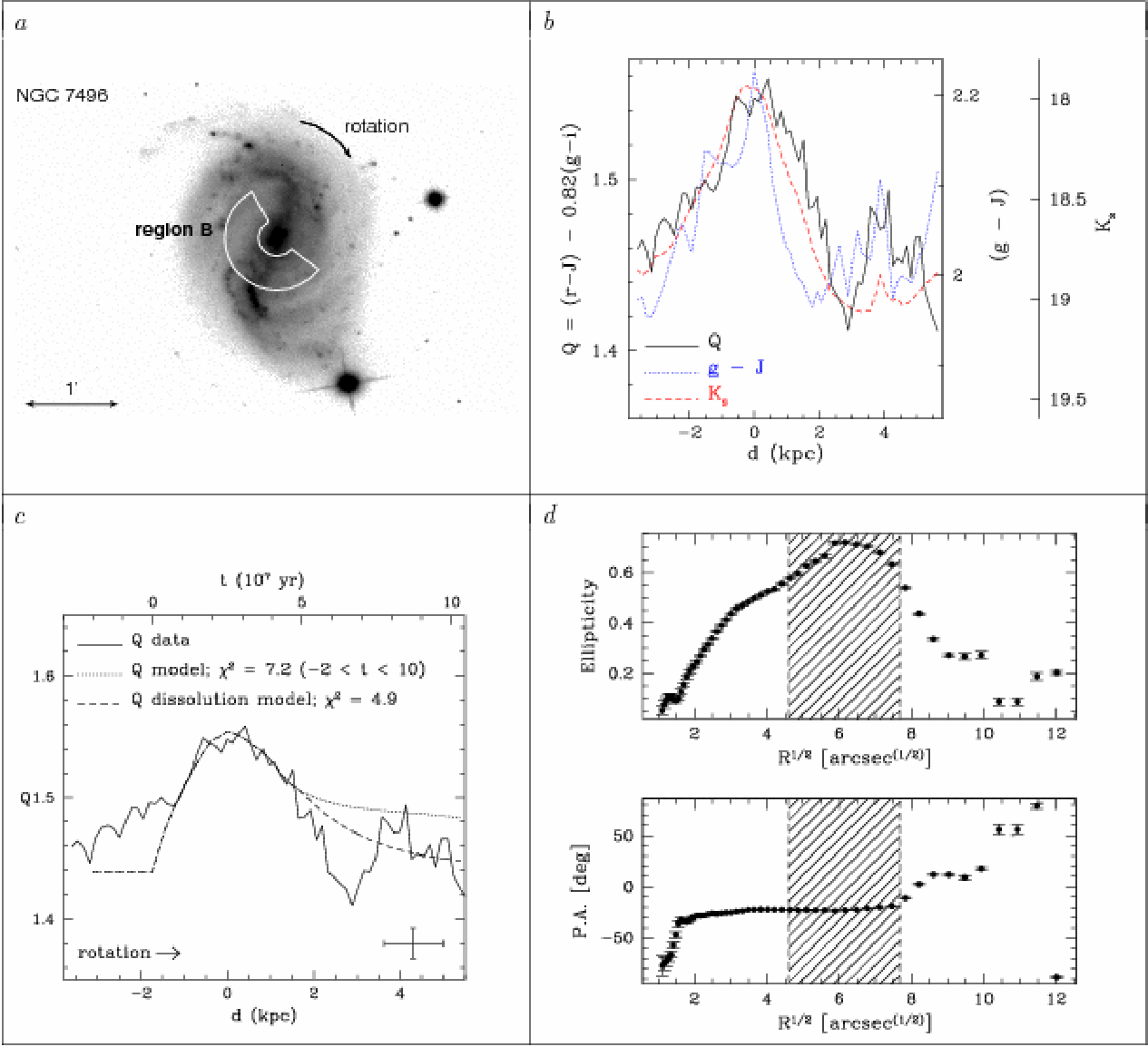}

\caption[f21.eps]{NGC~7496, region B.
({\it c.}) {\it Dotted and dashed lines:} stellar population models, IMF $M_{\rm upper} = 10 M_{\sun}$.
\label{REG_7496_B}}
\end{figure*}

\begin{figure*}
\centering
\includegraphics[scale=0.80]{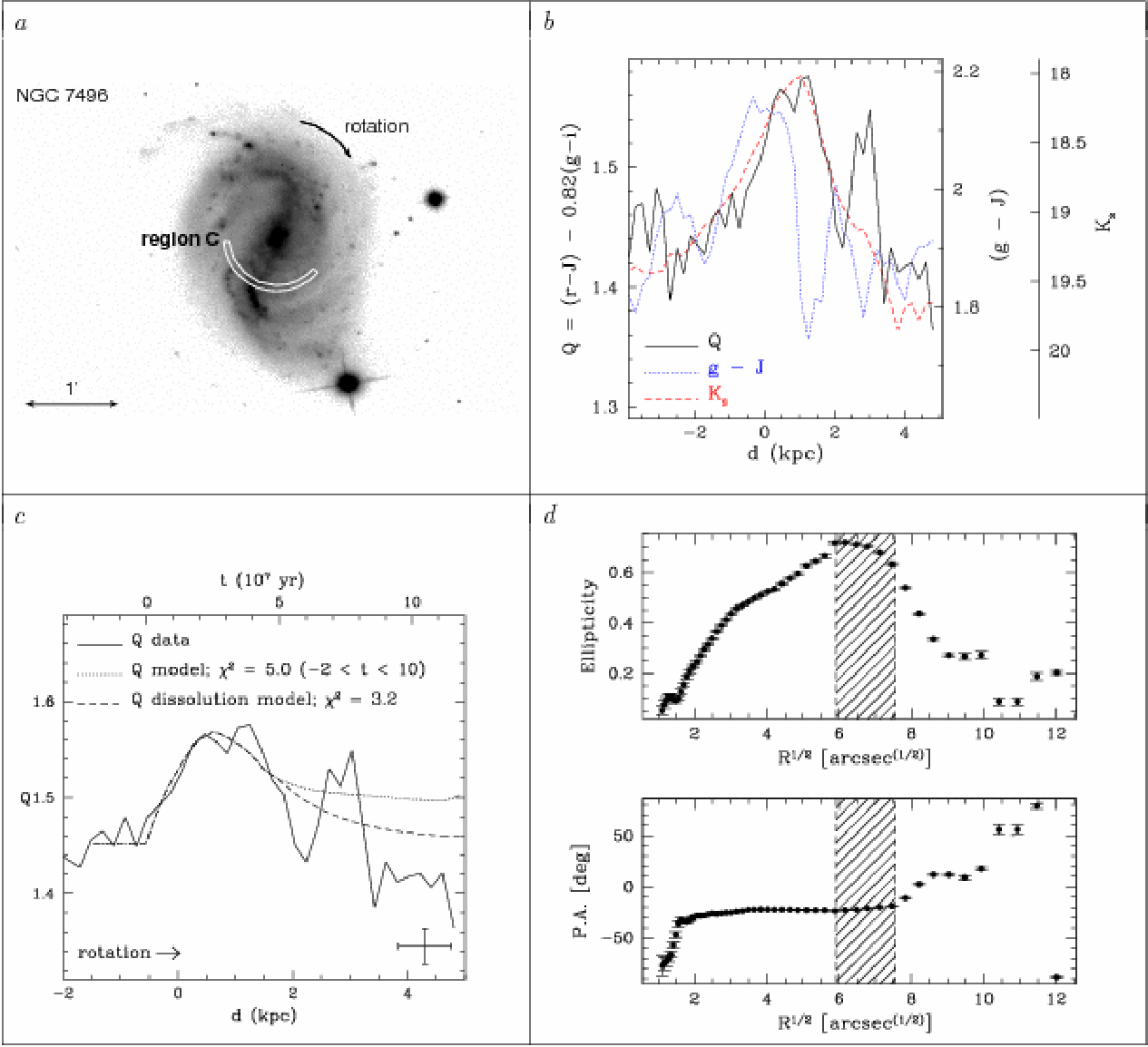}

\caption[f22.eps]{NGC~7496, region C.
({\it c.}) {\it Dotted and dashed lines:} stellar population models, IMF $M_{\rm upper} = 10 M_{\sun}$.
\label{REG_7496_C}}
\end{figure*}

\begin{figure*}
\centering
\includegraphics[scale=0.80]{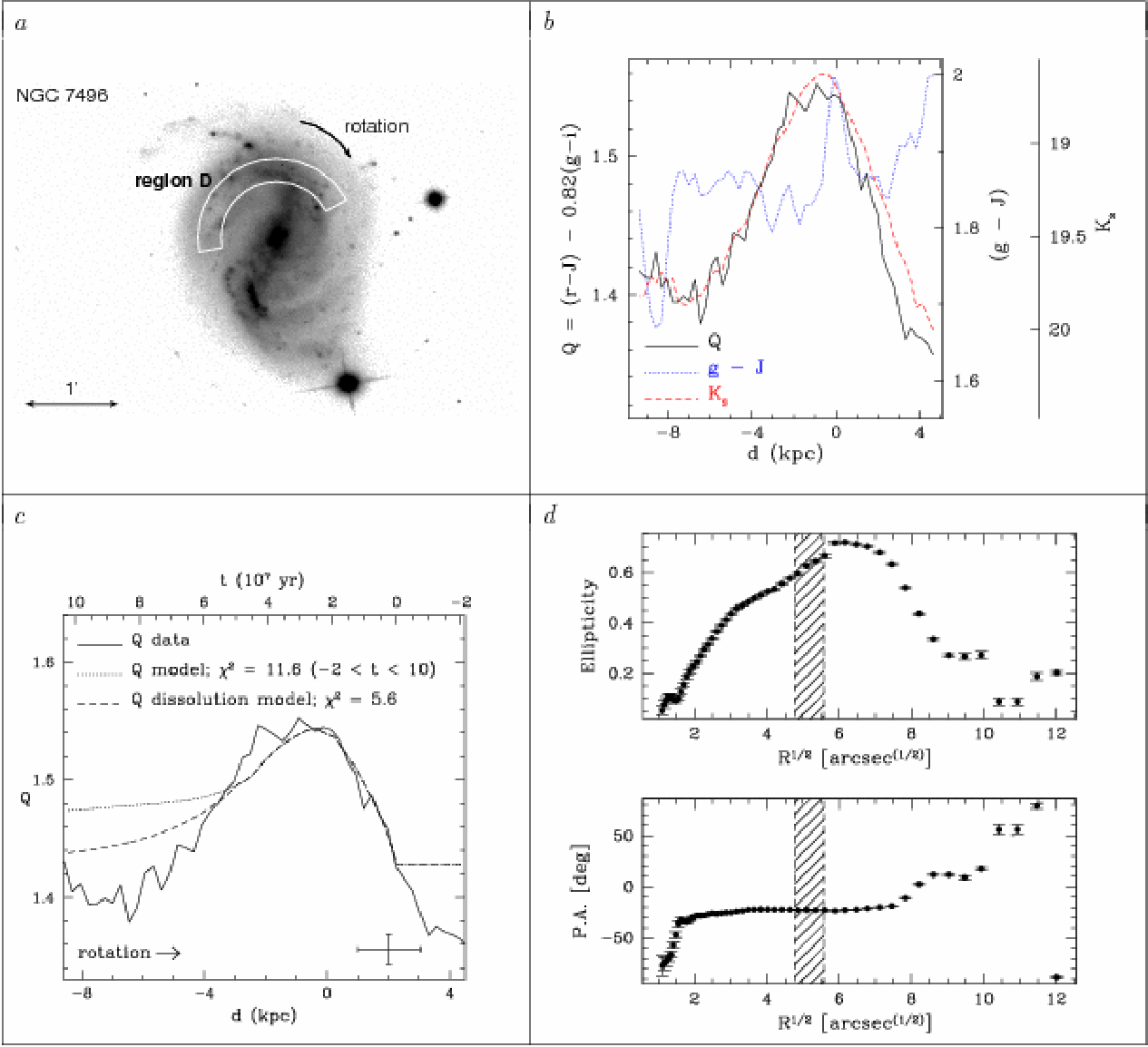}

\caption[f23.eps]{NGC~7496, region D.
({\it c.}) {\it Dotted and dashed lines:} stellar population models, IMF $M_{\rm upper} = 10 M_{\sun}$.
\label{REG_7496_D}}
\end{figure*}

\begin{figure*}
\centering
\includegraphics[scale=0.80]{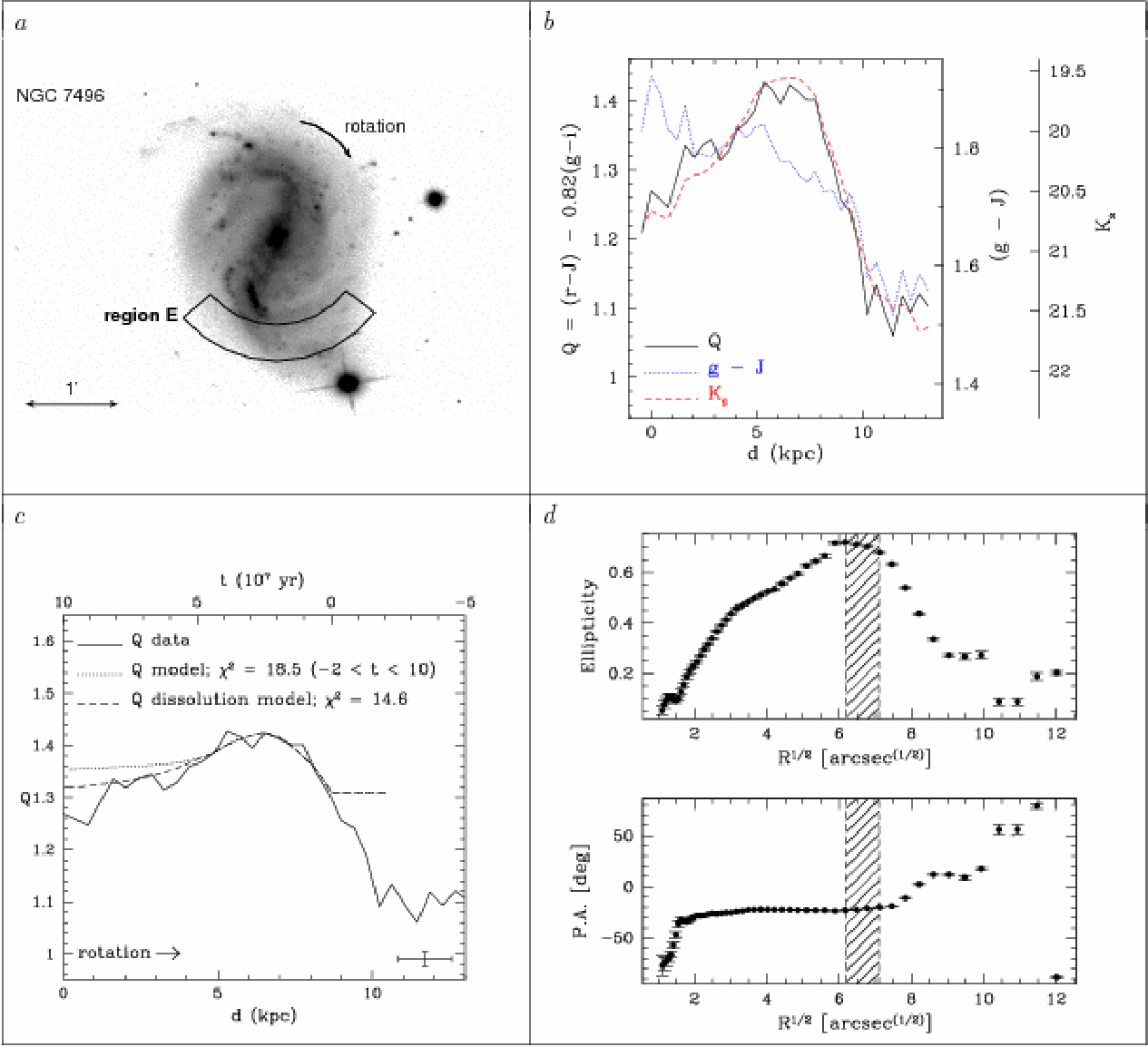}

\caption[f24.eps]{NGC~7496, region E.
({\it c.}) {\it Dotted and dashed lines:} stellar population models, IMF $M_{\rm upper} = 10 M_{\sun}$.
\label{REG_7496_E}}
\end{figure*}

\begin{figure*}
\centering
\includegraphics[scale=0.80]{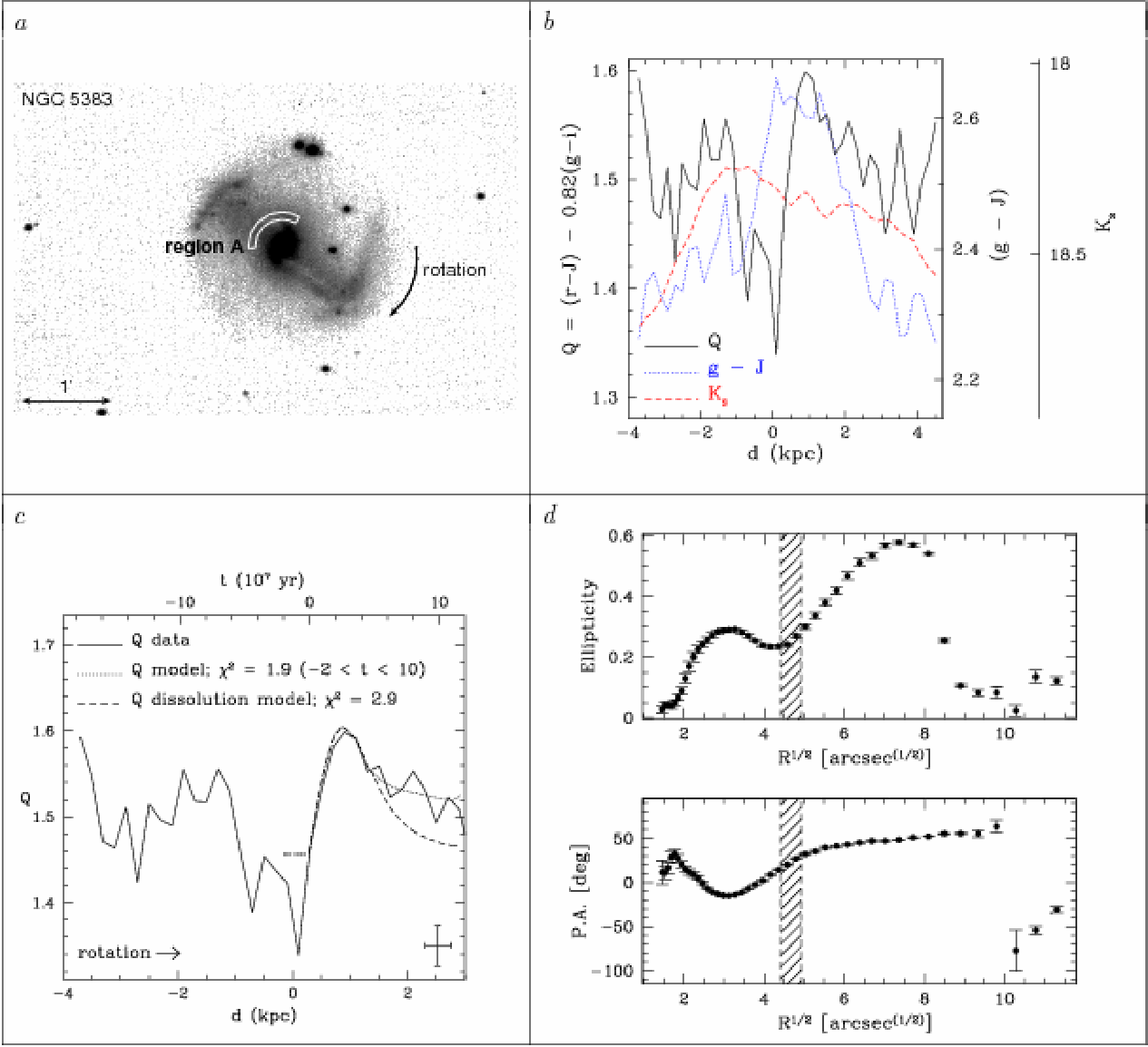}

\caption[f25.eps]{NGC~5383, region A.
({\it c.}) {\it Dotted and dashed lines:} stellar population models, IMF $M_{\rm upper} = 10 M_{\sun}$.
\label{REG_5383_A}}
\end{figure*}

\begin{figure*}
\centering
\includegraphics[scale=0.80]{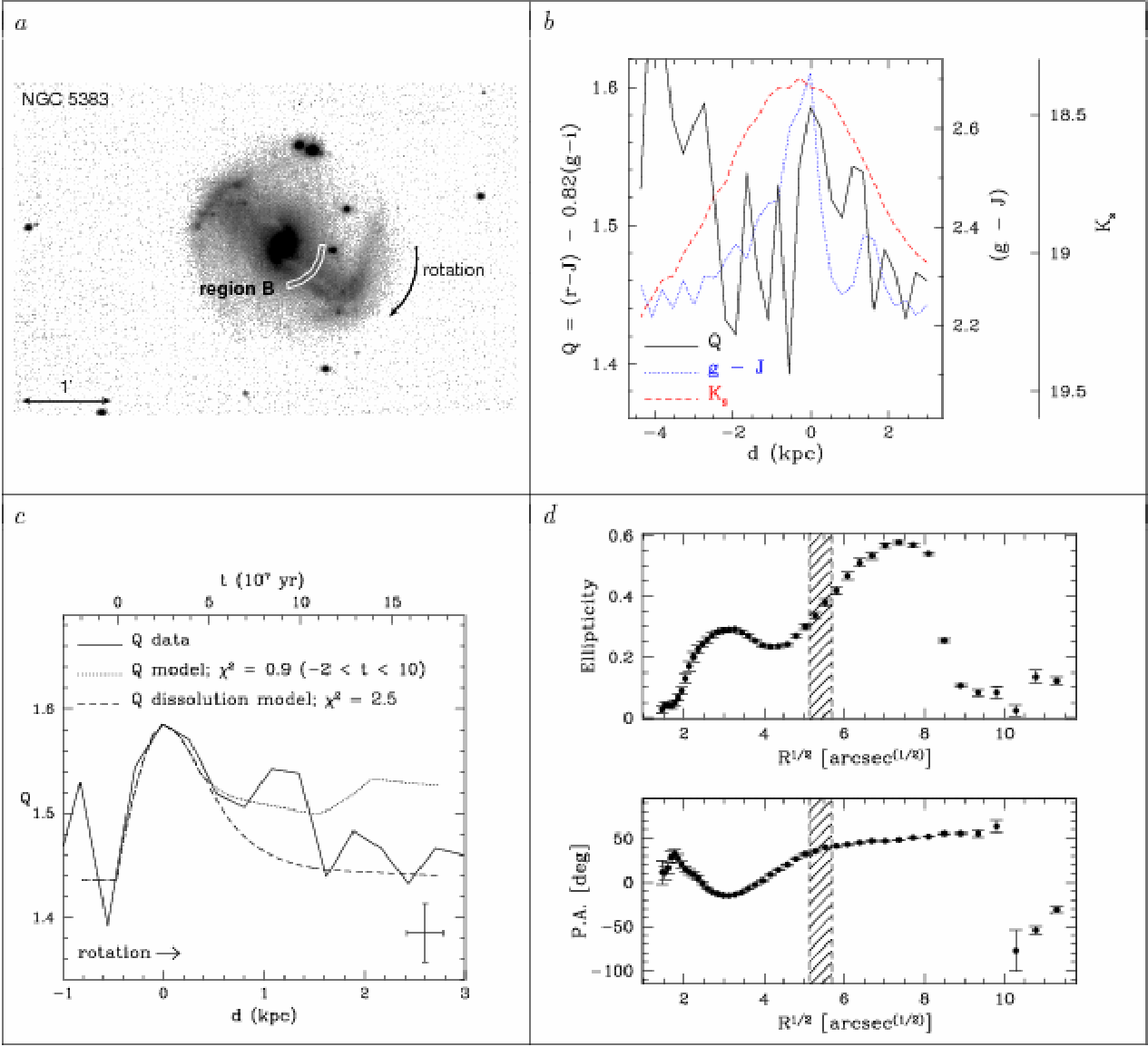}

\caption[f26.eps]{NGC~5383, region B.
({\it c.}) {\it Dotted and dashed lines:} stellar population models, IMF $M_{\rm upper} = 10 M_{\sun}$.
\label{REG_5383_B}}
\end{figure*}

\begin{figure*}
\centering
\includegraphics[scale=0.80]{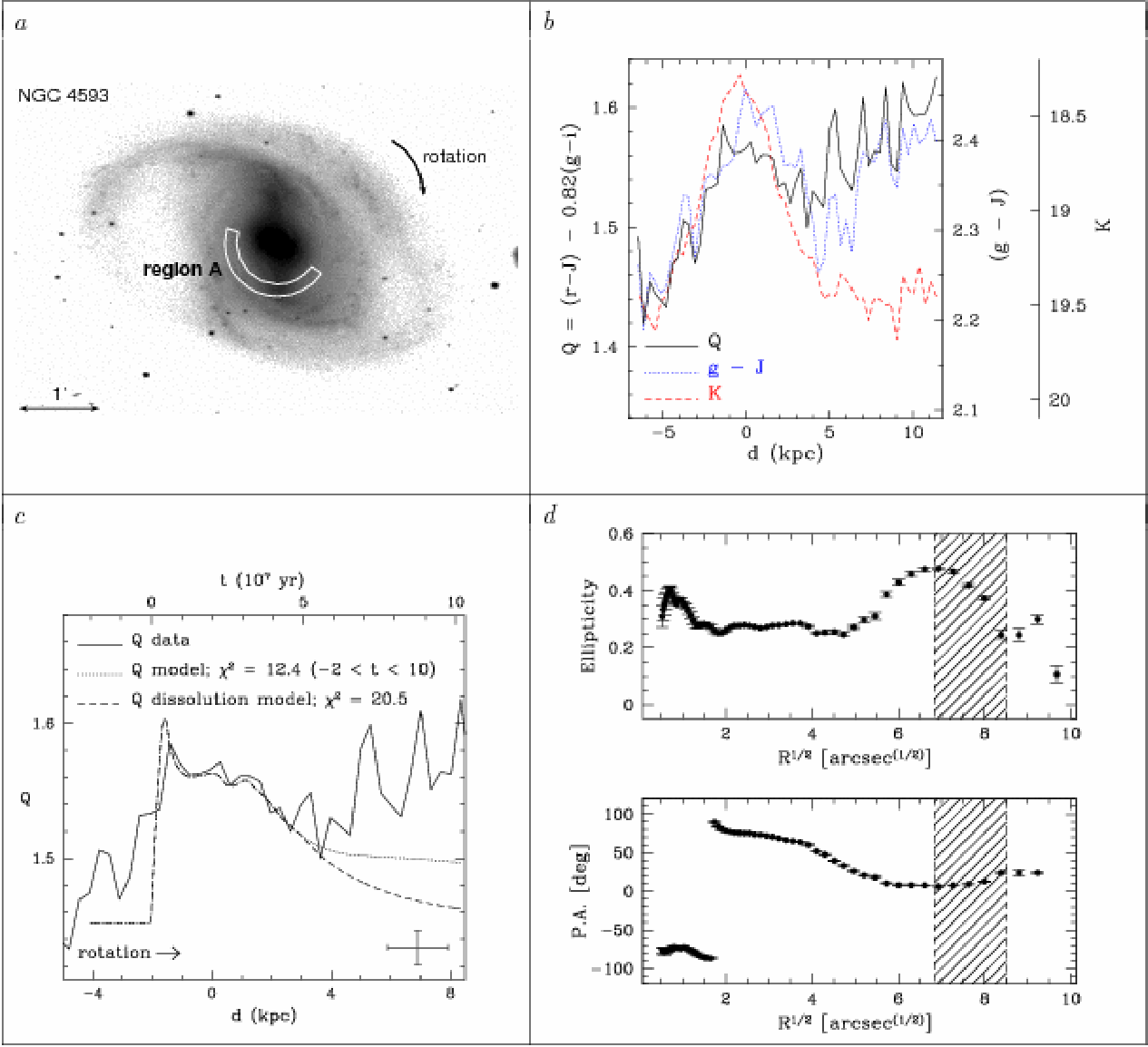}

\caption[f27.eps]{NGC~4593, region A.
({\it c.}) {\it Dotted and dashed lines:} stellar population models, IMF $M_{\rm upper} = 100 M_{\sun}$.
\label{REG_4593_A}}
\end{figure*}

\begin{figure*}
\centering
\includegraphics[scale=0.80]{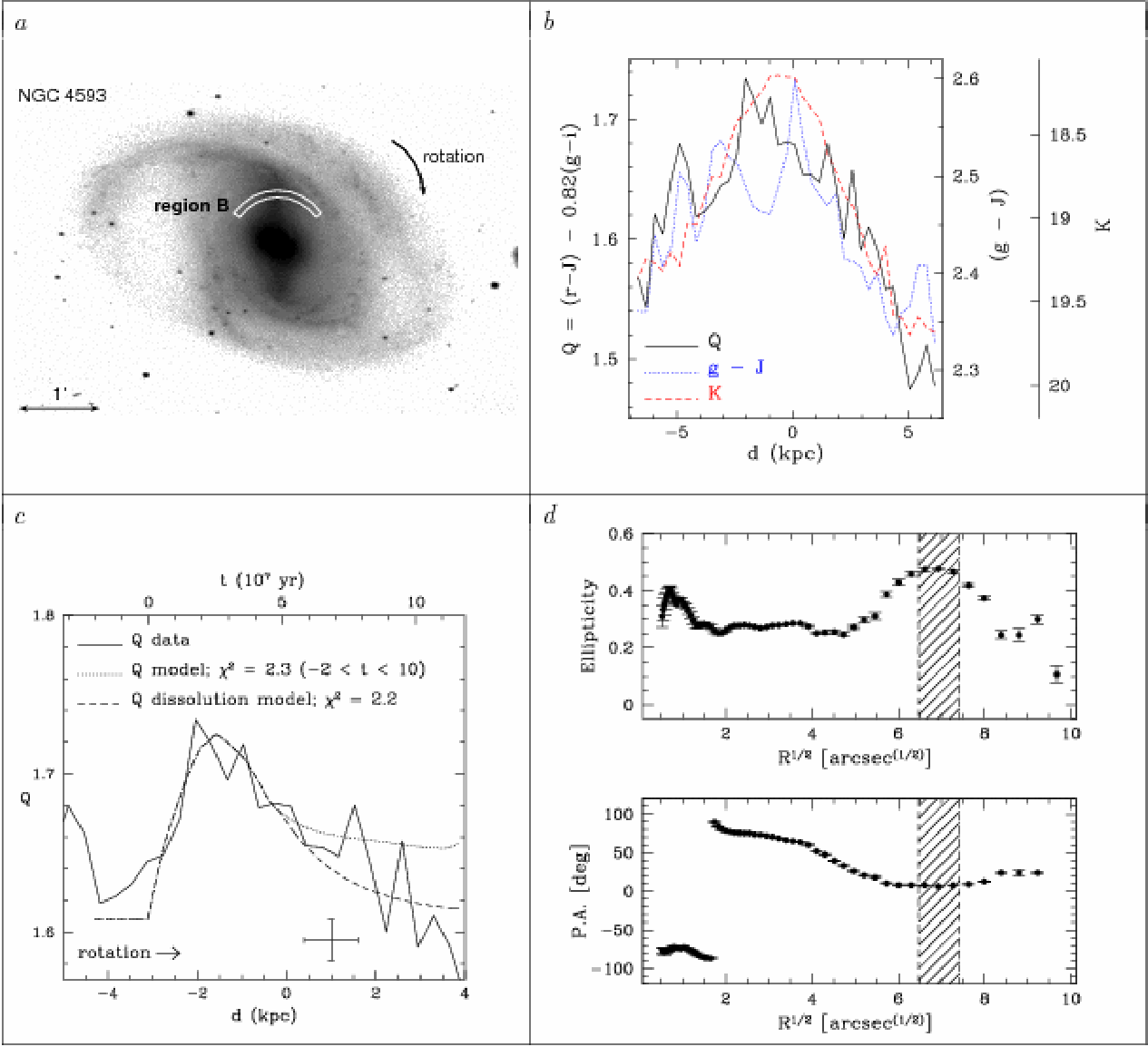}

\caption[f28.eps]{NGC~4593, region B.
({\it c.}) {\it Dotted and dashed lines:} stellar population models, IMF $M_{\rm upper} = 10 M_{\sun}$.
\label{REG_4593_B}}
\end{figure*}

\begin{figure*}
\centering
\includegraphics[scale=0.80]{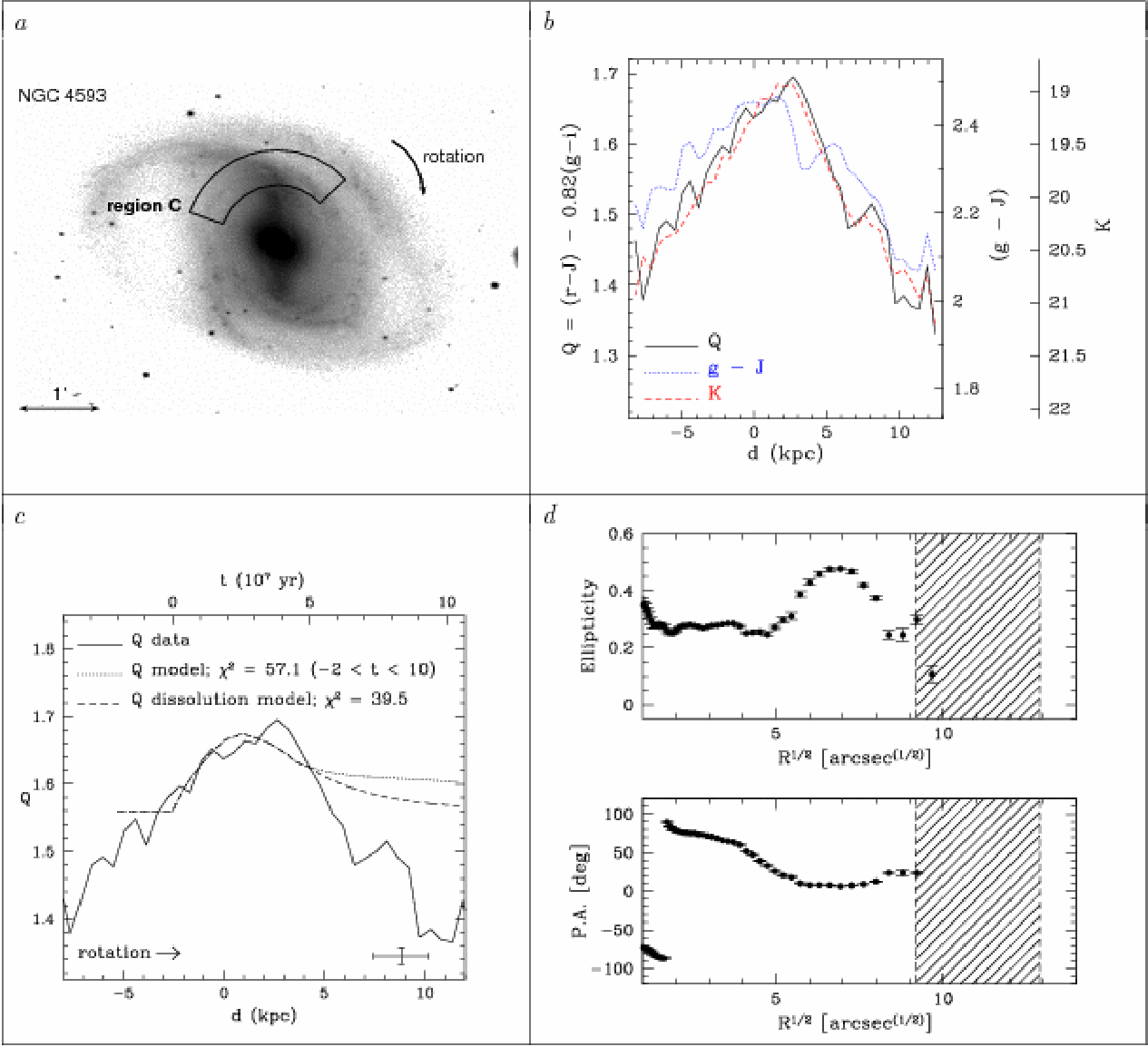}

\caption[f29.eps]{NGC~4593, region C.
({\it c.}) {\it Dotted and dashed lines:} stellar population models, IMF $M_{\rm upper} = 10 M_{\sun}$.
\label{REG_4593_C}}
\end{figure*}

\begin{figure*}
\centering
\includegraphics[scale=0.80]{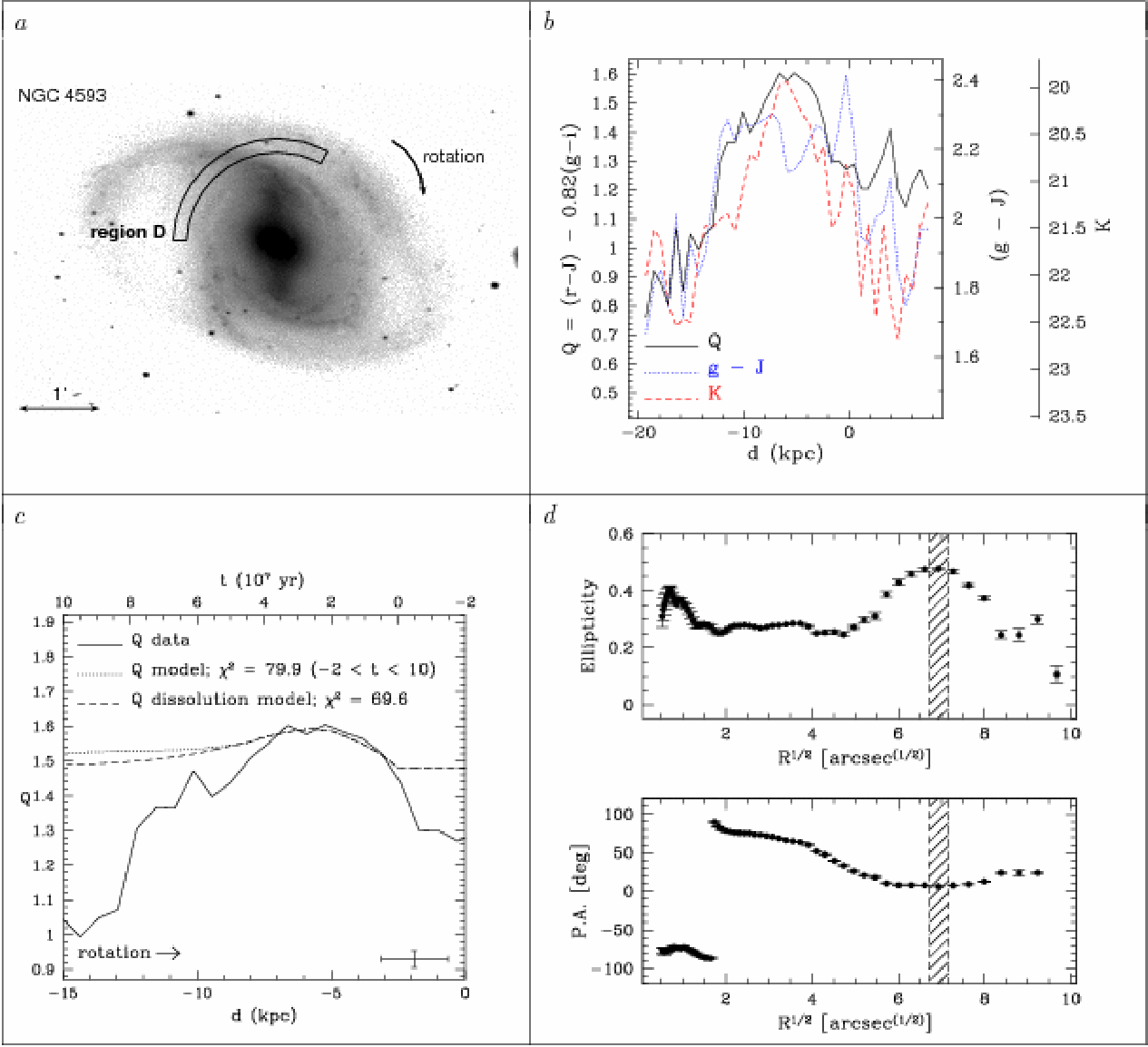}

\caption[f30.eps]{NGC~4593, region D.
({\it c.}) {\it Dotted and dashed lines:} stellar population models, IMF $M_{\rm upper} = 10 M_{\sun}$.
\label{REG_4593_D}}
\end{figure*}

\begin{figure*}
\centering
\includegraphics[scale=0.80]{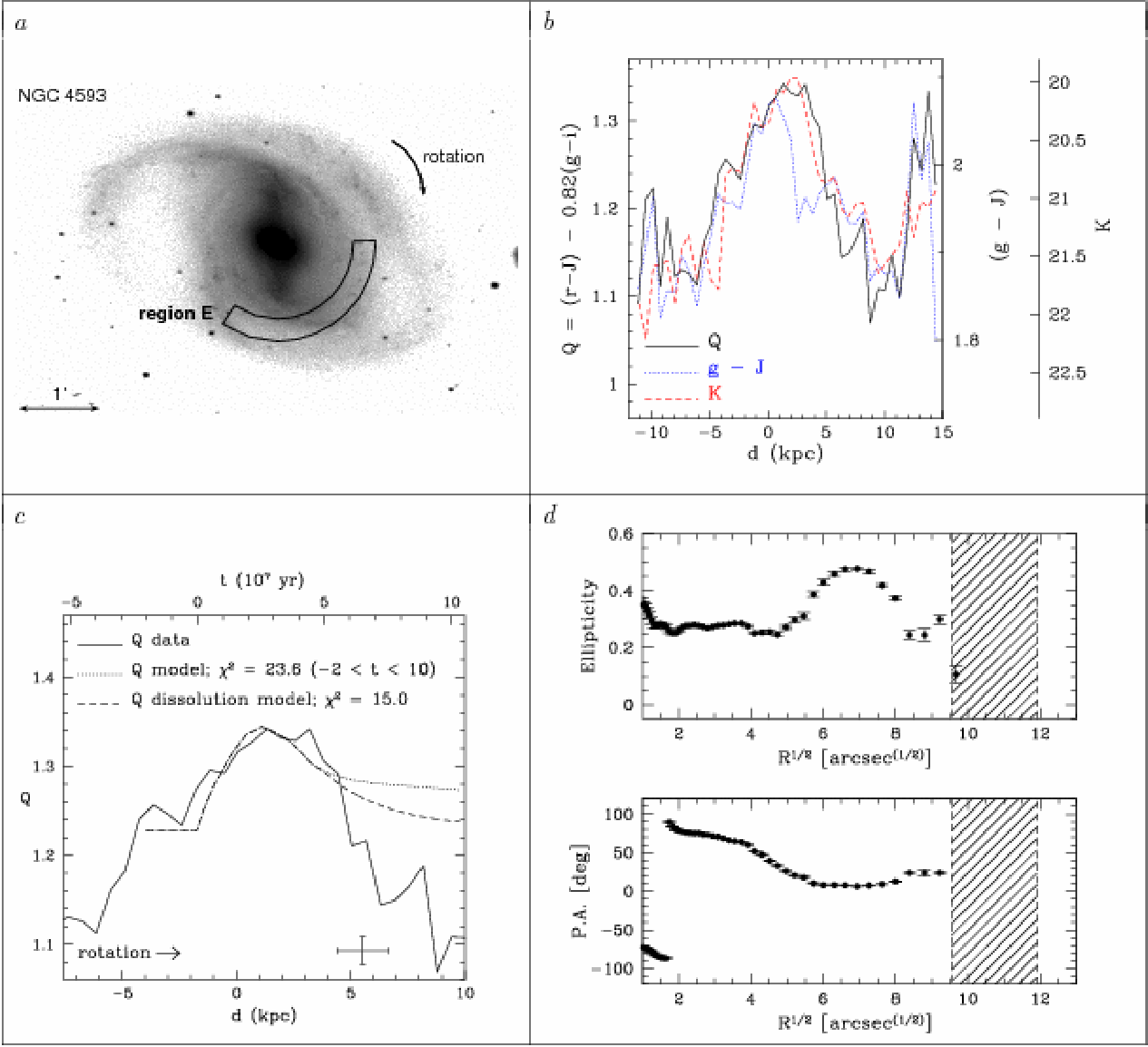}

\caption[f31.eps]{NGC~4593, region E.
({\it c.}) {\it Dotted and dashed lines:} stellar population models, IMF $M_{\rm upper} = 10 M_{\sun}$.
\label{REG_4593_E}}
\end{figure*}

\begin{figure*}
\centering
\includegraphics[scale=0.80]{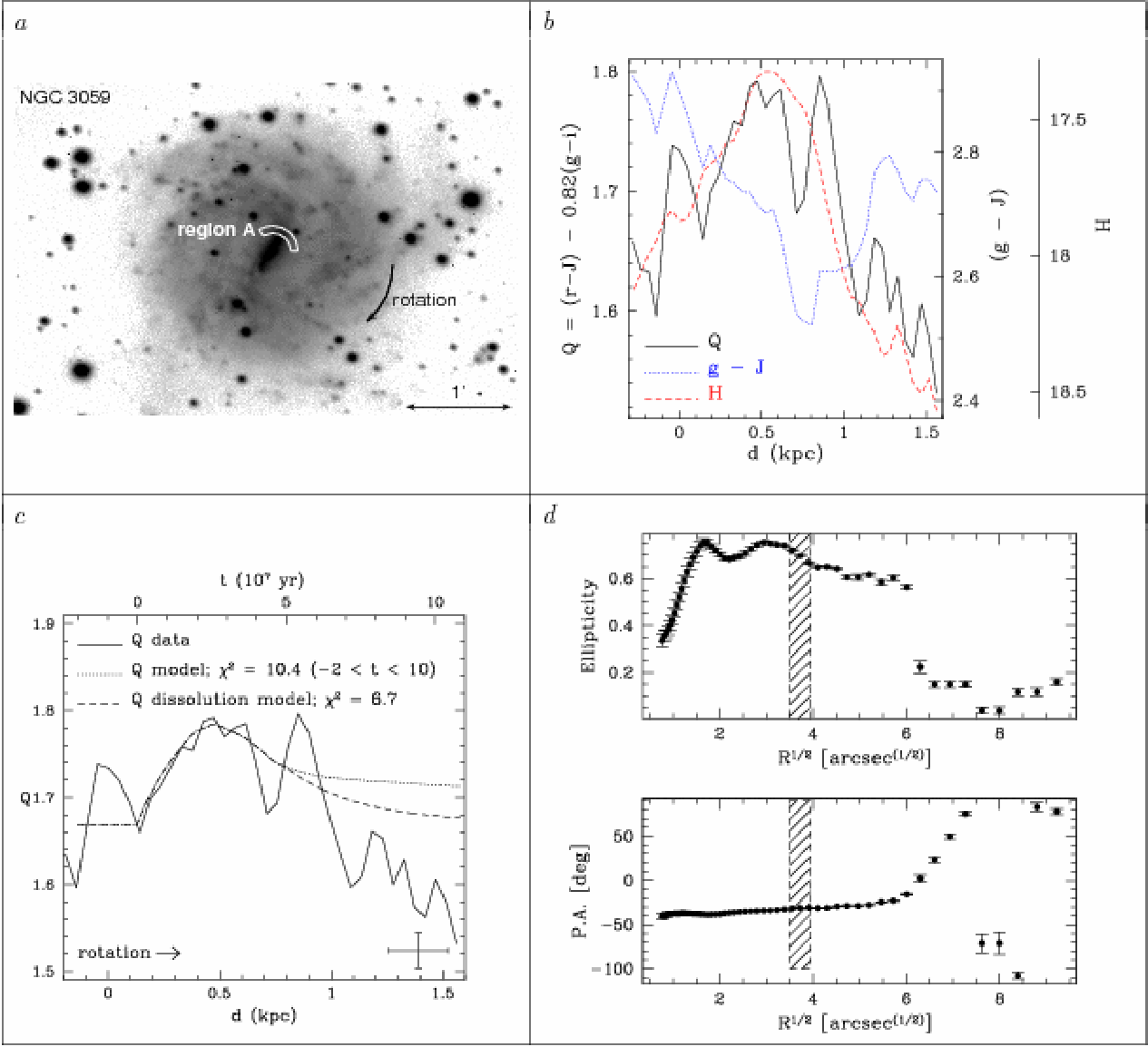}

\caption[f32.eps]{NGC~3059, region A. ({\it d}): isophotes from $H$-band mosaic.
({\it c.}) {\it Dotted and dashed lines:} stellar population models, IMF $M_{\rm upper} = 10 M_{\sun}$.
~\label{REG_3059_A}}
\end{figure*}

\begin{figure*}
\centering
\includegraphics[scale=0.80]{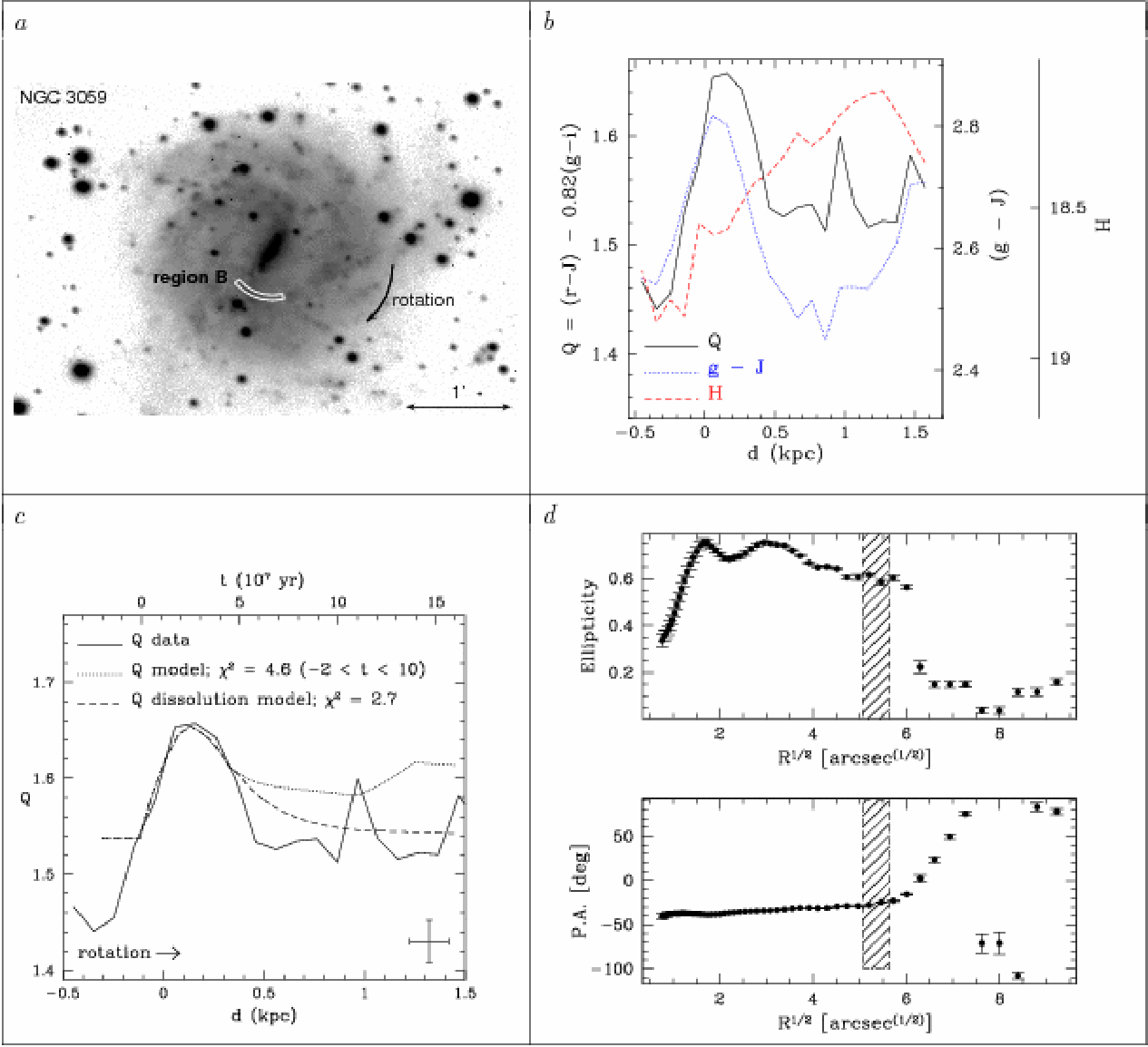}

\caption[f33.eps]{NGC~3059, region B. ({\it d}): isophotes from $H$-band mosaic.
({\it c.}) {\it Dotted and dashed lines:} stellar population models, IMF $M_{\rm upper} = 10 M_{\sun}$.
~\label{REG_3059_B}}
\end{figure*}

\begin{figure*}
\centering
\includegraphics[scale=0.80]{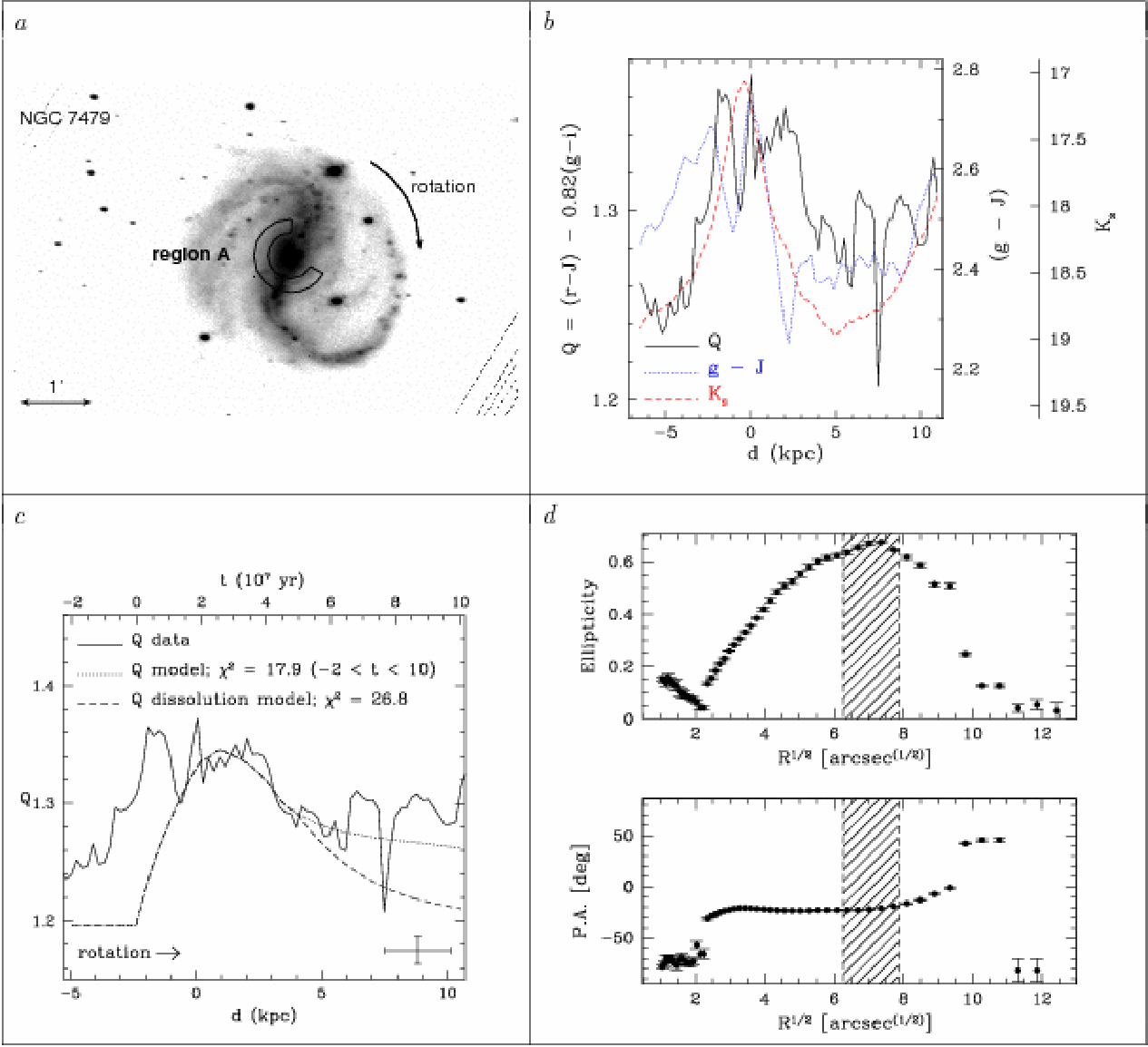}

\caption[f34.eps]{NGC~7479, region A.
({\it c.}) {\it Dotted and dashed lines:} stellar population models, IMF $M_{\rm upper} = 10 M_{\sun}$.
\label{REG_7479_A}}
\end{figure*}

\begin{figure*}
\centering
\includegraphics[scale=0.80]{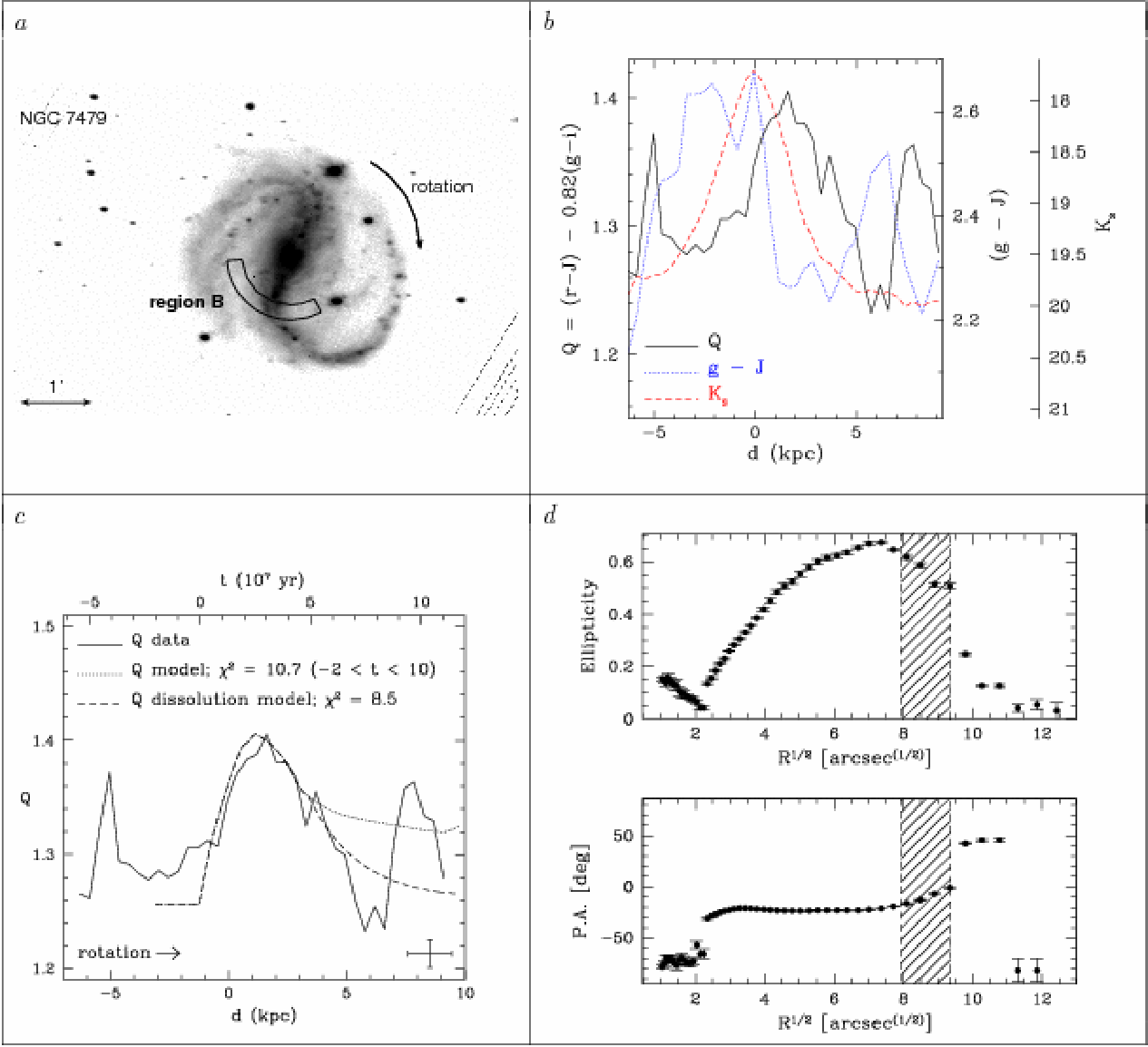}

\caption[f35.eps]{NGC~7479, region B.
({\it c.}) {\it Dotted and dashed lines:} stellar population models, IMF $M_{\rm upper} = 10 M_{\sun}$.
\label{REG_7479_B}}
\end{figure*}

\begin{figure*}
\centering
\includegraphics[scale=0.80]{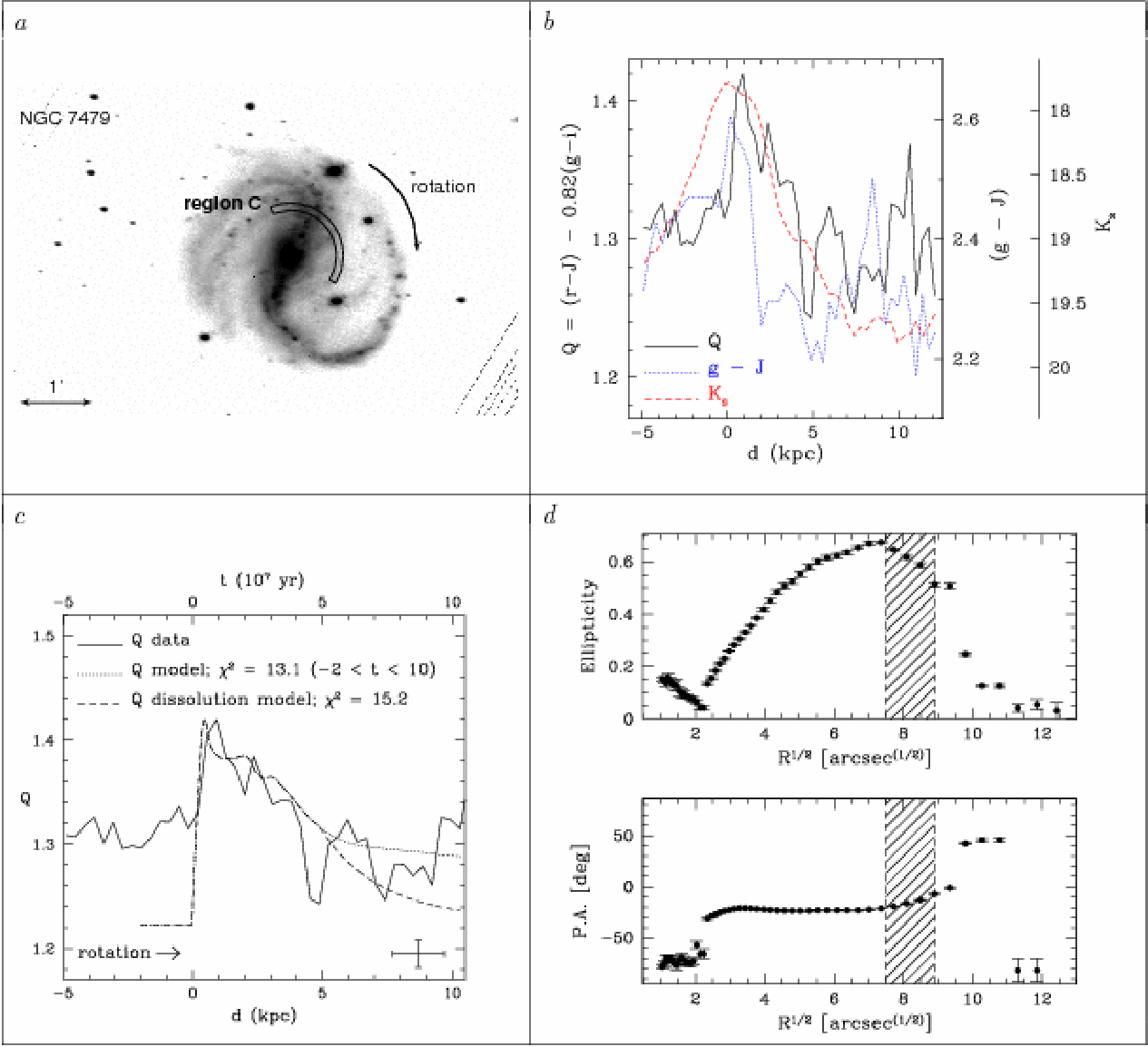}

\caption[f36.eps]{NGC~7479, region C.
({\it c.}) {\it Dotted and dashed lines:} stellar population models, IMF $M_{\rm upper} = 100 M_{\sun}$.
\label{REG_7479_C}}
\end{figure*}

\begin{figure*}
\centering
\includegraphics[scale=0.80]{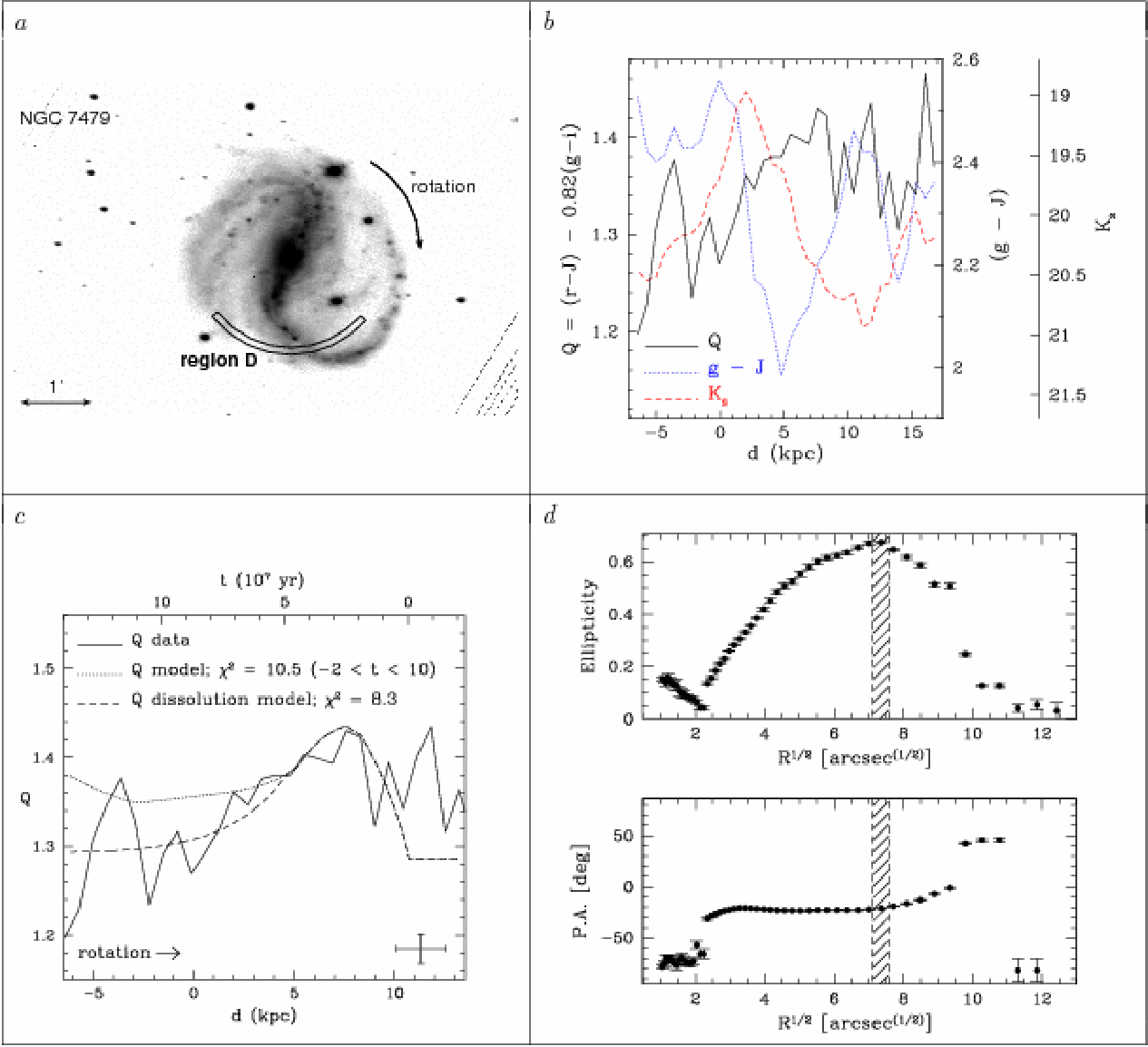}

\caption[f37.eps]{NGC~7479, region D.
({\it c.}) {\it Dotted and dashed lines:} stellar population models, IMF $M_{\rm upper} = 10 M_{\sun}$.
\label{REG_7479_D}}
\end{figure*}

\begin{figure*}
\centering
\includegraphics[scale=0.80]{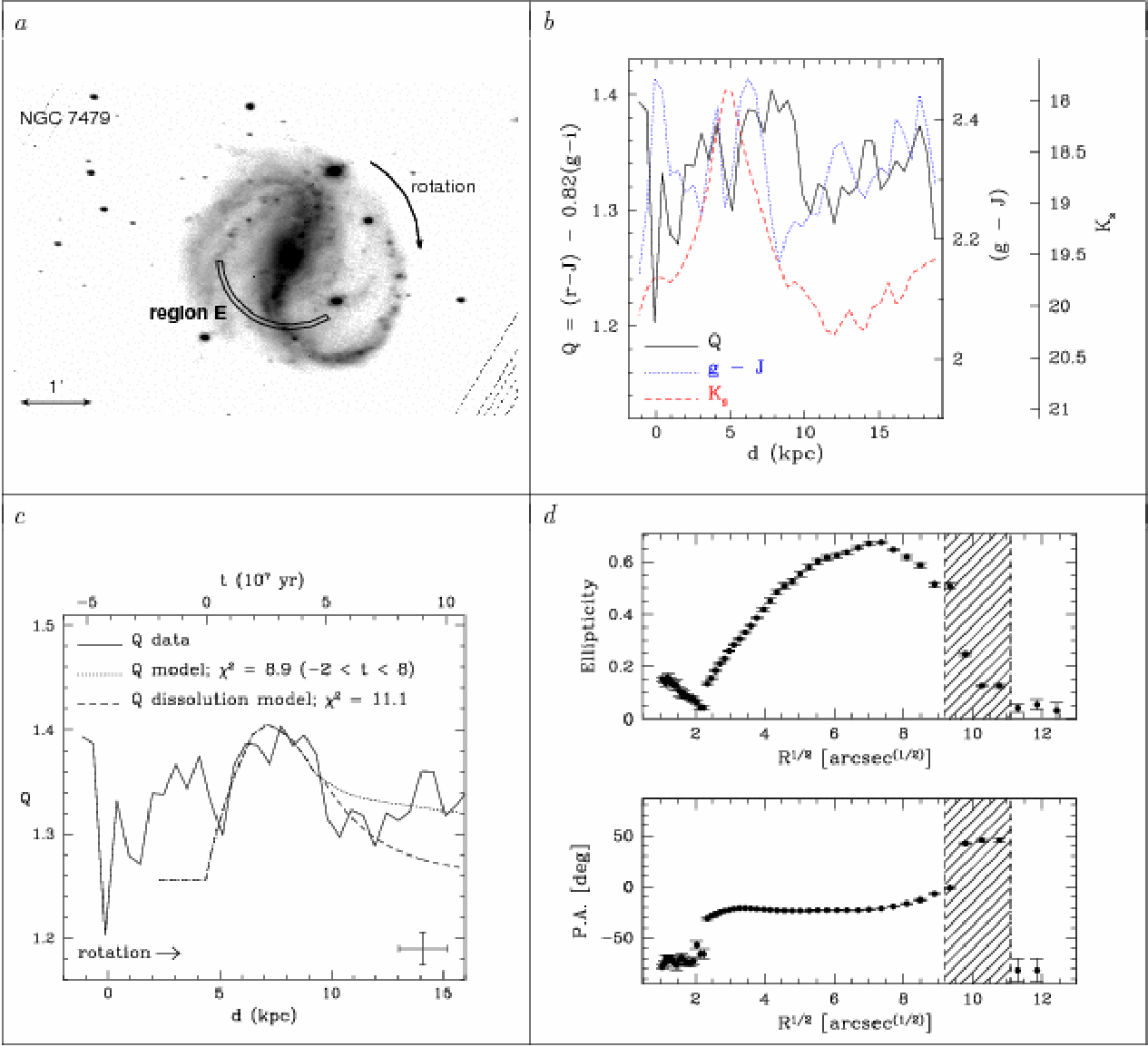}

\caption[f38.eps]{NGC~7479, region E.
({\it c.}) {\it Dotted and dashed lines:} stellar population models, IMF $M_{\rm upper} = 10 M_{\sun}$.
\label{REG_7479_E}}
\end{figure*}

\begin{figure*}
\centering
\includegraphics[scale=0.80]{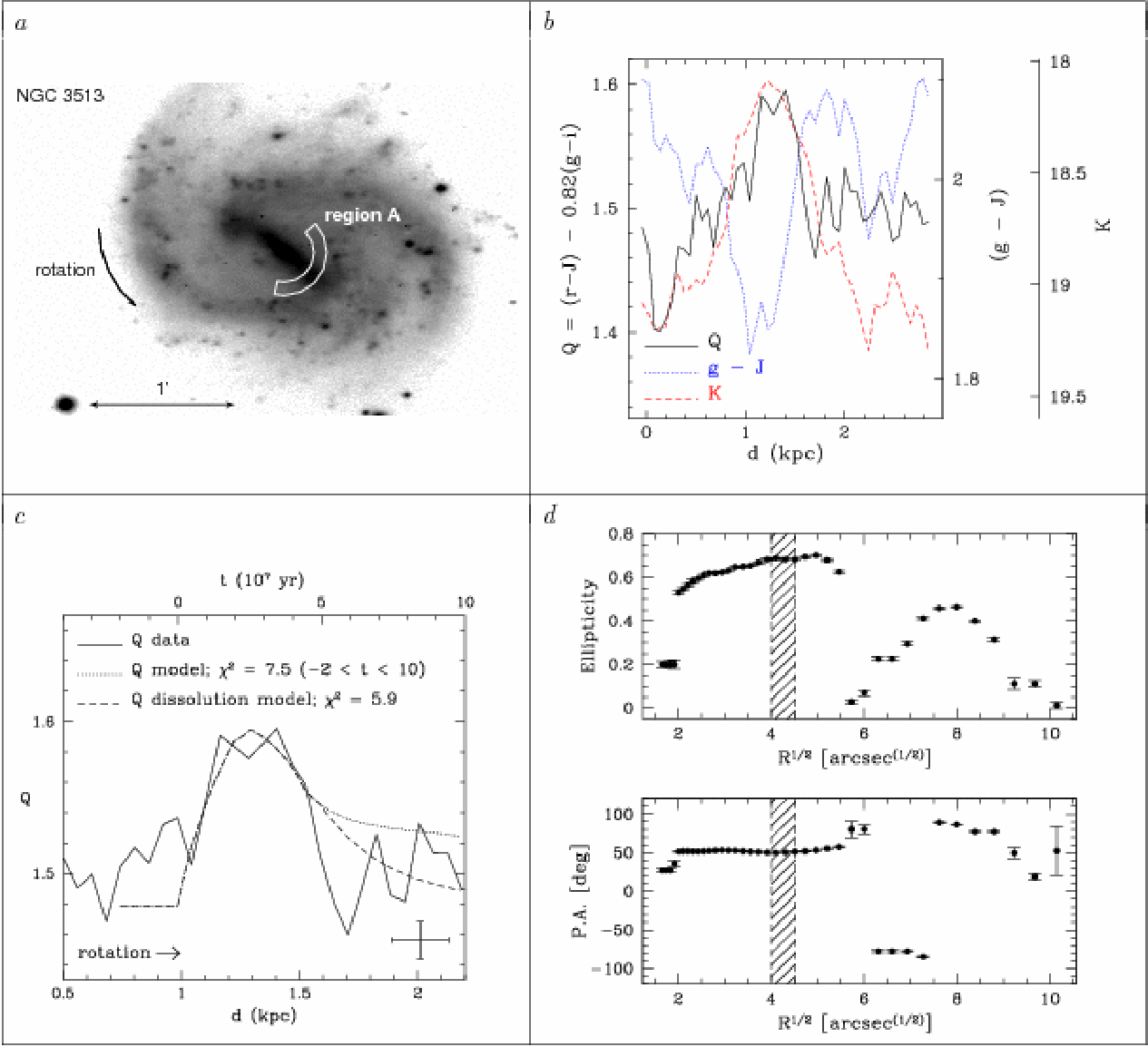}

\caption[f39.eps]{NGC~3513, region A. ({\it d}): isophotes from $H$-band mosaic.
({\it c.}) {\it Dotted and dashed lines:} stellar population models, IMF $M_{\rm upper} = 10 M_{\sun}$.
~\label{REG_3513_A}}
\end{figure*}

\begin{figure*}
\centering
\includegraphics[scale=0.80]{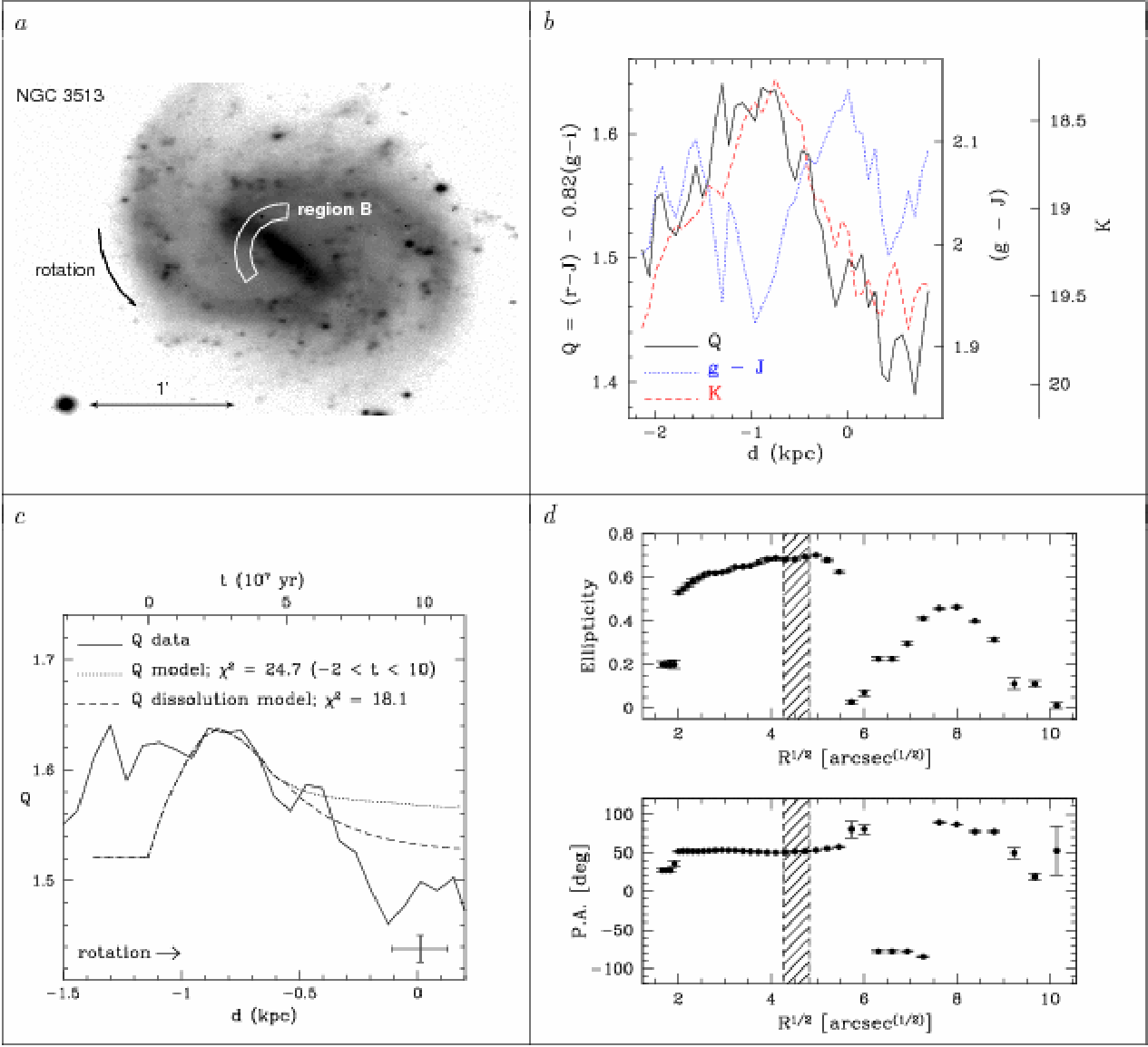}

\caption[f40.eps]{NGC~3513, region B. ({\it d}): isophotes from $H$-band mosaic.
({\it c.}) {\it Dotted and dashed lines:} stellar population models, IMF $M_{\rm upper} = 10 M_{\sun}$.
~\label{REG_3513_B}}
\end{figure*}

\begin{figure*}
\centering
\includegraphics[scale=0.80]{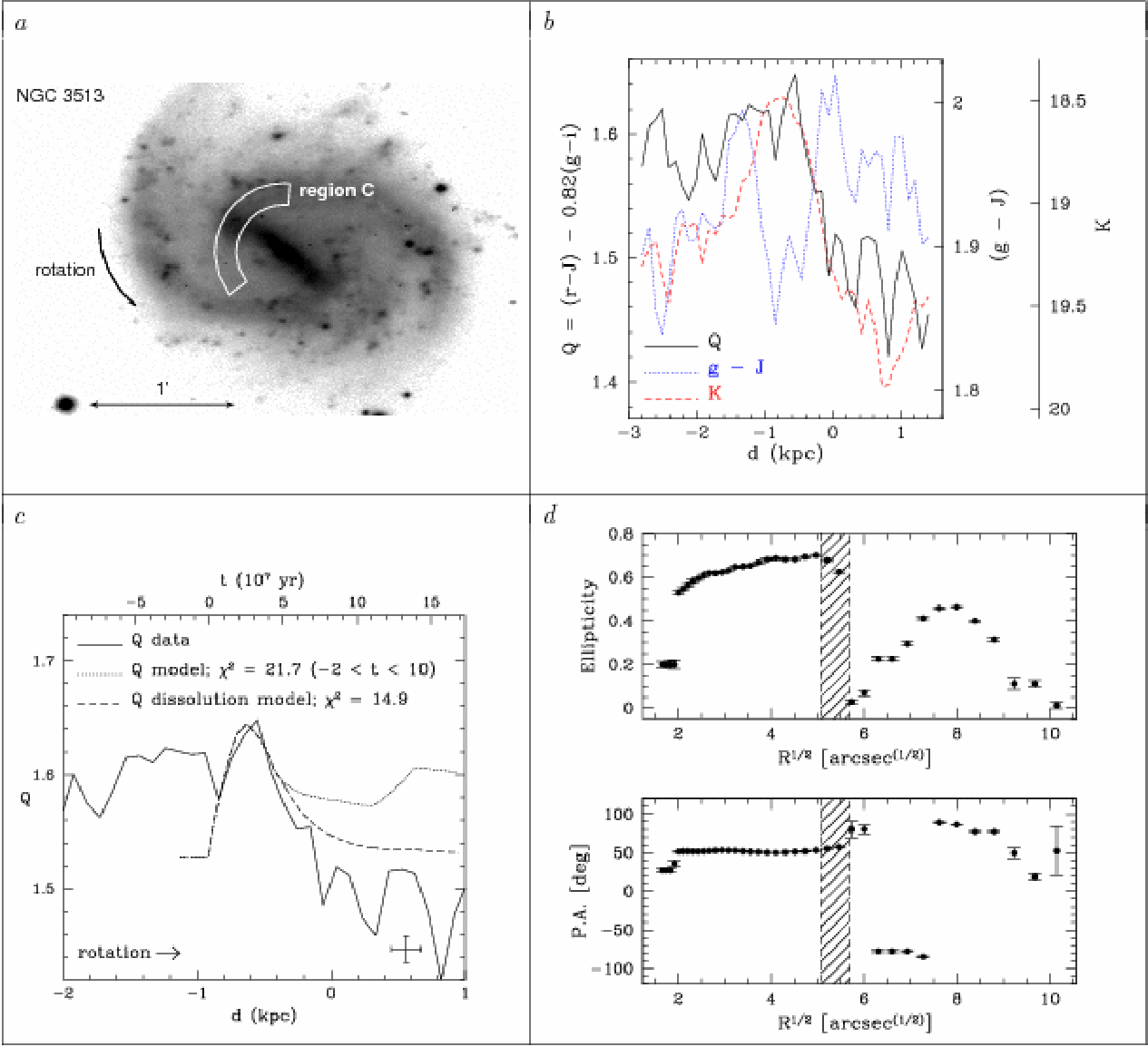}

\caption[f41.eps]{NGC~3513, region C. ({\it d}): isophotes from $H$-band mosaic.
({\it c.}) {\it Dotted and dashed lines:} stellar population models, IMF $M_{\rm upper} = 10 M_{\sun}$.
~\label{REG_3513_C}}
\end{figure*}

\begin{figure*}
\centering
\includegraphics[scale=0.80]{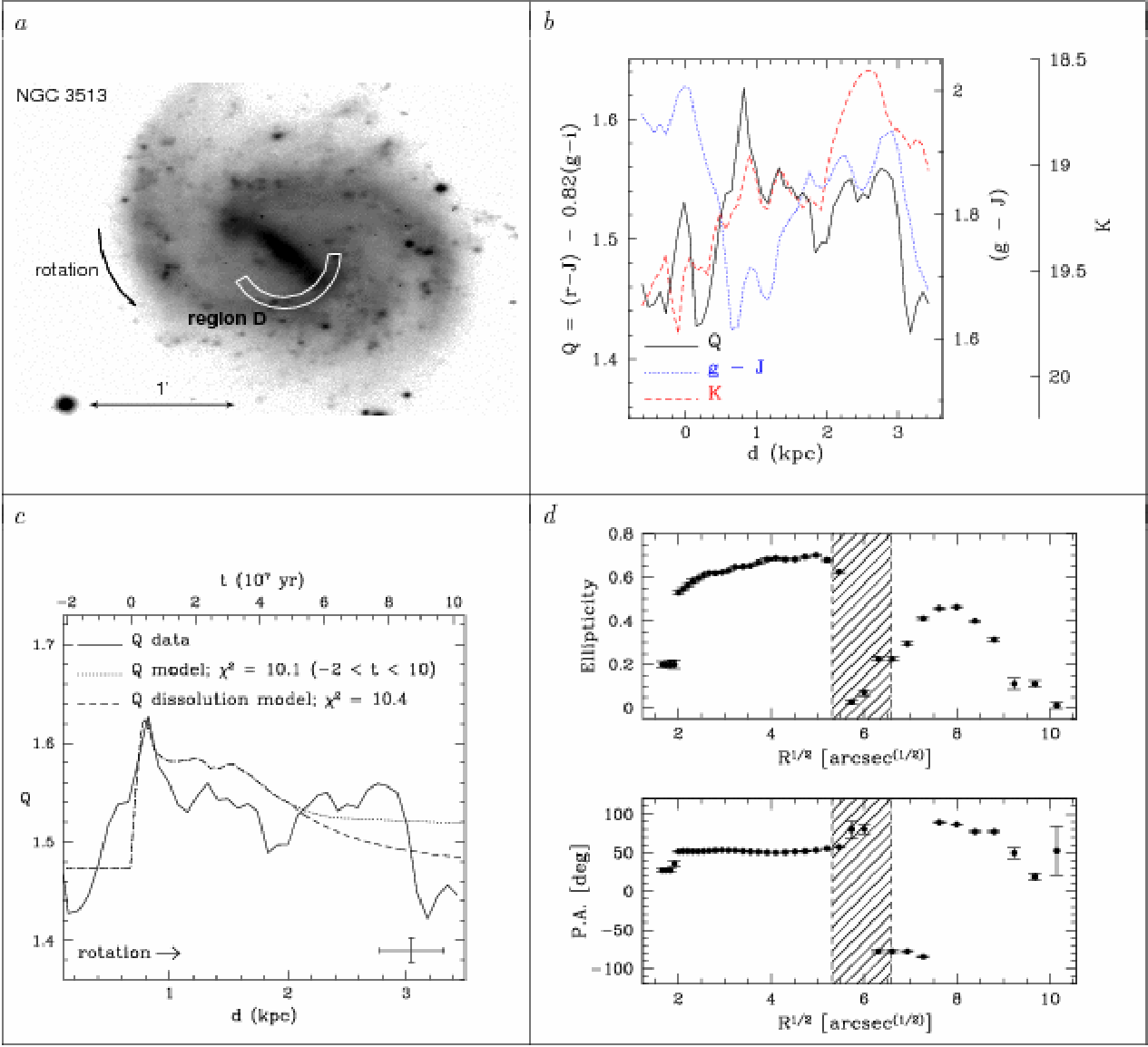}

\caption[f42.eps]{NGC~3513, region D. ({\it d}): isophotes from $H$-band mosaic.
({\it c.}) {\it Dotted and dashed lines:} stellar population models, IMF $M_{\rm upper} = 100 M_{\sun}$.
~\label{REG_3513_D}}
\end{figure*}

\begin{figure*}
\centering
\includegraphics[scale=0.80]{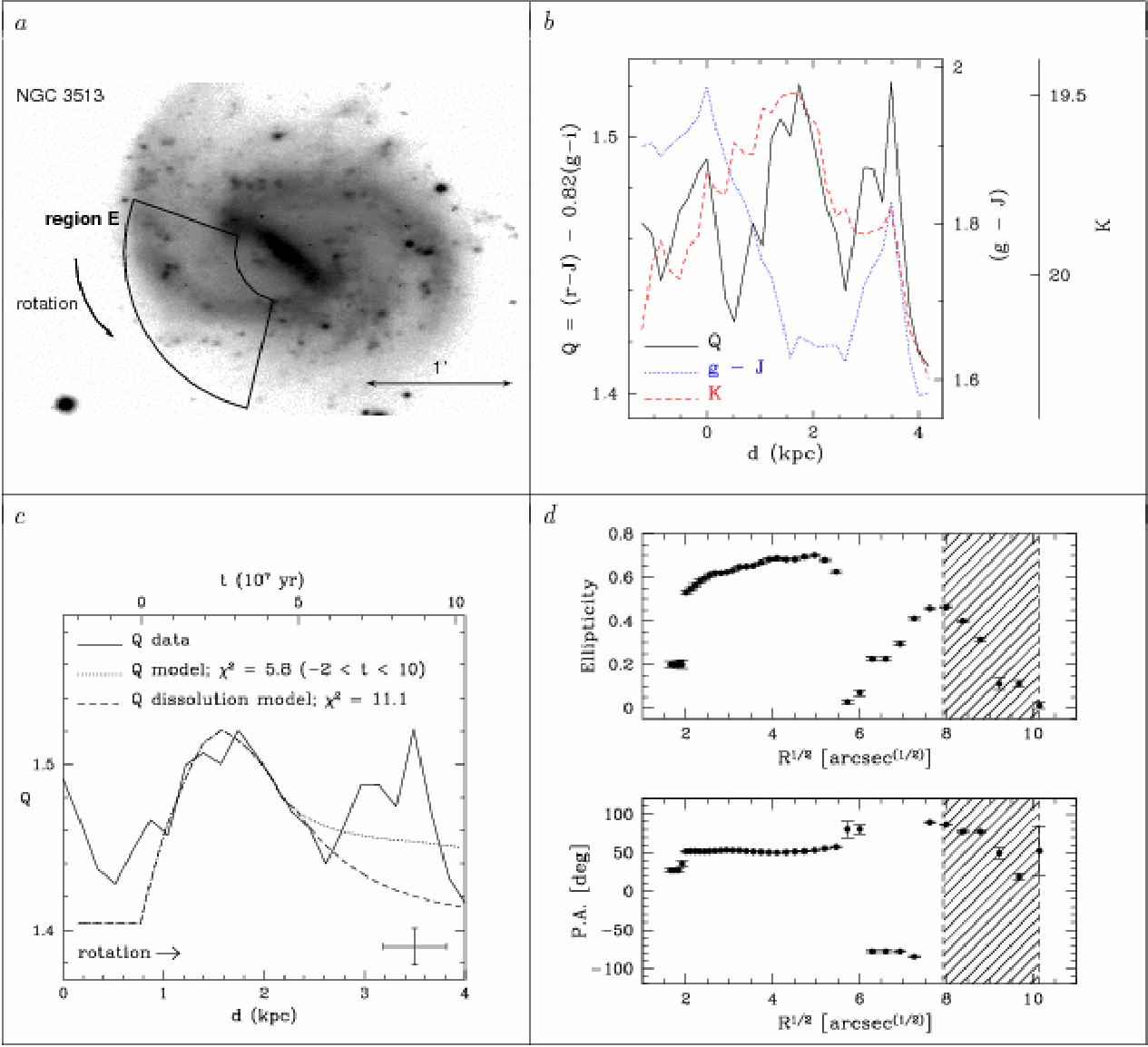}

\caption[f43.eps]{NGC~3513, region E. ({\it d}): isophotes from $H$-band mosaic.
({\it c.}) {\it Dotted and dashed lines:} stellar population models, IMF $M_{\rm upper} = 10 M_{\sun}$.
~\label{REG_3513_E}}
\end{figure*}

\begin{figure*}
\centering
\includegraphics[scale=0.80]{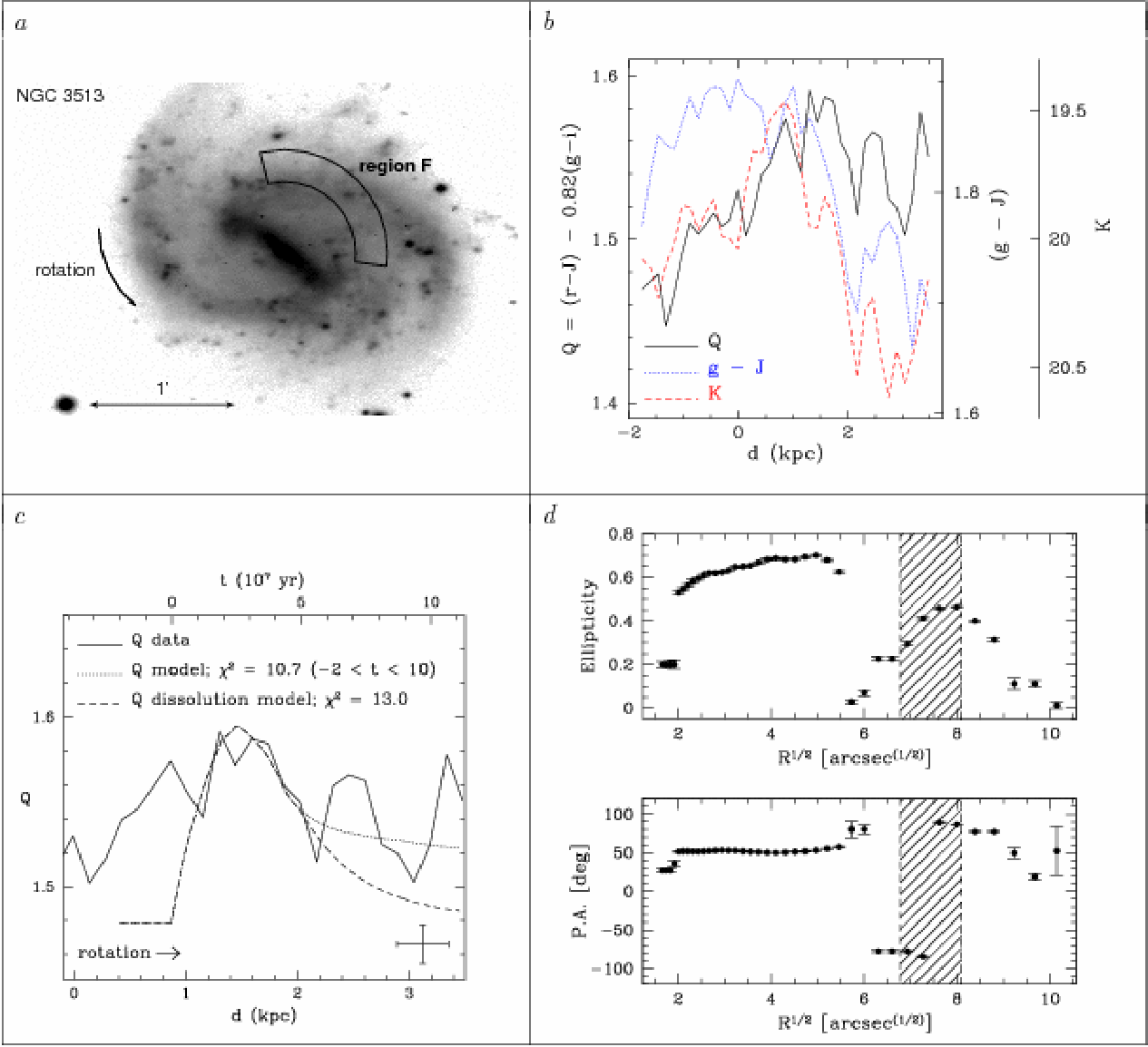}

\caption[f44.eps]{NGC~3513, region F. ({\it d}): isophotes from $H$-band mosaic.
({\it c.}) {\it Dotted and dashed lines:} stellar population models, IMF $M_{\rm upper} = 10 M_{\sun}$.
~\label{REG_3513_F}}
\end{figure*}

\begin{figure*}
\centering
\includegraphics[scale=0.80]{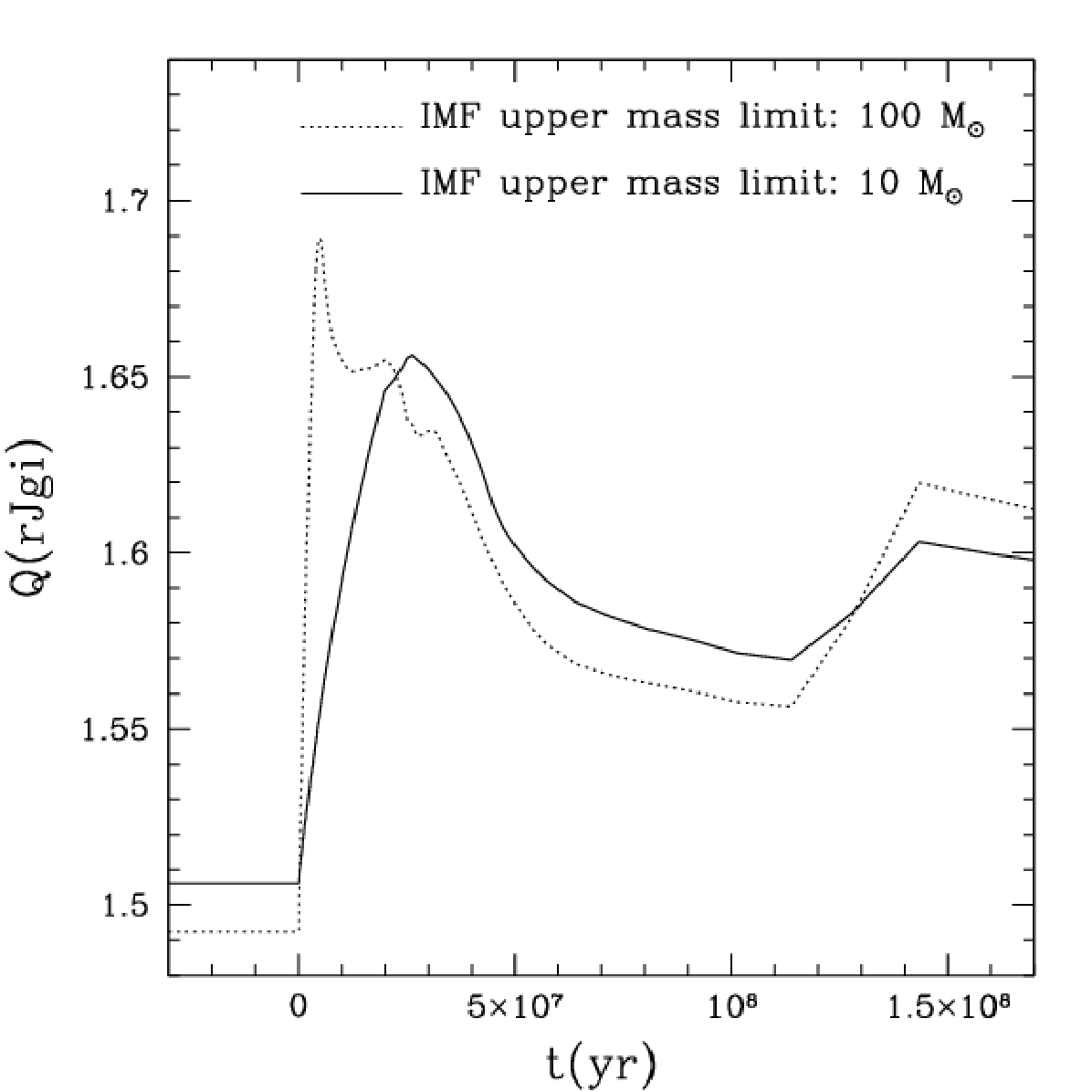}

\caption[f45.eps]{$Q(rJgi)$ profiles vs.\ time, stellar population synthesis models of S.\ Charlot \& G.\ Bruzual
with solar metallicity (2007, private communication), for the Lick system response curves.
{\it Dotted line:} IMF upper mass limit $M_{\rm upper} = 100 M_{\sun}$;
{\it solid line:}  $M_{\rm upper} = 10 M_{\sun}$. A star formation burst with duration of $2 \times 10^7$ yr
was superimposed on a background population $5 \times 10^9$ yr old. Young stars constitute
2\% by mass.~\label{fig_Models}}
\end{figure*}

In figures~\ref{REG_718_A}-\ref{REG_3513_F}, ordered by 
galaxy Hubble type, we show results for
42 regions with color gradient candidates. Of these, 22 are located
along the bar, and 20 are found in the spiral arms.
For all objects, the direction of disk rotation was established assuming that spiral arms trail.
Dust lane locations were inferred from the ($g-J$) color, and bar or spiral perturbations
were traced in the NIR. 
We fit the data with the stellar population synthesis models of S. Charlot \& G. Bruzual 
(2007, private communication);
only models with solar metallicity are considered.
The duration of the star formation burst is taken to be $2 \times 10^7$ yr; 
a fraction of young stars of 2\% by mass is mixed with 
a background of old stars $5 \times 10^9$ yr old. 
We use a Salpeter initial mass function (IMF) with a lower mass limit $M_{\rm lower} = 0.1 M_{\sun}$, and
we try two different IMF upper mass 
limits $M_{\rm upper}$: 10 and 100 $M_{\sun}$ (see figure~\ref{fig_Models}). 
It is important to have in mind that real data may have $M_{\rm upper}$ 
somewhere in between these values. 
Models with $M_{\rm upper} = 100 M_{\sun}$ exhibit a sharp peak shortly after
$t_{\rm age}=0$ yr that is absent in those with $M_{\rm upper} = 10 M_{\sun}$. 
In real data this peak may be easily lost in low signal to noise
regions or low resolution data, or confused with unnoticed artifacts or cosmic rays.
Also, there appears to be an
inverse correlation between gradient detectability and $H_\alpha$ emission
\citep[GG96,][]{mar09a},  
that could be explained if contamination from bright emission lines produced in
HII regions around the most massive stars masks the color gradients. 
We are able to fit a model with $M_{\rm upper} = 100 M_{\sun}$ only in 7 out of
42 regions (17\%).

For bar regions, we assume that stars age in the direction of
rotation.\footnote{This is the aging direction expected from
observations~\citep{zur08} and simulations~\citep{dob10}.}
In spiral regions, we search for dust lanes upstream of
star formation, and (or) establish the aging direction from 
the``match'' between the asymmetric profiles of the 
observations and the stellar models (see figure~\ref{fig_Models}).

Each one of figures ~\ref{REG_718_A}-\ref{REG_3513_F} corresponds to an analyzed region.
A deprojected mosaic of the galaxy  
(in the $g$-band, unless otherwise indicated),
in logarithmic scale, is shown in the top left panel ({\it a}).
The top right panel ({\it b}) displays the observed $Q(rJgi)$ profile
vs.\ azimuthal distance, $d$, in
kpc ({\it solid line and left y-axis}); the observed ($g-J$) color vs.\ $d$
({\it dotted line and second-from-right y-axis}),
with high values indicating dust lane locations;  
and the observed $K_s$ surface brightness (in mag arcsec$^{-2}$, lower
values correspond to higher densities of old stars) vs.\ $d$ 
({\it dashed line and rightmost y-axis}). 
In the bottom left panel ({\it c}),   
a zoomed-in version of the $Q(rJgi)$ vs.\ $d$ profile ({\it solid line}) is compared with
a stellar population model ({\it dotted line}) that has been 
``stretched'' in $t_{\rm age}$ to fit the data. 
Model stellar age, $t_{\rm age}$, in units of $10^7$ yr, is shown in the upper $x$-axis. 
The vertical error bars show the size of photometric random errors, excluding the systematic calibration
error; horizontal error bars represent possible deformations of the $Q(rJgi)$ profiles,
due to density and metallicity variations in the young and old populations~\citep[see][]{mar09a}.
We also compare the data with a model that considers the dissolution of stellar groups \citep{wie77}
after 50 Myr ({\it dashed line}, see~\S~\ref{down_falls}). Reduced values of 
\begin{equation}
    \chi^{2} = \Sigma \left( \frac{Q_{\rm data} - Q_{\rm model}}{\sigma_{Q(rJgi)}} \right)^{2}
\end{equation}
\noindent were calculated for both models in the time interval
$-20 < t_{\rm age} \rm{(Myr)} < 100$.\footnote{
Although for some of the regions this time interval may include structures not
related to the color gradient (see, e.g., regions NGC 266 A \& D, and NGC 3513 D;
figures~\ref{REG_266_A},~\ref{REG_266_D}, and~\ref{REG_3513_D}, respectively).}
The bottom right panel ({\it d}) exhibits 
isophote ellipticity vs.\ $R^{1/2}$ in the upper plot, 
and isophotal PA vs.\ $R^{1/2}$ in the lower one. Error bars were 
obtained from the ELLIPSE task in IRAF.
Hatched regions highlight the bar corotation region, as derived from the comparison between 
photometric data and stellar models. 

In table~\ref{tbl-omegas}, we show the pattern speeds and resonance radii inferred 
from the comparison of stellar population synthesis models with observed color gradients.

Remarks for each object:

\underline{NGC~718 } 
(Figures~\ref{REG_718_A} -~\ref{REG_718_C}).
Regions A and B belong to the bar region. 
For both regions, the origin ($t_{\rm age} = 0$) of the stellar model that bests fits
the observations is located in the leading side of the bar;
the two regions yield a similar bar CR radius, within the errors.
An inverse gradient is observed in spiral region NGC~718~C;
this region was fit with $M_{upper} = 100 M_{\sun}$. 
The arm and bar pattern speeds are similar in this object.

\underline{NGC~864 }
(Figures~\ref{REG_864_A} -~\ref{REG_864_D}).
Although the positions of color gradient candidates A and B, in the
bar region, are quite different, relative to the bar $K_s$ surface brightness and dust 
($g-J$) profiles, their analysis provides a similar CR position, within the errors.
This barred galaxy has spiral arms with a ``ragged'' structure.
Region C, located in the beginning of the eastern arm (left arm in the deprojected image), 
gives a corotation position near the end of the  spiral arms, at $\sim 76.5\pm0.5$ arcsec.
Region D, situated in an arm structure apparently decoupled from the main pattern, gives a corotation position
similar to that of region C, within the errors. 
The two main arms of this object have different $Q$ mean values,
possibly owing to different levels of star formation activity\footnote{
This behavior was dubbed ``$Q$ effect'' in~\citet{mar09a}.}

\underline{NGC~4314 }
(Figures~\ref{REG_4314_A} -~\ref{REG_4314_B}).
Two color gradient candidates were found in the bar of this object. 
The one in region A is located in the leading side of the bar;
the CR radius inferred from the comparison between theoretical and observed $Q(rJgi)$ profiles
is near the location of maximum ellipticity.
The color gradient candidate in region B is located in the trailing side of the bar;
the inferred CR agrees, within the errors, with the result from region A.
No color gradients were found in the arms. 

\underline{NGC~266 }
(Figures~\ref{REG_266_A} -~\ref{REG_266_D}).
Unfortunately, the error due to the inclination angle, $\alpha$, is higher than the
$\alpha$ value itself (see table~\ref{tbl-param}). Although this 
produces very large errors in the computed
$\Omega_{p}$ values, the errors in the resonance positions are
reasonable, when equations A1 and A3 from~\citet{mar09a} are used.
Region A is the only one detected in the bar of this object, and its analysis yields
a CR position close to the maximum of isophotal ellipticity;
this region was fitted with a model with $M_{\rm upper} = 100 M_\sun$.
Regions B (presumably just before corotation, $M_{upper} = 10 M_{\sun}$)
and D (presumably just after corotation, $M_{upper} = 100 M_{\sun}$)
give similar resonance positions that also agree with the results of region A, within the errors.
The resonance position derived from region C 
is within 1 $\sigma$ of the result of region A, and 
within 2 $\sigma$ of the position found from regions B and D. 
Regions B and C appear to be associated with a
``ring'' feature, rather than with the spiral arms.

\underline{NGC~986 }
(Figures~\ref{REG_986_A} -~\ref{REG_986_E}).
This galaxy shows an important activity in the bar region when observed in the
$Q(rJgi)$ index. The bar's end is made evident more by the change in PA 
of the isophotes, than by the location of their maximum ellipticity.
Region A harbors a color gradient candidate in the bar whose analysis results in a CR radii
close to the place where the isophotes change PA.
Region B is located near the end of the bar;
the fit of a stellar model (with $M_{\rm upper} = 100 M_{\sun}$) to its 
observed $Q(rJgi)$ profile gives a CR position that does not coincide with the bar's end.
For regions C and D, in one of the arms, the corotation position coincides (within the errors)
with the bar's end.

\underline{NGC~7496 }
(Figures~\ref{REG_7496_A} -~\ref{REG_7496_E}).
The northern spiral arm (lower arm in our deprojected image) displays 
a compact elongated region of high surface brightness, even in $K_s$; 
this arm is also more extended in radius, when compared to the southern arm
(upper arm in our deprojected image).
Regions A and B are located in the bar and give similar CR 
radii, that encompass the location of maximum ellipticity of the bar. 
Region C is probably located near the bar's CR, a fact that makes
the determination of dynamic parameters uncertain.
The observations, and the fit of the stellar model to regions D and E indicate
inverse color gradients. Corotation is close to
the ellipticity maximum of the bar; all the derived positions 
agree within 1 $\sigma$, except for the determination from
region D.

\underline{NGC~5383 }
(Figures~\ref{REG_5383_B} -~\ref{REG_5383_B}).
This barred galaxy has very short spiral arms that do not reach 
the outer disk.
The analysis of color gradient candidates in
regions A and B results in similar CR positions before the end of the bar.
This may be due to strong non-circular motions that lead to an overestimate
of $\Omega_{p}$, and hence to an underestimation of the CR radius. 

\underline{NGC~4593 }
(Figures~\ref{REG_4593_A} -~\ref{REG_4593_E}).
Regions A ($M_{\rm upper} = 100 M_{\sun}$) and B ($M_{\rm upper} = 10 M_{\sun}$), in the
bar region, give a similar CR position near the maximum of ellipticity. Region C is
located near the bar's end, and is probably associated with a ``ring'' feature.
Region D, inverse color gradient in the spiral arm region,
gives a similar corotation position in accordance with regions A and B.
Region E (``direct'' color gradient)
yields a corotation position further away from the galaxy center than the one inferred from region D.
The eastern and western sides of the bar (respectively, upper and lower sides in
the deprojected frame) have different $Q(rJgi)$ mean values.

\underline{NGC~3059 }
(Figures~\ref{REG_3059_A} -~\ref{REG_3059_B}).
This object is likely a double-bar system, and
it is difficult to assess the bar's end 
from the ellipticity and PA vs.\ radius plots. 
However, the CR position derived from the bar region A
lies close to the bar endpoint 
determined from visual inspection ($\sim 19.2 \pm 1.4$ arcsec).
Region B in the arms, gives a corotation position at radii larger than the bar's end. 

\underline{NGC~7479 }
(Figures~\ref{REG_7479_A} -~\ref{REG_7479_E}).
According to~\citet{woz95}, the HII regions located along the bar, near the dust lanes,
show a ``stretched" appearance probably due to strong gas flows that may trigger star formation.
Regions A, B, ($M_{\rm upper} = 10 M_{\sun}$), and C ($M_{\rm upper} = 100 M_{\sun}$)
show candidate color gradients in the bar. 
The comparison of stellar models with observations renders CR 
positions close to the ellipticity maximum (i.e., close to the bar's end). 
Region D, situated in one of the spiral arms and after the
bar's corotation, features an inverse color gradient.
The computed resonance positions agree,
within the errors, for regions A through D. 
Region E lies near the bar's end; even though no color gradients are expected at this position,\footnote{
At corotation the pattern and the rotating material have the same angular velocity,
hence shocks should not occur (at least for non-barred spirals).} the $Q(rJgi)$ profile
indicates that some star formation is taking place. 

\underline{NGC~3513}
(Figures~\ref{REG_3513_A} -~\ref{REG_3513_F}).
Regions B, C, E, and F exhibit several adjacent $Q(rJgi)$ profiles, so it is hard to define
a unique color gradient candidate. Regions A and B belong to the bar,
and their analysis locates CR just inside its end. 
Region C is located very close to the bar's end. 
Region D, in the spiral arm region, is presumably located inside corotation;
the best fit of the models to the observed $Q(rJgi)$ profiles 
suggests $M_{\rm upper} = 100 M_{\sun}$ for this region,
and yields a corotation radius close to the change in ellipticity and PA.
In contradiction with the region D result,
regions E and F indicate that the bar and spiral perturbations may rotate with different
pattern speeds.
Once again, $Q(rJgi)$ has different mean values in both arms.

\clearpage

\begin{deluxetable}{cccccccccc}
\tabletypesize{\scriptsize}
\rotate
\tablecaption{Observed and derived dynamic parameters.\label{tbl-omegas}}
\tablewidth{0pt}
\tablehead{
\colhead{Galaxy and region}
& \colhead{Figure}
& \colhead{Location}
& \colhead{$R_{\rm mean}$ (arcsec)}
& \colhead{$R_{\rm mean}$ (kpc)}
& \colhead{$R_{\rm bar}$ (arcsec)}
& \colhead{$R_{\rm bar}$ (kpc)} 
& \colhead{$\Omega_p$ (km s$^{-1}$ kpc$^{-1}$) }
& \colhead{$R_{\rm CR}$ (arcsec)}
& \colhead{$R_{\rm CR}$ (kpc)}
}

\startdata
NGC718 A	&~\ref{REG_718_A}~ &	Bar	& $	20.7	\pm	0.25	$ & $	2.4	\pm	0.2	$ &	22.4	$\pm$	2.0	($K_{s}$)	 &	2.6	$\pm$	0.3	&  $	37.9	\pm	24.2	$     & $	26.7	\pm	5.5	$      & $	3.1	\pm	0.6	$ \\
NGC718 B	&~\ref{REG_718_B}~ &	Bar	& $	21.2	\pm	0.25	$ & $	2.5	\pm	0.2	$ &	22.4	$\pm$	2.0	($K_{s}$)	 &	2.6	$\pm$	0.3	&  $	41.4	\pm	23.6	$     & $	24.4	\pm	3.6	$      & $	2.9	\pm	0.4	$ \\
NGC718 C	&~\ref{REG_718_C}~ &	Spiral	& $	41.4	\pm	0.25	$ & $	4.9	\pm	0.4	$ &	22.4	$\pm$	2.0	($K_{s}$)	 &	2.6	$\pm$	0.3	&  $	33.6	\pm	13.2	$     & $	30.1	\pm	3.0	$      & $	3.5	\pm	0.4	$ \\
\tableline
NGC864 A	&~\ref{REG_864_A}~ &	Bar	& $	26.0	\pm	0.25	$ & $	2.8	\pm	0.2	$ &	36.1	$\pm$	3.5	($K_{s}$)	 &	3.8	$\pm$	0.5	&  $	38.6	\pm	6.5	$     & $	36.5	\pm	4.7	$      & $	3.9	\pm	0.5	$ \\
NGC864 B	&~\ref{REG_864_B}~ &	Bar	& $	27.7	\pm	0.25	$ & $	3.0	\pm	0.3	$ &	36.1	$\pm$	3.5	($K_{s}$)	 &	3.8	$\pm$	0.5	&  $	44.1	\pm	6.6	$     & $	31.9	\pm	3.4	$      & $	3.4	\pm	0.4	$ \\
NGC864 C	&~\ref{REG_864_C}~ &	Spiral	& $	40.4	\pm	0.25	$ & $	4.3	\pm	0.4	$ &	36.1	$\pm$	3.5	($K_{s}$)	 &	3.8	$\pm$	0.5	&  $	18.6	\pm	4.1	$     & $	75.5	\pm	13.5	$      & $	8.1	\pm	1.4	$ \\
NGC864 D	&~\ref{REG_864_D}~ &	Spiral	& $	61.1	\pm	0.25	$ & $	6.5	\pm	0.6	$ &	36.1	$\pm$	3.5	($K_{s}$)	 &	3.8	$\pm$	0.5	&  $	17.0	\pm	3.1	$     & $	82.6	\pm	11.0	$      & $	8.8	\pm	1.2	$ \\
\tableline
NGC4314 A	&~\ref{REG_4314_A}~&	Bar	& $	44.6	\pm	0.25	$ & $	3.8	\pm	0.5	$ &	70.3	$\pm$	17.0	($K_{s}$)	 &	6.0	$\pm$	1.6	&  $	30.6	\pm	10.4	$     & $	59.3	\pm	11.0	$      & $	5.1	\pm	0.9	$ \\
NGC4314 B	&~\ref{REG_4314_B}~&	Bar	& $	30.4	\pm	0.25	$ & $	2.6	\pm	0.3	$ &	70.3	$\pm$	17.0	($K_{s}$)	 &	6.0	$\pm$	1.6	&  $	42.4	\pm	15.5	$     & $	42.8	\pm	8.8	$      & $	3.7	\pm	0.8	$ \\
\tableline
NGC266 A	&~\ref{REG_266_A}~ &	Bar	& $	16.0	\pm	0.14	$ & $	5.0	\pm	0.4	$ &	17.5	$\pm$	1.7	($K^\prime$)	 &	5.5	$\pm$	0.7	&  $	193.1	\pm	452.9	$     & $	17.0	\pm	2.4	$      & $	5.3	\pm	0.8	$ \\
NGC266 B	&~\ref{REG_266_B}~ &	Spiral\tablenotemark{r}	& $	19.6	\pm	0.14	$ & $	6.2	\pm	0.5	$ &	17.5	$\pm$	1.7	($K^\prime$)	 &	5.5	$\pm$	0.7	&  $	158.3	\pm	372.7	$     & $	20.7	\pm	3.3	$      & $	6.5	\pm	1.0	$ \\
NGC266 C	&~\ref{REG_266_C}~ &	Spiral\tablenotemark{r}	& $	15.1	\pm	0.14	$ & $	4.7	\pm	0.4	$ &	17.5	$\pm$	1.7	($K^\prime$)	 &	5.5	$\pm$	0.7	&  $	213.0	\pm	485.6	$     & $	15.4	\pm	1.8	$      & $	4.8	\pm	0.6	$ \\
NGC266 D	&~\ref{REG_266_D}~ &	Spiral	& $	22.9	\pm	0.14	$ & $	7.2	\pm	0.6	$ &	17.5	$\pm$	1.7	($K^\prime$)	 &	5.5	$\pm$	0.7	&  $	158.6	\pm	318.1	$     & $	20.7	\pm	2.7	$      & $	6.5	\pm	0.8	$ \\
\tableline
NGC986 A	&~\ref{REG_986_A}~ &	Bar	& $	26.9	\pm	0.15	$ & $	3.7	\pm	0.3	$ &	50.5	$\pm$	4.6	($K_{s}$)\tablenotemark{a}	 &	6.8	$\pm$	0.9	&  $	11.2	\pm	4.3	$     & $	43.9	\pm	9.2	$      & $	6.0	\pm	1.3	$ \\
NGC986 C	&~\ref{REG_986_C}~ &	Spiral	& $	52.4	\pm	0.15	$ & $	7.1	\pm	0.6	$ &	50.5	$\pm$	4.6	($K_{s}$)\tablenotemark{a}	 &	6.8	$\pm$	0.9	&  $	20.5	\pm	2.5	$     & $	23.8	\pm	2.8	$      & $	3.2	\pm	0.4	$ \\
NGC986 D	&~\ref{REG_986_D}~ &	Spiral	& $	68.9	\pm	0.15	$ & $	9.4	\pm	0.8	$ &	50.5	$\pm$	4.6	($K_{s}$)\tablenotemark{a}	 &	6.8	$\pm$	0.9	&  $	11.4	\pm	1.9	$     & $	42.8	\pm	4.0	$      & $	5.8	\pm	0.5	$ \\
NGC986 E	&~\ref{REG_986_E}~ &	Spiral	& $	83.7	\pm	0.15	$ & $	11.4	\pm	1.0	$ &	50.5	$\pm$	4.6	($K_{s}$)\tablenotemark{a}	 &	6.8	$\pm$	0.9	&  $	10.1	\pm	1.6	$     & $	48.2	\pm	4.7	$      & $	6.5	\pm	0.6	$ \\
\tableline
NGC7496 A	&~\ref{REG_7496_A}~&	Bar	& $	21.5	\pm	0.15	$ & $	2.5	\pm	0.2	$ &	38.0	$\pm$	3.5	($K_{s}$)	 &	4.3	$\pm$	0.5	&  $	39.6	\pm	32.5	$     & $	35.0	\pm	14.4	$      & $	4.0	\pm	1.6	$ \\
NGC7496 B	&~\ref{REG_7496_B}~&	Bar	& $	23.0	\pm	0.15	$ & $	2.6	\pm	0.2	$ &	38.0	$\pm$	3.5	($K_{s}$)	 &	4.3	$\pm$	0.5	&  $	34.6	\pm	30.2	$     & $	40.1	\pm	18.9	$      & $	4.6	\pm	2.2	$ \\
NGC7496 C	&~\ref{REG_7496_C}~&	Bar	& $	33.0	\pm	0.15	$ & $	3.8	\pm	0.3	$ &	38.0	$\pm$	3.5	($K_{s}$)	 &	4.3	$\pm$	0.5	&  $	30.1	\pm	20.2	$     & $	46.0	\pm	11.2	$      & $	5.3	\pm	1.3	$ \\
NGC7496 D	&~\ref{REG_7496_D}~&	Spiral	& $	44.1	\pm	0.15	$ & $	5.0	\pm	0.4	$ &	38.0	$\pm$	3.5	($K_{s}$)	 &	4.3	$\pm$	0.5	&  $	51.4	\pm	13.8	$     & $	27.0	\pm	4.2	$      & $	3.1	\pm	0.5	$ \\
NGC7496 E	&~\ref{REG_7496_E}~&	Spiral	& $	68.4	\pm	0.15	$ & $	7.8	\pm	0.7	$ &	38.0	$\pm$	3.5	($K_{s}$)	 &	4.3	$\pm$	0.5	&  $	31.1	\pm	9.3	$     & $	44.6	\pm	5.9	$      & $	5.1	\pm	0.7	$ \\
\tableline
NGC5383 A	&~\ref{REG_5383_A}~&	Bar	& $	20.1	\pm	0.25	$ & $	3.8	\pm	0.3	$ &	54.2	$\pm$	5.0	($K_{s}$)	 &	10.3	$\pm$	1.3	&  $	65.2	\pm	19	$     & $	21.9	\pm	2.4	$      & $	4.2	\pm	0.5	$ \\
NGC5383 B	&~\ref{REG_5383_B}~&	Bar	& $	27.3	\pm	0.25	$ & $	5.2	\pm	0.5	$ &	54.2	$\pm$	5.0	($K_{s}$)	 &	10.3	$\pm$	1.3	&  $	48.8	\pm	13.6	$     & $	29.2	\pm	3.0	$      & $	5.6	\pm	0.6	$ \\
\tableline
NGC4593 A	&~\ref{REG_4593_A}~&	Bar	& $	34.8	\pm	0.14	$ & $	6.4	\pm	0.6	$ &	48.0	$\pm$	4.4	($K$)		 &	8.9	$\pm$	1.1	&  $	21.8	\pm	6.4	$     & $	59.5	\pm	12.9	$      & $	11.0	\pm	2.4	$ \\
NGC4593 B	&~\ref{REG_4593_B}~&	Bar	& $	36.7	\pm	0.14	$ & $	6.8	\pm	0.6	$ &	48.0	$\pm$	4.4	($K$)		 &	8.9	$\pm$	1.1	&  $	26.8	\pm	5.6	$     & $	48.5	\pm	6.6	$      & $	9.0	\pm	1.2	$ \\
NGC4593 C	&~\ref{REG_4593_C}~&	Spiral\tablenotemark{r}	& $	56.1	\pm	0.14	$ & $	10.4	\pm	0.9	$ &	48.0	$\pm$	4.4	($K$)		 &	8.9	$\pm$	1.1	&  $	10.3	\pm	4.1	$     & $	125.8	\pm	41.4	$      & $	23.3	\pm	7.7	$ \\
NGC4593 D	&~\ref{REG_4593_D}~&	Spiral	& $	72.6	\pm	0.14	$ & $	13.4	\pm	1.2	$ &	48.0	$\pm$	4.4	($K$)		 &	8.9	$\pm$	1.1	&  $	27.0	\pm	2.8	$     & $	48.1	\pm	3.0	$      & $	8.9	\pm	0.6	$ \\
NGC4593 E	&~\ref{REG_4593_E}~&	Spiral	& $	64.1	\pm	0.14	$ & $	11.9	\pm	1.0	$ &	48.0	$\pm$	4.4	($K$)		 &	8.9	$\pm$	1.1	&  $	11.1	\pm	3.3	$     & $	117.0	\pm	25.7	$      & $	21.7	\pm	4.8	$ \\
\tableline
NGC3059 A	&~\ref{REG_3059_A}~&	Bar	& $	12.4	\pm	0.14	$ & $	0.9	\pm	0.1	$ &	19.2	$\pm$	1.4	($H$)\tablenotemark{b}		 &	1.4	$\pm$	0.2	&  $	121.2	\pm	49.8	$     & $	13.9	\pm	1.6	$      & $	1.0	\pm	0.1	$ \\
NGC3059 B	&~\ref{REG_3059_B}~&	Spiral	& $	26.5	\pm	0.14	$ & $	1.9	\pm	0.2	$ &	19.2	$\pm$	1.4	($H$)\tablenotemark{b}		 &	1.4	$\pm$	0.2	&  $	58.6	\pm	23.6	$     & $	28.7	\pm	3.1	$      & $	2.1	\pm	0.2	$ \\
\tableline
NGC7479 A	&~\ref{REG_7479_A}~&	Bar	& $	25.0	\pm	0.25	$ & $	4.2	\pm	0.4	$ &	54.2	$\pm$	5.0	($K_{s}$)	 &	9.0	$\pm$	1.1	&  $	29.5	\pm	8.1	$     & $	50.5	\pm	11.3	$      & $	8.4	\pm	1.9	$ \\
NGC7479 B	&~\ref{REG_7479_B}~&	Bar	& $	47.6	\pm	0.25	$ & $	7.9	\pm	0.7	$ &	54.2	$\pm$	5.0	($K_{s}$)	 &	9.0	$\pm$	1.1	&  $	19.8	\pm	4.3	$     & $	75.2	\pm	12.3	$      & $	12.5	\pm	2.1	$ \\
NGC7479 C	&~\ref{REG_7479_C}~&	Bar	& $	41.4	\pm	0.25	$ & $	6.9	\pm	0.6	$ &	54.2	$\pm$	5.0	($K_{s}$)	 &	9.0	$\pm$	1.1	&  $	21.9	\pm	4.8	$     & $	68.2	\pm	11.6	$      & $	11.4	\pm	1.9	$ \\
NGC7479 D	&~\ref{REG_7479_D}~&	Spiral	& $	80.3	\pm	0.25	$ & $	13.4	\pm	1.1	$ &	54.2	$\pm$	5.0	($K_{s}$)	 &	9.0	$\pm$	1.1	&  $	27.5	\pm	2.2	$     & $	54.2	\pm	3.3	$      & $	9.0	\pm	0.6	$ \\
NGC7479 E	&~\ref{REG_7479_E}~&	Spiral	& $	60.1	\pm	0.25	$ & $	10.0	\pm	0.8	$ &	54.2	$\pm$	5.0	($K_{s}$)	 &	9.0	$\pm$	1.1	&  $	14.4	\pm	3.4	$     & $	103.7	\pm	19.1	$      & $	17.3	\pm	3.2	$ \\
\tableline
NGC3513 A	&~\ref{REG_3513_A}~&	Bar	& $	14.8	\pm	0.14	$ & $	1.1	\pm	0.1	$ &	24.6	$\pm$	2.5	($H$)		 &	1.9	$\pm$	0.3	&  $	45.0	\pm	10.1	$     & $	18.2	\pm	2.2	$      & $	1.4	\pm	0.2	$ \\
NGC3513 B	&~\ref{REG_3513_B}~&	Bar	& $	17.0	\pm	0.14	$ & $	1.3	\pm	0.1	$ &	24.6	$\pm$	2.5	($H$)		 &	1.9	$\pm$	0.3	&  $	39.5	\pm	8.9	$     & $	20.7	\pm	2.5	$      & $	1.6	\pm	0.2	$ \\
NGC3513 C	&~\ref{REG_3513_C}~&	Bar	& $	24.1	\pm	0.14	$ & $	1.9	\pm	0.2	$ &	24.6	$\pm$	2.5	($H$)		 &	1.9	$\pm$	0.3	&  $	28.2	\pm	6.1	$     & $	29.0	\pm	3.4	$      & $	2.3	\pm	0.3	$ \\
NGC3513 D	&~\ref{REG_3513_D}~&	Spiral	& $	20.8	\pm	0.14	$ & $	1.6	\pm	0.1	$ &	24.6	$\pm$	2.5	($H$)		 &	1.9	$\pm$	0.3	&  $	22.9	\pm	7.3	$     & $	35.8	\pm	7.5	$      & $	2.8	\pm	0.6	$ \\
NGC3513 E	&~\ref{REG_3513_E}~&	Spiral	& $	43.0	\pm	0.14	$ & $	3.3	\pm	0.3	$ &	24.6	$\pm$	2.5	($H$)		 &	1.9	$\pm$	0.3	&  $	9.9	\pm	3.7	$     & $	82.7	\pm	19.8	$      & $	6.4	\pm	1.5	$ \\
NGC3513 F	&~\ref{REG_3513_F}~&	Spiral	& $	35.8	\pm	0.14	$ & $	2.8	\pm	0.2	$ &	24.6	$\pm$	2.5	($H$)		 &	1.9	$\pm$	0.3	&  $	14.7	\pm	4.2	$     & $	55.6	\pm	9.8	$      & $	4.3	\pm	0.8	$ \\

\enddata

\tablecomments{
Col.\ (4) and (5). Mean radius of the studied regions, in arcsec and kpc, respectively.
Col.\ (6) and (7). Radius of the bar, in arcsec and kpc, respectively, and bandpass used to determine it.
Col.\ (8) Pattern speed.
Col.\ (9) and (10). Corotation radius, in arcsec and kpc, respectively.
}

\tablenotetext{a}{Obtained from the change of PA with radius of the bar isophotes.}
\tablenotetext{b}{Obtained from visual estimate.}
\tablenotetext{r}{These regions seem to be associated with rings, rather than with spirals.}

\end{deluxetable}

\section{Discussion}

\subsection{Characteristics of observed color gradients}~\label{chara_grads}

In a ``standard color gradient picture'', assuming circular
motions, one would expect an azimuthal sequence starting with compression of
gas, dust lanes, star formation onset and stellar drift, which would lead to color
gradients. When one adds an extended period of
star formation, dispersion velocities and post shock velocities,
the predicted color profiles may become rather broad~\citep[see, e.g.,][]{yua81}.

Although shocks and/or dust lanes are located
mainly upstream relative to the spiral potential minimum, in a more detailed scenario,
for certain models and relatively short radial intervals,
they can be found downstream (in the gas stream direction)
the spiral potential minimum~\citep{git04}. This is due to the fact that pitch angles
of the arms are expected to follow $i_{\mathrm{P}} > i_{\mathrm{D}} > i_{\mathrm{SF}}$,
where $i_{\mathrm{P}}$ is the pitch angle of the potential,
$i_{\mathrm{D}}$ is the pitch angle of the dust lane (the shock), and 
$i_{\mathrm{SF}}$ is the pitch angle of star formation.
According to this, the azimuthal relative location of potential
and dust lanes may change within the same galaxy and be radially dependent.
In principle, the local potential minimum could be determined
from the the $K$-band spiral arm.\footnote{As already mentioned in \S~\ref{sec_analysis}, although young
stellar populations may account for only $3\%$ of the global $K$-band flux \citep{rho98}, they may contribute up
to a third of the 2 $\mu$m emission in local features.}
Nevertheless, gravity is a long-range force,
and all the non-axisymmetric contributions must be taken
into account for local potential minimum determinations.
Thus, the local potential does not necessarily coincide with the local density
of old stars~\citep[see, e.g.,][]{zha96,ber03,git04,zha07,but09},
and the expected locations of the dust lanes must follow
the potential, rather than the $K$-band intensity.

For some of the bar regions presented in figures~\ref{REG_718_A}-\ref{REG_3513_F},
the $t_{\rm age}=0$ location of the gradient seems to be located before
the main dust lane peak. This can be attributed to the fact
that star formation in bars is supposed to begin
at the dust ``spurs'' of the bar itself (see \S~\ref{sec_SF}), which are located in the
``trailing'' side and upstream the main dust lanes.

For this investigation the $d=0$ distance (i.e., the assumed
shock location) is chosen where the maximum in the $g-J$ color is observed.
This is the case, even for regions where double peaks or
dust lanes are seen (e.g., region NGC 864 B, figure~\ref{REG_864_B}).
For most of the fits we do not have $t_{\rm age}=0$ for $d=0$ precisely
(e.g., region NGC 7496 A in figure~\ref{REG_7496_A}).
This is due to the fact that only the width of the model profile in $Q(rJgi)$
is fitted to the observations by stretching. 
Theoretically, for spiral regions
$t_{\rm age}=0$ should coincide or be located after $d=0$. In real galaxies, it is hard
to pinpoint where the onset of star formation really occurs. 
One important reason is the fact that a diffuse cloud of neutral gas has first to become a dense
cloud, then a molecular cloud, and finally a self-gravitating cloud
to achieve star formation; this process may take $∼\sim 10^7$
years~\citep[see, e.g.,][]{tam08,egu09}.

\subsubsection{Downstream decline of the gradients}~\label{down_falls}

\citet{mar09a} already noticed that for some spiral regions
there is a ``downstream decline'' (or  ``downstream fall'') of the gradients.
In such regions, the observed $Q(rJgi)$ profile declines below the model 
(or the value $Q(rJgi)\sim 1.57$, for solar metallicity, in figure~\ref{fig_Models}),
apparently returning to the ``old background population'' level, or lower, by
$t \approx 5 \times 10^{7}$ years, much faster than the theoretical expectations.
This is the case, for example, of regions NGC 864 C, NGC 7496 B \& C,
NGC 7479 B, NGC 3513 A \& C (figures~\ref{REG_864_C},~\ref{REG_7496_B},~\ref{REG_7496_C},
\ref{REG_7479_B},~\ref{REG_3513_A}, and~\ref{REG_3513_C}, respectively),
for which the fit between $Q(rJgi)$ data and {\it{dotted line}} model is not good, except for
the range $0 \lesssim t_{\rm age} \lesssim 50$ Myr.\footnote{Incidentally,
Bruzual \& Charlot models previously to 1997 agreed much better with this rapid decline of
$Q$; see GG96, their figure 19.}

\citet{mar09a} hypothesized that the decline of
the observed $Q(rJgi)$ profiles below the models (that assume pure circular orbits) might be
caused by stellar non-circular motions in the data. However,~\citet{mar09b} found that 
non-circular motions cannot explain the discrepancy.

\begin{figure*}
\centering
\includegraphics[scale=0.80]{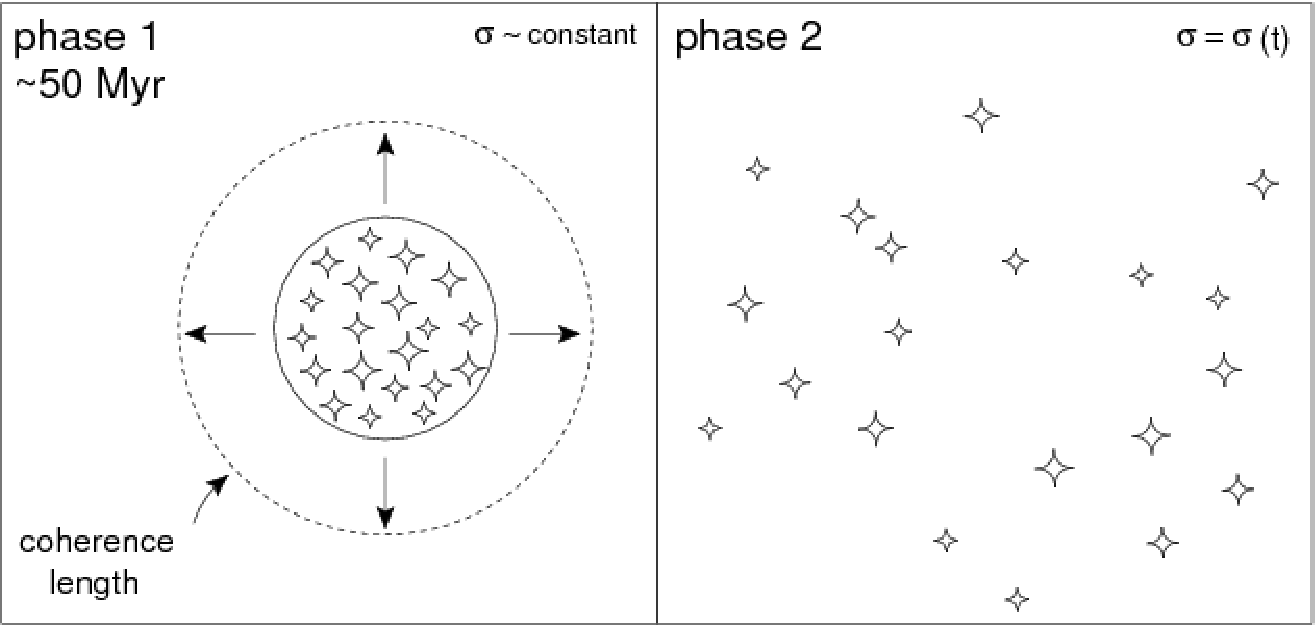}

\caption[f46.eps]{Dissolution of stellar groups scenario~\citep{wie77}.
During phase 1 ($\sim$ 50 Myr), the stellar group increases its diameter due to a $\sim$ constant
internal velocity dispersion, until the ``coherence length'' is reached.
During phase 2, the velocity dispersion increases with time and the ``diffusion
of stellar orbits'' causes the dissipation of the stellar group.
\label{dissol_fig}}
\end{figure*}

A solution to this problem may be provided by the
{\it{dissolution of stellar groups}} scenario proposed by~\citet[][see figure~\ref{dissol_fig}]{wie77}.
According to this author, the diffusion of stellar orbits can enhance
the dissolution of young stellar groups by increasing their internal
velocity dispersion. To explain the observed increase in the velocity
dispersion of stars with age,~\citet{wie77} proposes the existence of
local fluctuations of the gravitational field with a rather stochastic behavior.
This irregular field has the effect of creating a diffusion process in
the velocity space of stars.
The dissolution of stellar groups proceeds
in two phases. During the first phase, the internal velocity
dispersion, $\sigma_{\rm int}$, causes the  group to expand, until its
diameter, $D_{\rm sg}(t_{\rm age})$ is larger than the distance over which simultaneous
perturbations (for stars close together in space) are significantly correlated,
i.e., the coherence length, $L_{\rm co}$.\footnote{This length depends on the mechanism
that causes the irregular gravitational field, with the consequent disk heating.
Possible mechanisms may be provided by giant molecular clouds, spiral arms~\citep{lac84,lac91},
small-scale dark matter clumps~\citep{berez03,bar11},
or massive dark clusters~\citep{san99} in dark matter halos.}
During this first phase, the center of mass
of the group is affected by diffusion mechanisms, while the members of the
group only suffer small tidal effects, thus $\sigma_{\rm int} \sim$ constant.
The duration of the first phase is given by:
\begin{equation}
 t_{\rm phase1}= 0.975 (L_{\rm co}-D_{\rm sg}(0))/\sigma_{\rm int},
\end{equation}

\noindent for $t_{\rm phase1}$ in Myr,
$L_{\rm co}-D_{\rm sg}(0)$ in pc, and $\sigma_{\rm int}$ in km s$^{-1}$.
For typical values of $\sigma_{\rm int}=$10 km s$^{-1}$, and $L_{\rm co}-D_{\rm sg}(0)$=500 pc~\citep{wie77},
we have that:

\begin{displaymath}
 t_{\rm phase1} \sim 50 \rm{Myr}.
\end{displaymath}

During the second phase of the dissolution (after $t_{\rm phase1}$ years),
the perturbations over each star member of the group are not longer correlated,
and the group dissolves with time because of the diffusion mechanism.
During this phase the velocity dispersion increases with time, i.e.,
$\sigma = \sigma(t)$, where $t=t_{\rm age}-t_{\rm phase1}$.
 
In order to evaluate the relevance of the dissolution of stellar groups scenario~\citep{wie77}
for the color gradient picture, we have built a ``toy model'' for $Q(rJgi)$, consisting of two phases.
During the first phase, while $t_{\rm age} < 50$ Myr, the fractions by mass of young, $\beta_{\rm I}= 2 \%$, and old,
$\beta_{\rm II} = 98 \%$, stars are kept constant. 
During the second phase, we assume (1) that the fraction of young stars changes
as $\beta_{\rm I} \propto 1/t^{3}_{\rm age}$, and (2) that the surface
density of young stars behaves in the same way \citep[see eq.~10 in][]{wie77}.

\begin{figure*}
\centering
\includegraphics[scale=0.80]{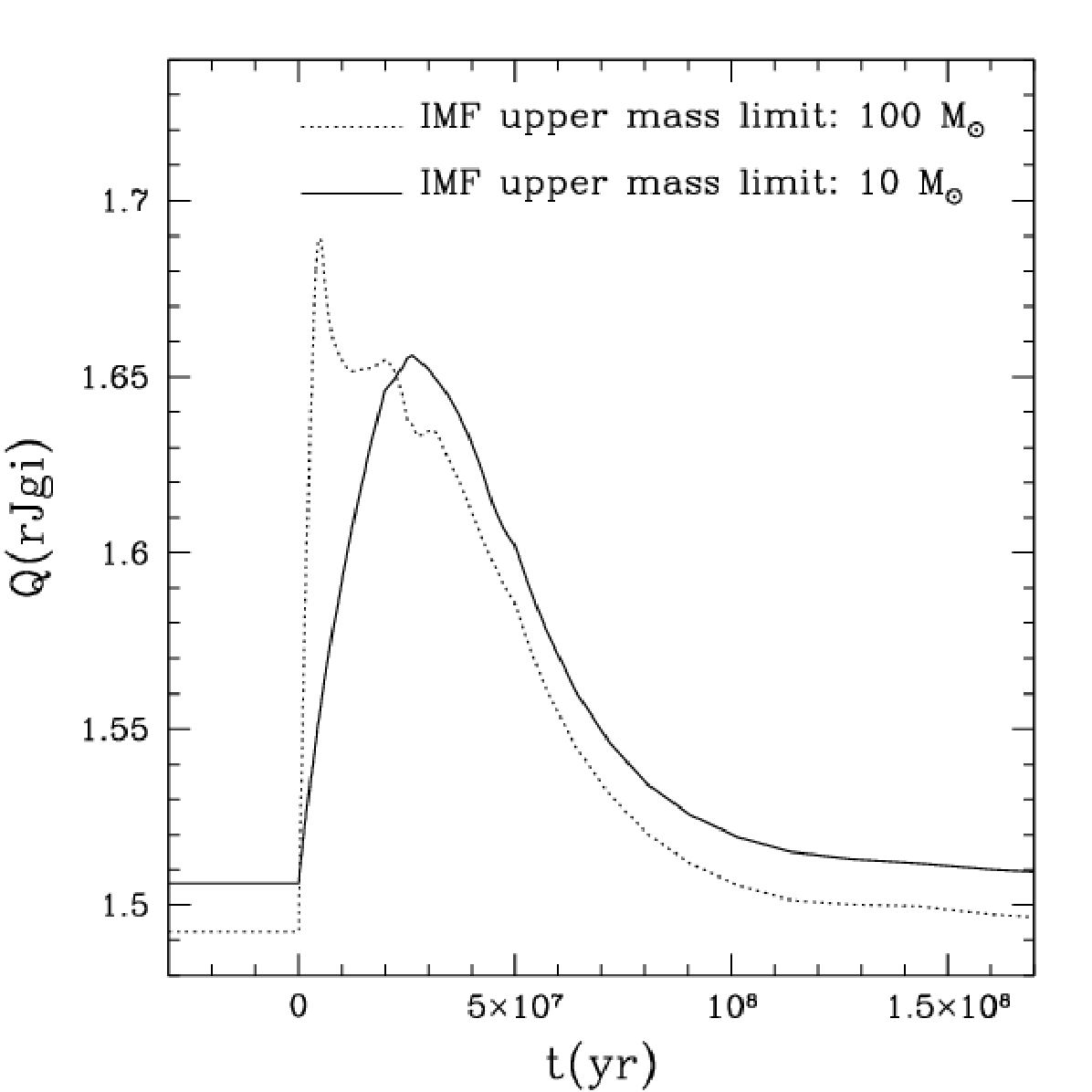}

\caption[f47.eps]{Same as figure~\ref{fig_Models}, including dissolution
of stellar groups~\citep{wie77} after 50 Myr (see \S~\ref{down_falls}).
~\label{fig_ModelsB}}
\end{figure*}

The models produced with this approximation are displayed in figure~\ref{fig_ModelsB},
with a continuous line for models with the IMF $M_{\rm upper} = 10 M_{\sun}$, and a dotted
line for those with $M_{\rm upper} = 100 M_{\sun}$ \citep[see also figures 8 and 9 in][]{mar09a}.
According to figure~\ref{fig_ModelsB}, the dissolution of stellar groups,
as a consequence of disk heating, may provide an efficient mechanism
to explain the observed ``downstream decline'' of the gradients
encountered in~\citet{mar09a}, and in this investigation.
It is important to mention that the dissolution effect operates also in the
presence of non-circular motions.

\vspace{\baselineskip}	
\noindent{\it{Other dissolution scenarios:}}
\vspace{\baselineskip}

Another effective disruption mechanism for star clusters comes from stellar winds
and supernovae explosions that remove gas (and dust) on short timescales. These perturb
the potential and cause young star clusters to become unbound~\citep[see, e.g.,][]{bas06,giel08,lam05}. 
This ``infant mortality'' occurs during the first 10 Myr and affects the most luminous 
clusters (those with the most rapid color evolution). Nevertheless,
stars escape with the initial velocity dispersion of the cluster and
are physically associated with it for 10-40 Myr after gas removal~\citep{bas06}.
The color gradients presented in this investigation are probably formed by 
clusters that survive ``infant mortality'' at least for 50 Myr.

\subsubsection{Color gradients in dusty environments}~\label{Qdust}

One more issue in the ``standard color gradient picture'' relates to the 
$Q(rJgi)$ value expected for the old background population at $t_{\rm age}<0$. 
Also, $Q(rJgi)$ (or any color) should return to the background value on 
both sides of the bar, at the same spatial offsets.
In some of the observed color gradient candidates in this investigation,
the $Q(rJgi)$ index does not agree well with the models for
$t_{\rm age} < 0$ (see, e.g.,
regions NGC 718 C, NGC 986 A, NGC 7479 A; figures~\ref{REG_718_C},~\ref{REG_986_A},
and~\ref{REG_7479_A}, respectively).

Models with pure old background population have an approximately constant $Q(rJgi)$ value,\footnote{
$Q(rJgi)$ only stabilizes after a couple of Gyr. Although we are using a
background population 5 Gyr old, in actuality stars take at most a few hundred Myr
to go from one arm to the next, so that the background will have a contribution 
from several previous bursts between 1e8 and 1e9 yr old, for which 
$Q$ is higher and still slowly declining.}
regardless of surface brightness. However, the exact value
depends on the average age and metallicity of the local region~\citep[see][]{mar09a},
that cannot, especially metallicity, be determined unambiguously just from photometric data. 
A complementary possible explanation of why, for some regions, the $Q(rJgi)$ index does
not decline to the background value for $t_{\rm age} < 0$
is that the adopted method (see~\S~\ref{GG96_method}) involves averaging over radial 
annuli to increase the S/N ratio. Albeit we do so carefully, we may be combining zones 
with somewhat different background levels.

\begin{figure*}
\centering
\includegraphics[scale=0.80]{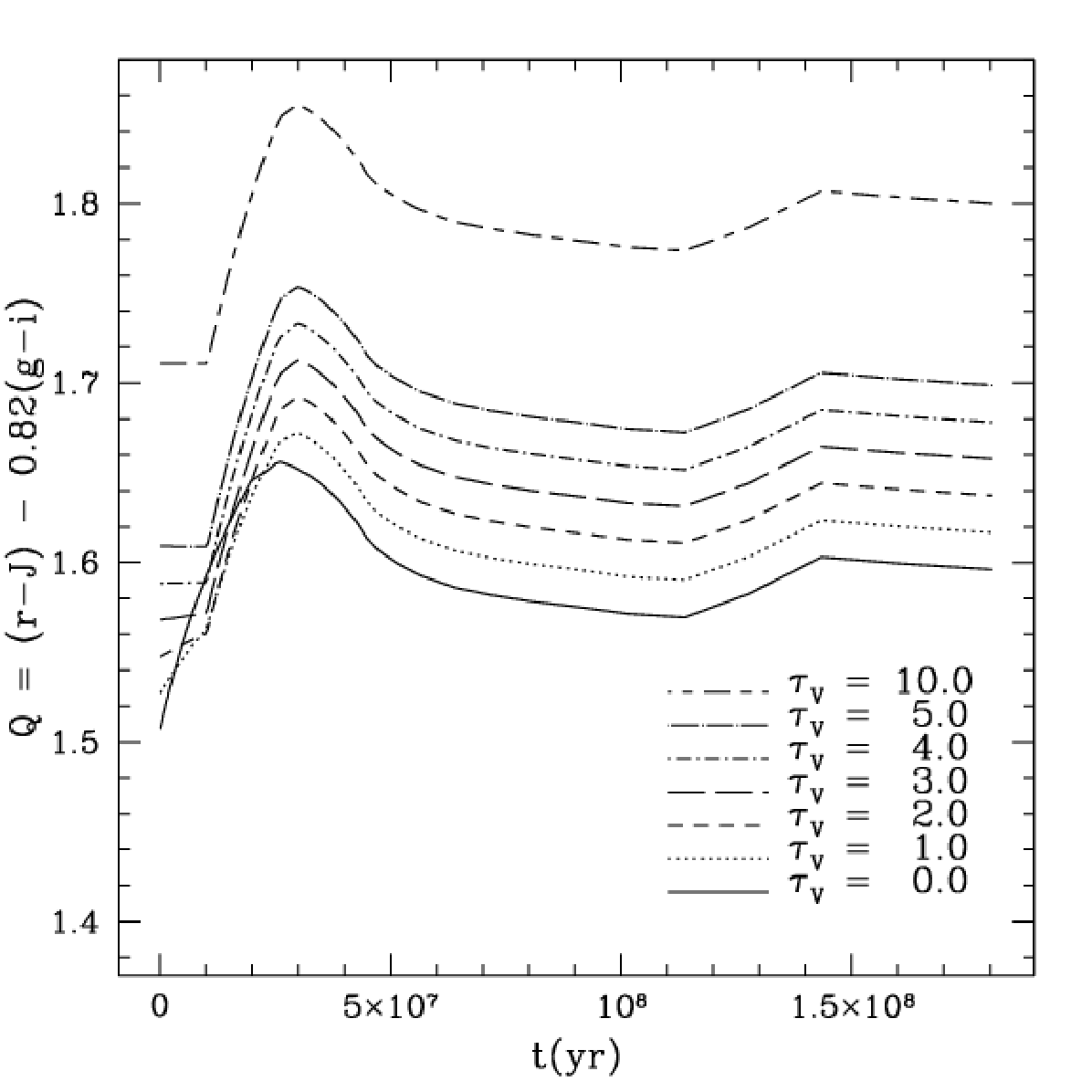}

\caption[f48.eps]{
Index $Q(rJgi)$ vs.\ time for CB07 models,
reddened as per the two-component dust model of~\citet{cha00}. The duration
of the burst is $2 \times 10^7$ yr, with a Salpeter IMF, $2\%$ by mass of young stars,
and solar metallicity.
Lower and upper mass limits are 0.1 and 10 $M_{\sun}$, respectively.
~\label{cfall_model}}
\end{figure*}

Yet another recurring concern is whether dust can mimic the $Q$ profile 
of a star formation burst, since $Q(rJgi)$ is reddening-insensitive for a foreground screen, 
but not exactly so for a mixture of dust and stars \citep{gon96,mar09a}. 
The main expected effect would be a higher $Q(rJgi)$ value (see figure~\ref{cfall_model}).

Assuming $\tau_{V} \sim 1.0$~\citep[e.g.,][]{xil99} for a nearly face-on disk,
given the inclination angles of the galaxy sample, our observations cover 
on average the range $0 < \tau_{V} < 2.0$. 
For regions where $t_{\rm age} < 0$, though, thick dust
lanes may be present with $\tau_{V}~\gg~2.0$. 

From the~\citet[][see also Bruzual \& Charlot 2003]{cha00}
dust model, we get the linear relation $Q(rJgi) = 0.02 \tau_{V} + 1.51$
(where 1.51 is the background $Q(rJgi)$ value; see figure~\ref{cfall_model} for $t_{\rm age} \sim 0$).\footnote{
The mean photometric error in $Q$ (excluding the zero point error) is 
$4\sigma_{Q(rJgi)} < 0.06$ mag.}
Thus, to get a $Q(rJgi)$ profile reminiscent of a density wave induced color
gradient in the absence of young stars, with a peak value $\sim 1.66$, 
the dust must have $\tau_{V} \sim 7.0$. But also, for this scenario to take 
place,  $\tau_{V}$ must increase for $Q(rJgi)$ to increase, and viceversa, i.e., 
the profile in a reddening sensitive index, like ($g - J$), must follow 
the $Q(rJgi)$ azimuthal distribution.

For $t_{\rm age} > 0$,
dust becomes progressively less important, because star formation processes (e.g., UV radiation) 
sweep away and destroy available material. 
Indeed,  {\it concentrations of dust that are higher downstream the shock than
at the shock are not observed in most real spiral arms.}
Fittingly, all of the observed color gradients indicate an 
inverse correlation between $Q(rJgi)$ and $(g - J)$, 
so that dust cannot be mimicking a star formation burst.\footnote{
In this regard, \citet{gon96b}~found that the gradient in M99 also follows the
models when mapped in colors sensitive to dust, such as $(g-i)$, $(g-K)$,
and $(g-J)$. This would not be the case if the profiles were contaminated by dust.}
Although the situation may be different for bar regions,
we do not observe a correlation between the dust and $Q(rJgi)$ profiles 
in any of the bar gradient candidates analyzed in this investigation.

Finally, because of the connection between the gradients (i.e., star formation) 
and disk dynamics (i.e., the orbital resonance positions; see~\S~\ref{dyn-conn}), we argue
that dust features are not affecting the analyzed gradients in any important way.

\subsection{Connection with dynamics}~\label{dyn-conn}

\begin{figure*}
\centering
\includegraphics[scale=0.80]{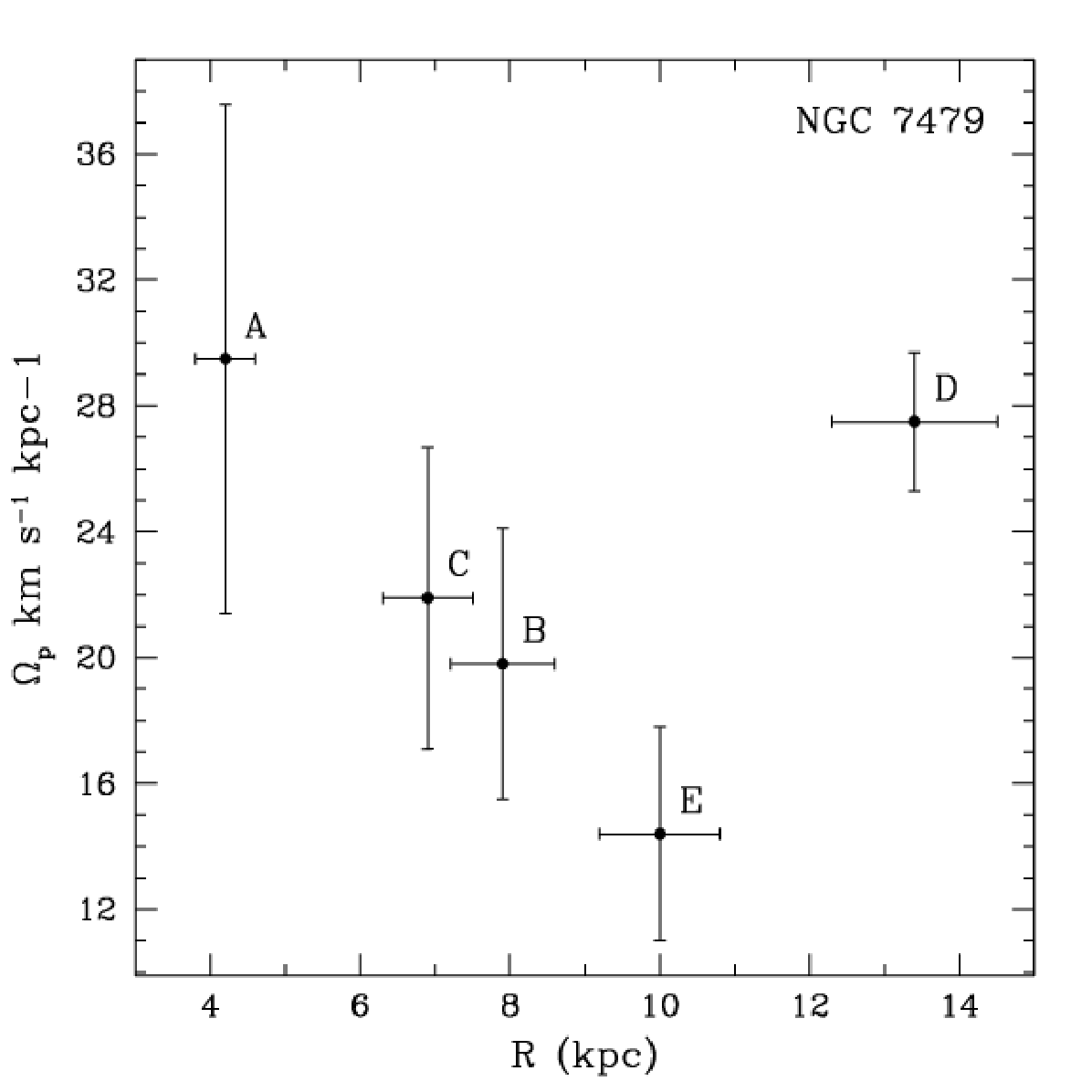}

\caption[f49.eps]{Pattern speed, $\Omega_{\rm p}$, for NGC~7479, obtained from
the comparison between
observations of color gradient candidates and 
stellar population synthesis models; the galactocentric
radii of the color gradients are indicated in the x-axis. 
Regions A, B, and C belong to the bar, while regions D and E are located
in the spiral arms. 
\label{OMvsR_N7479}}
\end{figure*}

\begin{figure*}
\centering
\includegraphics[scale=0.80]{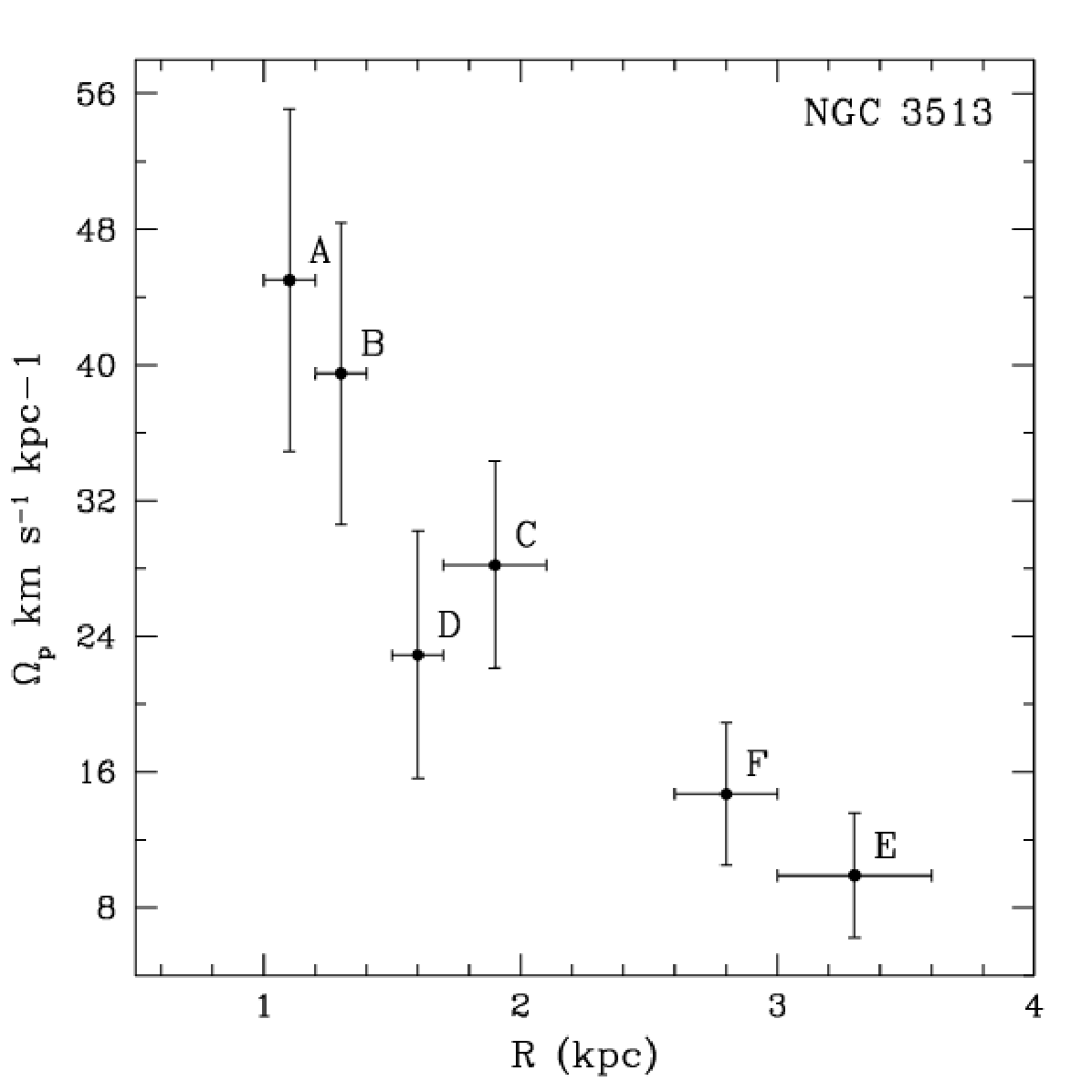}

\caption[f50.eps]{Pattern speed, $\Omega_{\rm p}$, for NGC~3513, obtained from 
the comparison between
observations of color gradient candidates and 
stellar population synthesis models; the galactocentric
radii of the color gradients are indicated in the x-axis. 
Regions A, B, and C belong to the bar, while regions D, E, and F are located
in the spiral arms. 
\label{OMvsR_N3513}}
\end{figure*}

In figure~\ref{OMvsR_N7479}, the pattern speed obtained from each region
in NGC 7479 is plotted vs.\ the mean radius of the region. Notice the tendency to derive 
slower $\Omega_{p}^{\rm bar}$ from bar regions (A, B, C) at larger radius.
Likewise, figure~\ref{OMvsR_N3513} shows the $\Omega_{p}^{\rm bar}$ obtained for 
NGC 3513, vs.\ the radius of each region. As in plot~\ref{OMvsR_N7479} (NGC 7479),
a trend whereby $\Omega_{p}^{\rm bar} \propto r^{-1}$ is observed for the bar regions (A, B, and C).
In~\citet{mar09b}, we found that this effect is observed in non-barred or weakly
barred galaxies, if non-circular motions
are present but color gradient data are analyzed assuming stars have purely
circular orbits.
In the case of barred galaxies, non-circular motions in the bar cause 
$\Omega_{p}^{\rm bar}$ 
to be overestimated inside the bar CR radius; once more, the size of the
discrepancy between the real and the measured pattern speed increases 
as the radius decreases.\footnote{Again, to explain a ``thinner" gradient
the observer has to invoke a smaller difference between the orbital and
the pattern speeds; since orbital speed increases inward, so must the 
measured $\Omega_{p}^{\rm bar}$.}  This signature does not mean that
$\Omega_{p}^{\rm bar}$ is changing with radius, in either kind of
galaxy.\footnote{If it were, it would not be a pattern speed!}

\begin{figure*}
\centering
\includegraphics[scale=0.80]{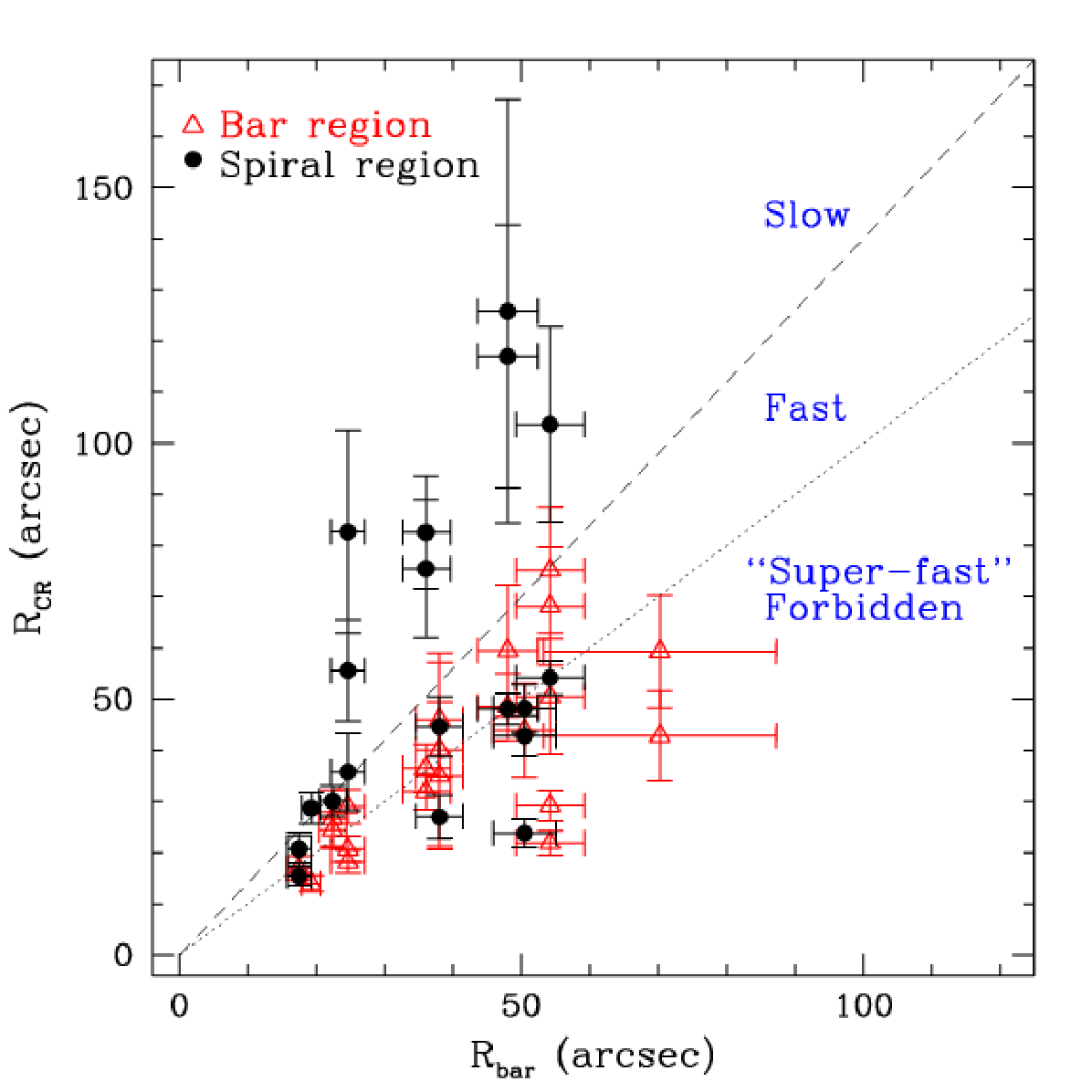}

\caption[f51.eps]{$R_{\rm {CR}}$ vs.\ $R_{\rm {bar}}$.
{\it Red open triangles:} bar regions; {\it black solid circles:} spiral regions.
{\it Dotted line:} $R_{\rm {CR}}=R_{\rm {bar}}$; 
{\it dashed line:} $R_{\rm {CR}}=1.4R_{\rm {bar}}$. Lines separate 
zones inhabited by ``slow", ``fast" and ``forbidden"
~\citep[or ``super-fast'',][]{but09} bars~\citep{ague03}.~\label{Rbar_vs_CR}}
\end{figure*}

Given that many of the computed resonance positions seem to be in accordance
with theoretical predictions (e.g., bars ending near their CR radius),
most of the analyzed color gradient candidates
may indeed have a relation with the dynamics of the disk. 
Figure~\ref{Rbar_vs_CR} shows a plot of bar extent, $R_{\rm {bar}}$ (see \S~\ref{sec_Bar_end}),
vs.\ bar CR radius, $R_{\rm {CR}}$, as inferred from the comparison between 
stellar population models,
and broad-band optical and NIR observations.
Red open triangles denote bar regions, whereas black
solid circles indicate color gradient candidates in the arms.
The plot is divided in three zones, i.e., the ``slow'', ``fast'',
and ``forbidden" bar areas; the latter corresponds to the ``super-fast'' bars of~\citet{but09}. 
The dotted line, $R_{\rm {CR}}=R_{\rm {bar}}$, and the dashed line, $R_{\rm {CR}}= 1.4 R_{\rm {bar}}$,
enclose the parameter space where most of the points should be if bars end near CR~\citep{ague03}.
If the spiral pattern speed, $\Omega_{p}^{\rm {spiral}}$, is similar to the bar's pattern speed, $\Omega_{p}^{\rm {bar}}$, then  
spiral region points should fall in this area, too. 
Unfortunately, with the GG96 method it is difficult to distinguish between ``super-fast'' bars
and the expected overestimation of $\Omega_{p}^{\rm {bar}}$, owing to non-circular motions~\citep{mar09b}.
All the points below the $R_{\rm {CR}}=R_{\rm {bar}}$ (dotted) line are more likely due to 
the latter effect. In order 
to discriminate between these two possibilities, we follow~\citet{mar09b} and
define: 

\begin{equation}
\delta \Omega_{p} =  \frac{\Omega_{\mathrm{data}}}{\Omega_{p}'} - 1.
\end{equation}

\noindent Here, $\Omega_{\mathrm{data}}$ is the pattern speed value
obtained with the GG96 method (either for the bar or the spiral),
and $\Omega_{p}'$ is the pattern speed of the (bar of spiral) perturbation
derived from the rotation curve once a resonance position is fixed. 

\begin{figure*}
\centering
\includegraphics[scale=0.80]{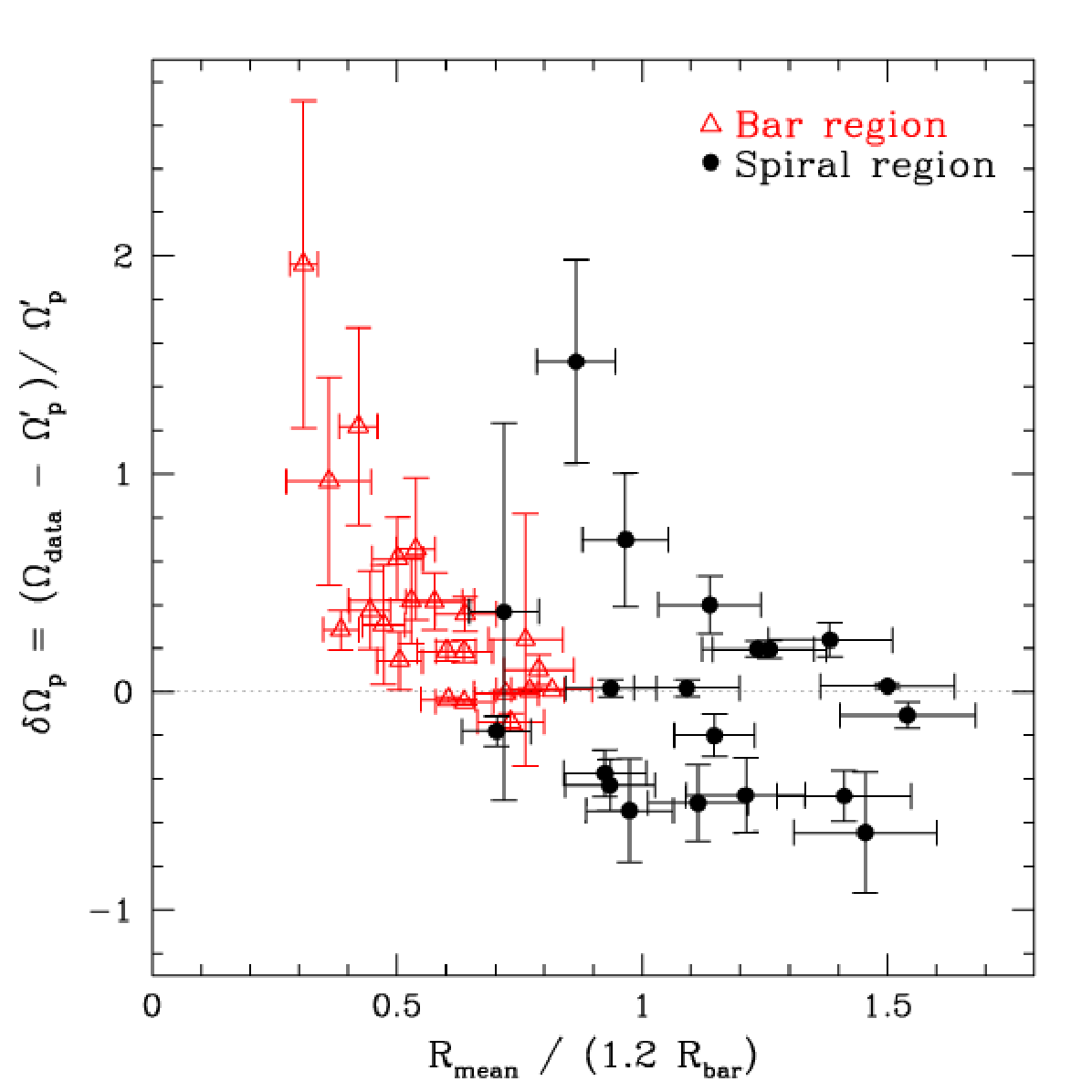}

\caption[f52.eps]{$\delta \Omega_{p}$ (see text) vs.\ 
$R_{\rm{mean}}/ 1.2 R_{\rm {bar}}$. 
{\it Red open triangles:} bar regions; {\it black solid circles:} spiral regions.
{\it Dotted line:}  $\delta \Omega_{p}=0$, i.e.,
the pattern speed inferred from the color gradient candidates, $\Omega_{\mathrm{data}}$,
equals the pattern speed derived from a flat rotation curve, if the bar 
CR is located at 1.2 $R_{\rm {bar}}$ (the bar end point).~\label{SB_grads}}
\end{figure*}

For barred spirals, in the case where bar and arms share
the same pattern speed (and thus the same CR radius), we have:

\begin{equation}
 \Omega_{p}' \sim \frac{v_{\mathrm{rot}}}{1.2 R_{\rm {bar}}},
\end{equation}

\noindent where we adopt $\frac{R_{\rm {CR}}}{R_{\rm {bar}}} \sim 1.2$,
the expected ratio for spirals with well defined bars~\citep[see, e.g.,][]{ath92,elm96,but09}.
In figure~\ref{SB_grads}, we show $\delta \Omega_{p}$ vs. $R_{\rm{mean}}/1.2R_{\rm {bar}}$;
this ratio is $\sim 1$ for regions near CR. Once again, bar regions are
shown with open red triangles to distinguish them from spiral regions (solid black circles). For the
bar regions, there is a systematic trend, whereby $\Omega_{p}^{\rm {bar}}$ is overestimated 
at small radii, the magnitude of the bias is inversely correlated with radius, and
the points converge to $\delta \Omega_{p} = 0$ as they approach the corotation zone.
This is the expected effect when non-circular motions are present, but
color gradients are interpreted with a dynamic model that considers purely  
circular motions~\citep{mar09b}. However, 
as a consequence of data processing that was already discerned by these authors,  
regions in the outer half of the bar yield $\Omega_{p}^{\rm {bar}}$ values
with less than 50\% error. The detectable trend, on the other hand, confirms
the link of the gradients to disk dynamics.

In figure~\ref{SB_inv_grads}, we plot $\delta \Omega_{p}$ for the spiral
arm regions vs.\ region radius, in the same units as figure~\ref{SB_grads}; we 
highlight those with ``inverse'' color gradients (i.e., the sense of
rotation is opposite to the observed aging of stars) with open red triangles.
Most of these regions lie over the dotted line, where the pattern speed obtained 
from the color gradient candidates, $\Omega_{\rm data}$, equals the pattern speed
derived from a flat rotation curve, if the bar CR is located at 1.2 $R_{\rm bar}$.
Conversely, regions where stars age in the direction of disk rotation (solid black
circles) sit below the dotted line. If such ``inverse" color gradients occur 
in objects where $\Omega_{p}^{\rm {bar}} \approx \Omega_{p}^{\rm {spiral}}$, 
then this plot may indicate the presence of non-circular motions close to
corotation, that cause the overestimation of the pattern speed with our method.

On the other hand, the existence of gradients where stellar aging follows
disk rotation, {\em at radii beyond the bar CR radius}, may be interpreted as
$\Omega_{p}^{\rm {bar}} \neq \Omega_{p}^{\rm {spiral}}$,
i.e., decoupled pattern speeds for the bar and the spiral; 
the CR radius of the spiral pattern would be larger than that of the bar.\footnote{
Two of these regions (NGC 266 B \& C) seem, at least visually, more associated with
rings rather than with spiral arms.}

\subsection{The origin of spiral arms.}~\label{spiral_origin}

The origin of spiral arms may be different in barred galaxies 
than in non-barred or weakly-barred spirals.
An alternative theory to density waves
propounds that manifold-driven chaotic orbits are the foundation of spirals and rings 
(nuclear rings excluded) in barred galaxies~\citep{pat06,rom06,rom07,atha09a,atha09b,
atha10,vog06a,vog06b,tso08,tso09,har09}.
In the view of~\citet{rom06,rom07} and~\citet{atha09a,atha09b,atha10},
invariant manifolds are ``tubes"
that guide orbits; they are associated with unstable Lagrangian points at corotation.
Material is trapped in the manifolds during disk evolution,
and circulates across the disk, inducing radial mixing~\citep{atha10}.
Another interpretation of the ``invariant manifold theory''
considers only apsidal (apocentric or pericentric) sections of unstable
manifolds~\citep{vog06a,vog06b,tso08,tso09}. In this latter version, there is no need
for constantly supplying material inside the manifolds~\citep{eft10}.

One important prediction of the ``manifold theories'' (both views) is that
the spiral arms and the bar must have the same pattern speed.
According to the age gradients found in the spiral arms of our sample,
the objects NGC~718, NGC~266, NGC~986, NGC~7496, NGC~4593, and NGC~7479 
are the most likely to have $\Omega_{p}^{\rm {bar}} \approx \Omega_{p}^{\rm {spiral}}$. 
Conversely, NGC~864, NGC~3059, and NGC~3513 appear to have different spiral 
and bar pattern speeds. No conclusion can be drawn for NGC~4314 and NGC~5383, 
since we were able to detect gradients only in the bars of these objects.
Quite interestingly, with the exception of NGC~7479, all objects in our sample 
with $\Omega_{p}^{\rm {bar}} \approx \Omega_{p}^{\rm {spiral}}$
are early Hubble types, while the objects with
$\Omega_{p}^{\rm {bar}} \neq \Omega_{p}^{\rm {spiral}}$ are late
Hubble types.

Both the density wave and the ``manifold" theories predict
a spiral potential that can produce shocks in the circulating
gas. If star formation is triggered by these shocks, it would be difficult
to discriminate between the two theories on the basis of the presence  
of color gradients alone. But, at least in the three objects in our
sample with uncoupled bar and spiral 
pattern speeds, the origin of the arms and/or their subsequent evolution
may follow different paths from the ones proposed by the ``manifold'' theory.

\begin{figure*}
\centering
\includegraphics[scale=0.80]{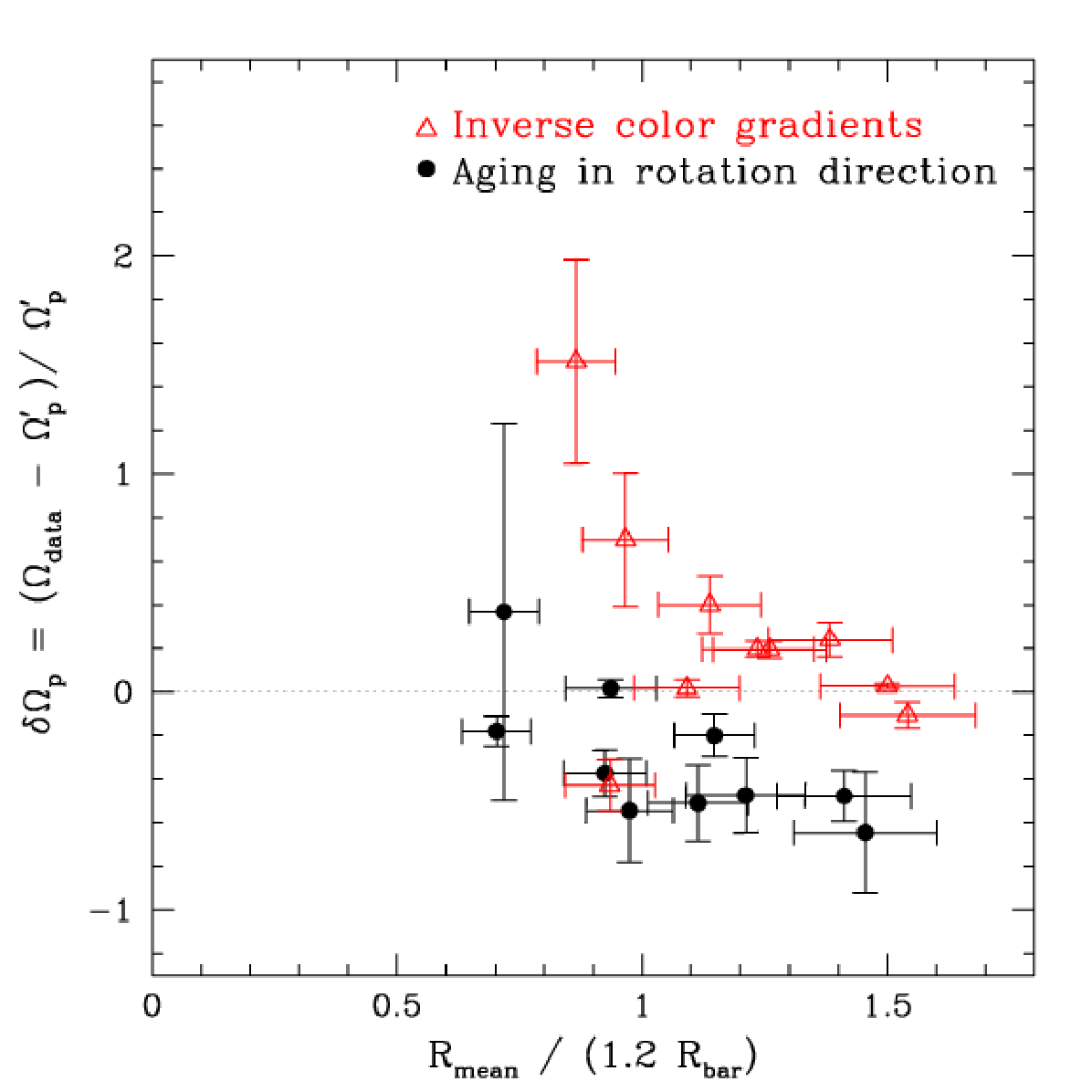}

\caption[f53.eps]{Same as figure~\ref{SB_grads}, for the spiral regions only. 
{\it Red open triangles:} ``inverse" color gradients; 
{\it black solid circles:} stellar aging and disk rotation have the same direction.
\label{SB_inv_grads}}
\end{figure*}

Concerning a different but related aspect, 
the observed trends for color gradients with radius
in figure~\ref{SB_inv_grads}
suggest that non-circular motions are important near the 
bar corotation in barred galaxies, and that their effect 
on the determination of the spiral pattern speed are significantly
stronger for inverse color gradients, i.e., when $\Omega_{p}^{\rm {bar}} 
\approx \Omega_{p}^{\rm {spiral}}$.\footnote{For inverse color gradients,
$\delta\Omega_{\rm p} = 1.55 (1.2 R_{\rm bar}/R_{\rm mean}) - 1.07 $
(pearson correlation coefficient, p = 0.51); 
for gradients with aging in the rotation direction (i.e., decoupled 
pattern speeds), $\delta\Omega_{\rm p} = 0.89 (1.2 R_{\rm bar}/R_{\rm mean}) - 1.19$ (p = 0.73).} 
The presence of non-circular motions is not surprising, given the
existence of the bar. If, however, the size of $\delta\Omega_{\rm p}$ 
in barred galaxies correlates with shock strength,   
the amplitudes of coupled spiral arms should also be stronger near CR
~\citep[see, e.g., NGC 1566 spiral Fourier amplitudes in][]{sal10}.
Strong shocks are neither expected nor observed \citep[e.g.,][]{mar09b} near the
pattern CR radius of non-barred or weakly-barred spiral galaxies.

After a re-analysis of the result in~\citet{butet09},
who find only a weak indication that some strong\footnote{Strong bars
or spirals may be defined by comparing the non-axisymmetric gravitational
perturbation (that induces a tangential force)
to the mean axisymmetric radial force~\citep[e.g.,][]{blo02}.}
bars may drive strong spirals,
\citet{sal10} conclude, contrariwise, that in a statistical sense spiral density waves may 
indeed be driven by bars.
More comparisons with observations~\citep[see, e.g.,][]{grou10}
are needed to test both the ``manifold'' and the density wave theories~\citep{atha10}.

\begin{figure*}
\centering
\includegraphics[scale=0.80]{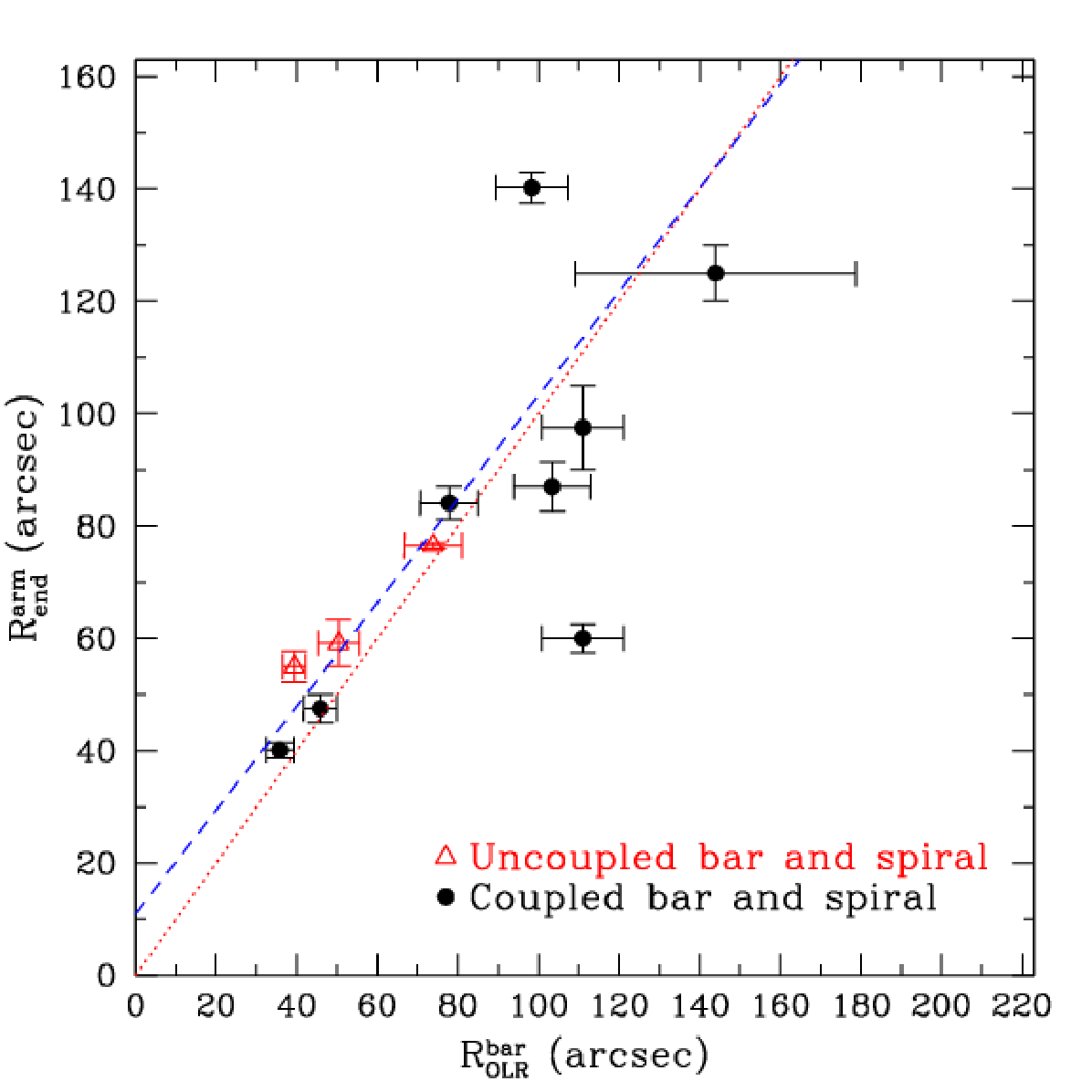}

\caption[f54.eps]{
Mean arm maximum extent, $R^{\rm arm}_{\rm end}$, obtained by eye from NIR data, 
vs.\ bar OLR, $R^{\rm bar}_{\rm OLR}$, assuming $R_{\rm CR} = 1.2 R_{\rm bar}$.
{\it Red open triangles:} $\Omega^{\rm bar}_p \neq \Omega^{\rm arm}_p$;
{\it black solid circles:} $\Omega^{\rm bar}_p \approx \Omega^{\rm arm}_p$.
{\it Dotted line:}
$R^{\rm arm}_{\rm end} = R^{\rm bar}_{\rm OLR}$; {\it dashed line:} OLS (weighted by errors) bisector line.
\label{OLR_graph}}
\end{figure*}

To confirm the link between the gradients and the
disk dynamics, it is also important to compare the location of
the spiral ``end points''\footnote{Or the maximum radial extent of the arms,
since spirals may fall back towards smaller radii~\citep{atha09b,atha10}.}
with the orbital resonance positions. \citet{sch85} shows that bar-induced gas spiral arms
may reach beyond the outer Lindblad resonance (OLR) of the bar.
In figure~\ref{OLR_graph} we plot the OLR of the bar, $R^{\rm bar}_{\rm OLR}$, assuming

\begin{equation}
   R^{\rm bar}_{\rm OLR} \sim 1.2 R_{\rm bar} \left(1+\frac{\sqrt{2}}{2}\right),
\end{equation}

\noindent vs.\ the spiral extent, $R_{\rm end}^{\rm arm}$, estimated by eye in the NIR, 
for all objects\footnote{This plot includes NGC 4314 and NGC 5383, although no gradients were found
in the arms of these objects.} in our sample (arms with decoupled pattern speeds 
are shown as red open triangles). $R_{\rm end}^{\rm arm}$ values are listed in
table~\ref{Spi_end}.

The dotted line indicates the identity, 
$R_{\rm end}^{\rm arm} =  R^{\rm bar}_{\rm OLR}$, 
while the dashed line is the OLS (ordinary least squares) bisector.\footnote{The bisector line 
was obtained by first
fitting the OLS(Y$\mid$X) and OLS(X$\mid$Y), weighted by the errors~\citep{bev03}, and then applying
\citet{iso90} formula for the OLS bisector slope.}
On average, for our whole barred spiral galaxy sample,
the radial extent of the spiral arms coincides with the OLR of the bar,
regardless of the pattern speeds. Interestingly, a separate fit to the 3 decoupled spirals
yields a bad match to all the end point locations expected from theory, that is,
to the bar OLR, the arm CR, and the arm OLR. Since 2 of the 
objects (NGC~864 and NGC~3513), though, 
are consistent with ending at the arm CR, more data are needed to better understand
the dynamics of spirals decoupled from the bar.

\begin{deluxetable}{lc}
\tabletypesize{\scriptsize}
\tablecaption{Spiral maximum extents~\label{Spi_end}}
\tablewidth{0pt}
\tablehead{
\colhead{Galaxy} & \colhead{$R^{\rm arm}_{\rm end}$ (arcsec)}
}
\startdata

NGC          718     &    47.5    $\pm$      2.5 \\
NGC          864     &    76.5    $\pm$      0.5 \\
NGC         4314     &    125.0   $\pm$      5.0 \\
NGC          266     &    40.1    $\pm$      1.4 \\
NGC          986     &    87.0    $\pm$      4.4 \\
NGC         7496     &    84.1    $\pm$      2.9 \\
NGC         5383     &    60.0    $\pm$      2.5 \\
NGC         4593     &    140.2   $\pm$      2.8 \\
NGC         3059     &    55.0    $\pm$      2.8 \\
NGC         7479     &    97.5    $\pm$      7.5 \\
NGC         3513     &    59.1    $\pm$      4.1 \\

\enddata
\end{deluxetable}

\section{Conclusions}

Our results show that a connection exists between bar/spiral dynamics and star formation.
We have found indications of the existence of azimuthal color (age) gradients across the bars
and spirals of disk galaxies (although different mechanisms of star formation triggering
may take place in both types of regions, see \S~\ref{sec_SF}).
Through the comparison of optical and NIR images with stellar population
synthesis models, a link
can be established between large-scale star formation in
the disks and bar/spiral dynamics. 

For the bar regions, we compare the
computed CR positions with the bar's end, and with results from
other authors, both theoretical and observational.
The calculated CR radii for the bar pattern speeds are 
close to the bars' end points, in agreement with theoretical expectations.
The analysis of azimuthal color (age) gradients shows that non-circular motions 
are important. In the case of bar regions, the use of a circular 
dynamic model produces a trend
to overestimate $\Omega_{p}^{\rm {bar}}$ for regions inside CR.
This trend is similar to the one encountered in non-barred and weakly-barred spirals and,
as already demonstrated by \citet{mar09b}, does not imply the absence of a pattern speed. 

For regions in the spiral arms of barred galaxies, we find that
``inverse'' color gradients (10 of 20) also follow a trend that can be attributed  
to non-circular motions. In this case, though, the overestimation of $\Omega_{p}^{\rm {spiral}}$
occurs near the CR radius of the bar and converges to zero at higher radii.
We also find gradients in the spiral arms where stellar aging follows the direction of rotation.
The  $\Omega_{p}^{\rm {spiral}}$ values derived from these regions are in general
lower than $\Omega_{p}^{\rm {bar}}$.

Out of 9 galaxies with detected gradients in both the bar and the arms,
6 appear to have $\Omega^{\rm bar}_p \approx \Omega^{\rm spiral}_p$; with
one exception, these are all galaxies with early Hubble types.
The remaining 3 galaxies are late Hubble types, and 
appear to have $\Omega^{\rm bar}_p \neq \Omega^{\rm spiral}_p$.

From the presence of azimuthal color (age) gradients alone,
it is difficult to discern between modern theories of the origin of spiral arms
in barred galaxies; the pattern speeds that we can obtain based on
the gradients, however, can provide significant clues.  

\acknowledgments
We are grateful to the anonymous referee for many important
remarks and helpful comments that have greatly improved this paper. 
E.~E. Mart\'inez-Garc\'ia acknowledges postdoctoral financial support from DGAPA (UNAM) and
from CIDA (Centro de Investigaciones de Astronom\'ia) in M\'erida, Venezuela during the earlier stages of this investigation.
We acknowledge the use of the HyperLeda database (http://leda.univ-lyon1.fr).
RAGL thanks DGAPA (UNAM) grant IN118110.




\clearpage

\clearpage

\clearpage

\clearpage

\clearpage

\clearpage

\clearpage

\clearpage

\clearpage

\clearpage

\end{document}